\documentclass[a4paper,11pt]{article}
\pdfoutput=1
\usepackage{jcappub} 
\usepackage{amssymb}
\usepackage{amsmath}
\usepackage{bm}
\usepackage[T1]{fontenc}
\usepackage{natbib}
\usepackage{xcolor}
\usepackage{ctable}
\usepackage[hypcap=false]{caption} 
\usepackage{subcaption}
\usepackage{hyperref}
\usepackage{xspace}
\usepackage{lineno}
\usepackage{makecell}
\usepackage{multirow}
\usepackage{ctable}
\usepackage{float}
\usepackage{anyfontsize}
\usepackage{bm}
\usepackage[compact]{titlesec}
\usepackage{cleveref}
\usepackage{newtxtext,newtxmath}
\usepackage{graphicx}	
\usepackage{aas_macros}
\usepackage{tabularx}
\newcommand{\myfigure}{Fig.}
\newcommand{\myfigures}{Figs.}
\newcommand{\myequation}{Eqn.}

\newcommand{\myappendix}{Appendix}
\newcommand{\mysection}{\S}

\newcommand{\paramonewidthfactor}{0.21}
\newcommand{\paramfivewidthfactor}{0.35}
\newcommand{\paramtwelvewidthfactor}{0.75}

\newcommand{\deeplenstronomy}{\texttt{deeplenstronomy}}
\newcommand{\sbi}{\texttt{sbi}}
\newcommand{\lenstronomy}{\texttt{lenstronomy}}
\newcommand{\einsteinrad}{$\theta_{\mathrm{E}}$}
\DeclareMathOperator\arctanh{arctanh}
\newcommand{\eccentricity}[2]{#1_{e#2}}
\newcommand{\tableembeddingnetnpe}{
\begin{table}
  \centering
  \noindent\begin{minipage}[b]{\columnwidth}
  \centering
    \caption{
    The architecture of the embedding network in the NPE model. 
    The first column lists the layer type, the second lists the dimensionality of the output from that layer, and the third column lists the parameters of that layer: $k$ is the convolution kernel size, $s$ is the stride, and $P$ is the number of physical model parameters --- with $P = 1,\, 5,\, 12$ for the 1-, 5-, and 12-parameter problems, respectively.
    The final layer outputs the summary features (4*$P$). 
    See \S\ref{sec:npe_implementation} for a detailed description of the design of this embedding network.
   }
  \label{table:embedding_net}
  \centering
  \begin{tabular}{l l c}
 \hline   Layer   &  Output shape   &  Parameters \\\hline \hline
  Conv2d    &  [-1, 8, 32, 32]  &  $k=3$, $s=1$\\ 
  \midrule
  BatchNorm2d & [-1, 8, 32, 32] & $k=3$, $s=1$\\ 
  \midrule
  Conv2d & [-1, 16, 32, 32] & $k=3$, $s=1$ \\
  \midrule
  BatchNorm2d & [-1, 16, 32, 32] & $k=3$, $s=1$\\ 
  \midrule
  MaxPool2d & [-1, 16, 16, 16] & $k=2$, $s=2$  \\
  \midrule
  Conv2d    &  [-1, 32, 16, 16]  &  $k=3$, $s=1$\\ 
  \midrule
  BatchNorm2d & [-1, 32, 16, 16] & $k=3$, $s=1$\\ 
  \midrule
  Conv2d & [-1, 32, 16, 16] & $k=3$, $s=1$\\
  \midrule
  BatchNorm2d & [-1, 32, 16, 16] & $k=3$, $s=1$\\ 
  \midrule
  MaxPool2d & [-1, 32, 8, 8] & $k=2$, $s=2$ \\
  \midrule
  Conv2d    &  [-1, 64, 8, 8]  &  $k=3$, $s=1$ \\ 
  \midrule
  BatchNorm2d & [-1, 64, 8, 8] & $k=3$, $s=1$\\ 
  \midrule
  Conv2d & [-1, 128, 8, 8] & $k=3$, $s=1$\\
  \midrule
  BatchNorm2d & [-1, 128, 8, 8] & $k=3$, $s=1$\\ 
  \midrule
  MaxPool2d & [-1, 128, 4, 4] & $k=2$, $s=2$ \\
  \midrule
  Flatten & [-1, 2048] & - \\
  \midrule
  Linear & [-1, 4*$P$] & - \\
  \hline
\end{tabular}
\end{minipage}
\end{table}
}

\newcommand{\tableembeddingnetbnn}{
\begin{table}
  \centering
  \noindent\begin{minipage}[b]{0.99\columnwidth}
  \centering
    \caption{
    The BNN architecture for the 5- and 12-parameter models. 
    The first column lists the layer type, the second lists the dimensionality of the output from that layer, and the third column lists the parameters of that layer: $k$ is the convolution kernel size, $s$ is the stride, and $P$ output normal distributions for $P$ parameters.
    See \S\ref{sec:bnn} for a detailed description of the design of this network. 
    }
  \label{table:BNN}
  \centering
  \begin{tabular}{llc}
 \hline   Layer   &  Output shape   &  Parameters \\\hline \hline
  Conv2dFlipout &  [-1, 16, 32, 32]  &  $k=3$, $s=1$\\ 
  \midrule
  MaxPool2d & [-1, 16, 16, 16] & $k=2$, $s=2$\\ 
  \midrule
  Conv2dFlipout &  [-1, 32, 16, 16]  &  $k=3$, $s=1$ \\
  \midrule
  Conv2dFlipout &  [-1, 32, 16, 16]  &  $k=3$, $s=1$\\ 
  \midrule
  MaxPool2d & [-1, 32, 8, 8] & k=2, $s=2$ \\
  \midrule
  Conv2dFlipout &  [-1, 48, 8, 8]  &  $k=3$, $s=1$\\ 
  \midrule
  Conv2dFlipout &  [-1, 48, 8, 8]  &  $k=3$, $s=1$\\ 
  \midrule
  MaxPool2d & [-1, 48, 4, 4] & k=2, $s=2$\\ 
  \midrule
  Conv2dFlipout &  [-1, 64, 4, 4]  &  $k=3$, $s=1$\\ 
  \midrule
  Conv2dFlipout &  [-1, 64, 4, 4]  &  $k=3$, $s=1$\\ 
  \midrule
  Conv2dFlipout &  [-1, 64, 4, 4]  &  $k=3$, $s=1$\\ 
  \midrule
  MaxPool2d & [-1, 64, 2, 2] & $k=2$, $s=2$\\ 
  \midrule
  Flatten & [-1, 256] & - \\
  \midrule
  DenseFlipout & [-1, 2048] & - \\
  \midrule
  DenseFlipout & [-1, 512] & - \\
  \midrule
  DenseFlipout & [-1, 64] & - \\
  \midrule
  Dense & [-1, 90] & - \\
  \midrule
  MNTriL & [-1, P] & - \\
  \hline
\end{tabular}

\end{minipage}
\end{table}
}

\newcommand{\tableembeddingnetbnnoned}{
\begin{table}
  \centering
  \noindent\begin{minipage}[b]{0.99\columnwidth}
  \centering
    \caption{
    The BNN architecture for the 1-parameter model. 
    The first column lists the layer type, the second lists the dimensionality of the output from that layer, and the third column lists the parameters of that layer: $k$ is the convolution kernel size, $s$ is the stride.
    See \S\ref{sec:bnn} for a detailed description of the design of this network. 
    }
  \label{table:BNN1d}
  \centering
  \begin{tabular}{llc}
 \hline   Layer   &  Output shape   &  Parameters \\\hline \hline
  Conv2dFlipout &  [-1, 16, 32, 32]  &  $k=3$, $s=1$\\ 
  \midrule
  MaxPool2d & [-1, 16, 16, 16] & $k=2$, $s=2$\\ 
  \midrule
  Conv2dFlipout &  [-1, 32, 16, 16]  &  $k=3$, $s=1$ \\
  \midrule
  Conv2dFlipout &  [-1, 48, 16, 16]  &  $k=3$, $s=1$\\ 
  \midrule
    Conv2dFlipout &  [-1, 64, 16, 16]  &  $k=3$, $s=1$\\ 
  \midrule
  MaxPool2d & [-1, 64, 8, 8] & k=2, $s=2$ \\
  \midrule
  Flatten & [-1, 4096] & - \\
  \midrule
  DenseFlipout & [-1, 768] & - \\
  \midrule
  DenseFlipout & [-1, 320] & - \\
  \midrule
  DenseFlipout & [-1, 64] & - \\
  \midrule
  Dense & [-1, 90] & - \\
  \midrule
  MNTriL & [-1, 1] & - \\
  \hline
\end{tabular}
\end{minipage}
\end{table}
}

\newcommand{\tabledatasetsizes}{
\begin{table}
\noindent\begin{minipage}[b]{0.99\columnwidth}
  \caption{
  Dataset sizes (training, validation, and test) for each model type --- BNN and NPE --- and for each level of model complexity --- 1, 5, and 12 physics parameters.
  The details of each dataset are described in \mysection{}~\ref{sec:simulations}. 
  \label{table:dataset_size}
  }
  \centering
  \begin{tabular}{lllll}
 \hline   Model  & Training &  BNN Validation & NPE Validation & Test \\
 \midrule
   1-parameter      &    $200,000$   &    $50,000$  & $20,000$ &  $1000$   \\ 
    5-parameter      &   $400,000$     &   $100,000$   & $40,000$ & $1000$    \\ 
    12-parameter     &   $800,000$        & $200,000$   & $80,000$ & $1000$      \\ 
  \hline
\end{tabular}
\end{minipage}
\end{table}
}

\newcommand{\tableparameterdistributionmain}{
\begin{table*}
\noindent
\begin{minipage}[b]{0.99\linewidth}
\caption{
Parameter distributions used to generate training and test sets. 
The Models column lists the models that use a given parameter.  
The full list of model parameters is as follows. 
For the lens mass, there are the Einstein radius \einsteinrad, eccentricity components ($\eccentricity{l}{1}$, $\eccentricity{l}{2}$), and lens-source offset positions ($x_c$, $y_c$). 
For the lens environment, there are two components of the external shear ($\gamma_1$, $\gamma_2$). 
For the source light, there are Sersic profile with apparent magnitude $m_s$, half-light radius $R$, Sersic index $n$, and eccentricity components ($\eccentricity{s}{1}$, $\eccentricity{s}{2}$). 
Uniform distributions are denoted by $\mathcal{U}\mathrm{(minimum, maximum)}$. 
} 
\label{table:params}
\centering
\begin{tabular}{llrll}
\hline   
Parameter           & Name                  & Models        &  Training Priors          & Test Priors                   \\ 
\toprule  \hline
\multicolumn{5}{c}{Lens Parameters}\\
\midrule
\einsteinrad\  ($''$)  & Einstein Radius       & 1,5,12        & $\mathcal{U}(0.3,4.0)$    & $\mathcal{U}(0.5,3.0)$        \\ 
$\eccentricity{l}{1}$            & lens eccentricity comp. 1    & 5,12          & $\mathcal{U}(-0.8,0.8)$   & $\mathcal{U}(-0.2,0.2)$       \\ 
$\eccentricity{l}{2}$            & lens eccentricity comp. 2    & 5,12          & $\mathcal{U}(-0.8,0.8)$   & $\mathcal{U}(-0.2,0.2)$       \\ 
$x_\mathrm{c}$ ($''$)  & x lens position       & 5,12          & $\mathcal{U}(-2,2)$       & $\mathcal{U}(-1,1)$           \\ 
$y_\mathrm{c}$ ($''$)  & y lens position       & 5,12          & $\mathcal{U}(-2,2)$       & $\mathcal{U}(-1,1)$           \\
\midrule
\multicolumn{5}{c}{Lens Environment Parameters}\\
\midrule
$\gamma_1$          & shear component 1               & 12            & $\mathcal{U}(-0.8,0.8)$   &  $\mathcal{U}(-0.05,0.05)$    \\ 
$\gamma_2$          & shear component 2               & 12            & $\mathcal{U}(-0.8,0.8)$   & $\mathcal{U}(-0.05,0.05)$     \\ 
\midrule
\multicolumn{5}{c}{Source Light Parameters}\\
\midrule
$m_s$               & magnitude             & 12            & $\mathcal{U}(18,25)$      & $\mathcal{U}(19,24)$          \\ 
$R$ ($''$)             & half-light radius     & 12            & $\mathcal{U}(0.1,3.0)$    & $\mathcal{U}(0.5,1.0)$        \\ 
$n$                 & Sersic index          & 12            & $\mathcal{U}(0.5,8.0)$    & $\mathcal{U}(2,4)$            \\ 
$\eccentricity{s}{1}$              & source eccentricity comp. 1  & 12            & $\mathcal{U}(-0.8,0.8)$   & $\mathcal{U}(-0.2,0.2)$       \\ 
$\eccentricity{s}{2}$              & source eccentricity comp. 2  & 12            & $\mathcal{U}(-0.8,0.8)$   & $\mathcal{U}(-0.2,0.2)$       \\ 
\hline
\end{tabular}
\end{minipage}
\end{table*}
}

\newcommand{\tableresultssummary}{
{\renewcommand{\arraystretch}{1.5}
\begin{table}
\footnotesize
\noindent
\begin{minipage}[b]{0.99\linewidth}
\caption{
Summary of results comparing the NPE and BNN model performance for the 1-, 5- and 12-parameter models for the test set.
The table values are the average central values and the 68\% scatter in the residuals for each parameter across 1000 lenses.
}
\label{table:12paramuncertainties}
\centering
\begin{tabular}{l*{6}{c}}
\hline   
                        &  \multicolumn{2}{c}{1-param Model}                        & \multicolumn{2}{c}{5-param Model}                             & \multicolumn{2}{c}{12-param Model} \\
\toprule  \hline 
Param               &  NPE                        &  BNN                        &  NPE                          &  BNN                          &  NPE                      &  BNN                      \\ 
\toprule  \hline
$\Delta$\einsteinrad\ ($''$)       & $-0.0004 \pm{0.0011}$     & $0.004 ^{+0.015}_{-0.012}$     & $0.002 ^{+0.013}_{-0.011}$    & $0.002 ^{+0.040}_{-0.035}$   & $-0.046^{+0.13}_{-0.078}$   & $0.14^{+0.17}_{-0.13}$   \\   
$\Delta$ $\eccentricity{l}{1}$   & -                           & -                           & $0.003 ^{+0.012}_{-0.017}$              & $0.002 \pm 0.066$             & $0.03^{+0.16}_{-0.20}$            & $0.02\pm 0.16$           \\
$\Delta$$\eccentricity{l}{2}$   & -                           & -                           & $0.001 \pm 0.030$     & $0.028 ^{+0.034}_{-0.047}$     & $0.00\pm 0.20$   & $0.02\pm 0.16$         \\ 
$\Delta$$x_\mathrm{c}$ ($''$)      & -                           & -                           & $0.004 ^{+0.018}_{-0.030}$      & $0.043 ^{+0.093}_{-0.14}$         & $-0.01\pm 0.24$   & $-0.03\pm 0.42$          \\ 
$\Delta$$y_\mathrm{c}$ ($''$)      & -                           & -                           & $-0.006 ^{+0.031}_{-0.026}$               & $-0.02 \pm 0.20$      & $-0.03^{+0.14}_{-0.12}$          & $-0.11^{+0.42}_{-0.34}$            \\  
$\Delta$$\gamma_1$              & -                           & -                           & -                             & -                             & $-0.01\pm 0.12$           & $-0.0095\pm 0.079$         \\   
$\Delta$$\gamma_2$              & -                           & -                           & -                             & -                             & $-0.01\pm 0.12$           & $-0.020\pm 0.094$           \\
$\Delta$$m_s$                   & -                           & -                           & -                             & -                             & $-0.08^{+0.25}_{-0.17}$  & $-0.24^{+0.29}_{-0.23}$            \\    
$\Delta$$R$ ($''$)                 & -                           & -                           & -                             & -                             & $0.041^{+0.090}_{-0.19}$    & $0.43\pm 0.18$             \\ 
$\Delta$$n$                     & -                           & -                           & -                             & -                             & $-0.23^{+0.59}_{-1.4}$     & $ 2.38^{+0.69}_{-0.63} $           \\ 
$\Delta$$\eccentricity{s}{1}$   & -                           & -                           & -                             & -                             & $0.03\pm 0.17$          & $0.01^{+0.15}_{-0.17}$            \\ 
$\Delta$$\eccentricity{s}{2}$   & -                           & -                           & -                             & -                             & $0.02\pm 0.17$           & $0.03^{+0.17}_{-0.20}$           \\ 
\hline 
\end{tabular}
\end{minipage}
\end{table}
}
}

\title{Deep inference of simulated strong lenses in ground-based surveys}

\author[a]{Jason Poh, \note{Corresponding author.}}
\author[b]{Ashwin Samudre,}
\author[a, c]{Aleksandra \'{C}iprijanovi\'{c},}
\author[a, d]{Joshua Frieman,}
\author[e]{Gourav Khullar,}
\author[a, c, d]{Brian D. Nord}

\affiliation[a]{Department of Astronomy and Astrophysics, University of Chicago, IL 60637, USA}
\affiliation[b]{School of Computing Science, Simon Fraser University, Canada}
\affiliation[c]{Fermi National Accelerator Laboratory, Batavia, IL 60510, USA}
\affiliation[d]{Kavli Institute for Cosmological Physics, University of Chicago, Chicago, IL 60637, USA}
\affiliation[e]{Department of Astronomy, University of Washington, WA 98195, USA}

\emailAdd{jasonpoh@uchicago.edu}
\emailAdd{ashwin.samudre@gmail.com}
\emailAdd{aleksand@fnal.gov}
\emailAdd{Jfrieman@uchicago.edu}
\emailAdd{gkhullar@uw.edu}
\emailAdd{nord@fnal.gov}

\abstract{
The large number of strong lenses discoverable in future astronomical surveys will likely enhance the value of strong gravitational lensing as a cosmic probe of dark energy and dark matter. 
However, leveraging the increased statistical power of such large samples will require further development of automated lens modeling techniques. 
We show that deep learning and simulation-based inference (SBI) methods produce informative and reliable estimates of parameter posteriors for strong lensing systems in ground-based surveys. 
We present the examination and comparison of two approaches to lens parameter estimation for strong galaxy-galaxy lenses --- Neural Posterior Estimation (NPE) and Bayesian Neural Networks (BNNs).
We perform inference on 1-, 5-, and 12-parameter lens models for ground-based imaging data that mimics the Dark Energy Survey (DES).
We find that NPE outperforms BNNs, producing posterior distributions that are more accurate, precise, and well-calibrated for most parameters.
For the 12-parameter NPE model, the calibration is consistently within $<$10\% of optimal calibration for all parameters, while the BNN is rarely within 20\% of optimal calibration for any of the parameters.
Similarly, residuals for most of the parameters are smaller (by up to an order of magnitude) with the NPE model than the BNN model.
This work takes important steps in the systematic comparison of methods for different levels of model complexity.
}

\begin{document}
\maketitle
\flushbottom

\section{Introduction}
\label{sec: introduction}

When light from a distant source is deflected by a massive object along the line-of-sight to an observer, causing the source image to appear multiply imaged or distorted to the observer, we have what is known as {strong gravitational lensing}. 
Strong gravitational lensing systems are well-established observational probes of astrophysics and cosmology; 
on the astrophysical side, the amplification of the source galaxy or quasar light due to gravitational lensing allows for the observation and analysis of high-redshift sources that would otherwise be too faint to detect  \cite[e.g.,][]{Bayliss2014, Hainline2009, Quider2009, Smit2017, Dessauges-Zavadsky2009, Christensen2012, James2013, Yuan2009, Finkelstein2009, Iwata2008, Jones2013a}. 
Additionally, the morphology of lensed images provides information about the distribution of dark and baryonic matter in the lens systems \cite[e.g.,][]{Auger2010a, Barnabe2011, Newman2015, Newman2012, Newman2012a}. 
On the cosmological front, strong lensing systems with multiple images of a time-varying source (e.g., quasars and supernovae) \cite[e.g.,][]{Suyu2010, Suyu2013, Shajib2020, Jee_2016, Jee_2019}, and strong lensing systems with multiple source planes \cite[e.g.,][]{Gavazzi2008, Collett2014, Collett2012}, can be used to constrain the Hubble constant and dark energy.

Achieving these science objectives requires precise and accurate modeling of statistical samples of lensing systems. 
Conventional lens modeling involves reconstructing the lens galaxy mass distribution and source galaxy light distribution from imaging (and perhaps ancillary) data. 
A well-established class of modeling methods involves Bayesian parametric forward modeling, where both the source light and the lens mass distributions are approximated by parametric models  \cite[e.g.,][]{BA2018, BS2021, Kneib2011, Suyu2010, Suyu2013, Lefor_2013}. 
An explicit likelihood function is constructed from the residual between the observed image and the reconstructed model image. 
This likelihood function is then maximized, often through random sampling methods like Markov Chain Monte Carlo sampling. 
The posterior model parameter distribution is then used for lensing science applications. 

Approximately a thousand galaxy-galaxy strong lensing systems have been discovered in large cosmic imaging surveys~\cite{Zaborowski_2023, Jacobs_2019, Jacobs_2017, Nord2016} and the catalogs referenced in \cite{Chen_2019, Moustakas2012} and the Gravitationally Lensed Quasar Database \cite{Lemon}.
The Vera Rubin Observatory Legacy Survey of Space and Time~\cite[LSST;][]{IK2019} is expected to find about 100,000 such systems, and space-based surveys such as Euclid \cite{troja2022euclidnutshell, Racca_2016} and the Nancy Grace Roman Space Telescope~\cite{sanderson2024recommendationsearlydefinitionscience, Wenzl_2022} will additionally find tens of thousands more \cite{Collett2015, Nord2016, Weiner_2020}.

While the current approach to strong lens modeling described above has been successful for samples of order 100 or less \cite{tan2023project}, it is labor- and time-intensive; scaling it to the projected LSST strong lens sample would require an estimated $\sim 1000$ person-years of effort for the analysis, which is infeasible.
While there have been efforts to develop automated lens modeling techniques \cite[e.g.,][]{Nightingale2016}, this is still a nascent field with many approaches that have not yet been explored. 

Explicit inference --- where $p(X | \theta)$ is approximated by an analytic likelihood function --- is the traditional approach to estimating parameter posteriors in Bayesian inference and is the most common method used in lens modeling. In traditional Bayesian inference, the posterior $p(\theta | X)$ is given by Bayes' theorem as:
\begin{equation}
\begin{gathered}
p(\theta | X) = \frac{p(X | \theta) * p(\theta)}{p(X)},
\end{gathered}
\end{equation}
where $X$ denotes the data vector (e.g., flux values in an image), $\theta$ is the set of model parameters to be inferred, 
$p(X | \theta)$ denotes the likelihood, $p(\theta)$ is the prior on the parameters before acquiring the data $X$, $p(X)$ is the evidence (or marginal likelihood), and we are implicitly working in the context of a particular model characterized by the parameters $\theta$.
Examples of explicit likelihood inference methods include Markov Chain Monte Carlo (MCMC) methods, which are now ubiquitous in astronomical and cosmological inference. 
However, in cases where the likelihood function is unknown or computationally intractable (e.g., in cases where the model contains hundreds or even thousands of parameters), explicit likelihood methods may not be feasible. 

To address the challenges in explicit likelihood calculations, Simulation-Based Inference (SBI) --- a.k.a. Implicit Likelihood Inference (ILI) or Likelihood-Free Inference (LFI) --- methods have gained popularity in recent years \cite{CB2019}. 
SBI methods leverage simulators at the core of a statistical inference procedure that can accurately infer the likelihood or posterior distributions of scientific analysis for the experiment or observation at hand. Traditional SBI methods include Approximate Bayesian Computation~\cite[ABC;][]{Rubin1984,tavare1997inferring,Beaumont2002} and its frequentist analog, Approximate Frequentist Computation~\cite[AFC;][]{brehmer2018guide,Brehmer_2018}. 

ABC has a long history and has been applied to a number of astronomical problems \cite{cameron2012, Jennings_2017}. 
ABC starts by simulating realizations of model observables $X$ (synthetic data) that are drawn from points in the model parameter space $\theta$.
The points are then accepted or rejected based on the model observables' similarity to the observed data up to some tolerance $\epsilon$ in some specified metric. 
Through this rejection sampling, the distributions of accepted points in parameter space will approximate the desired posterior distribution while circumventing explicit likelihood calculations. 
ABC suffers from two drawbacks. 
First, the inference is not amortized --- for every new observation, the potentially computationally expensive inference process has to be repeated; this is also true of traditional explicit likelihood methods like MCMC.
Second, there is an unavoidable trade-off between inference quality and sampling efficiency. 
Accurate and precise 
inference of the posterior occurs in the limit $\epsilon \rightarrow 0$, in which case the rate of acceptance can be so low that the computational expense is prohibitively high even for low-dimensional (and certainly for high-dimensional) parameter spaces.   

Recent advancements in deep learning methods and SBI techniques have largely addressed these drawbacks. 
They have led to the adoption of novel neural network surrogates for modeling the conditional probability densities of the likelihood or posterior distribution in a given inference problem from simulations \cite{CB2019}. 
These {neural density estimators} are more flexible than conventional parametric density estimators and scale better to higher-dimensional parameter spaces. 
They are also amortized --- once the upfront computational cost of training the density estimator is paid, the subsequent inference is fast and cheap. 
Advances in SBI help address the constraining trade-off between inference quality and sampling efficiency and overcome the curse of dimensionality for problems with a large parameter space. 

Next-generation SBI methods have been used in a variety of astronomical tasks, such as the analysis of supernova data, tomographic cosmic shear,  inference of the HI ionization rate from high-redshift Lyman-$\alpha$ forests in~\cite{AC2019}, dark matter substructure inference in galaxy-galaxy strong lenses~\cite{Brehmer_2019, Coogan2022, Anau2023}, inference of variability properties of astrophysical objects from dead time-affected light curves in X-ray observations~\cite{HB2021}, and the inference of the Hubble constant from binary neutron star mergers ~\cite{GF2021}. 

For strong lensing, there have been efforts to automatically infer strong lens parameters with deep learning:  Neural Posterior Estimation \cite[NPE;][]{Legin2021, Legin2023, WagnerCarena2022},  Bayesian Neural Networks \cite[BNNs;][]{levasseur2017uncertainties, pearson2021strong, wagner2021hierarchical, park2021large}, Neural Ratio Estimation \cite[NRE;][]{Coogan2022, Anau2023}, vision transformers \cite{huang2022strong}. 

In this paper, we use NPE to model strong lensing systems in synthetic imaging data designed to emulate ground-based surveys like the Dark Energy Survey \cite[DES;][]{DES2016}. 
Our work differs from and complements previous studies applying NPE to strong lensing \cite[]{Legin2021, Legin2023, WagnerCarena2022}.
First, previous studies focused on performing inference on simulated imaging data that are either idealized ~\cite[]{Legin2021, Legin2023}, or are emulating space-based imaging data~\cite{WagnerCarena2022}, which has a small pixel scale and no image atmosphere-induced blurring. 
Since the majority of new lens discoveries will come from ground-based survey data, it is important to investigate the suitability (and limitations) of SBI methods when applied to noisier ground-based strong lensing images. 
To this end, we use \deeplenstronomy~\cite{MN2021}, which is built on \lenstronomy~\cite{BA2018, BS2021}, to emulate the observing conditions of the DES survey, which are informed by the physical specifications of the DES experiment and empirical atmospheric data collected by the survey. 

We emulated DES imaging data because the full 6-year DES dataset is publicly available \cite{Abbott_2021}, and a natural progression of this line of work is to evaluate the performance of these models on real observed survey data. 
Many lenses in the DES survey footprint have previously been modeled using conventional MCMC techniques, and we can use the lens parameter values and uncertainties from those studies as a future benchmark for our models. 
The takeaways from such a study can help inform the direction of future work on survey data from the Rubin Observatory. 

Another way in which this work differs from previous SBI studies is that it uses masked autoregressive flows (MAFs)~\cite{papamakarios2017masked} as the density estimator. 
To our knowledge, this is one of the first application of MAFs to the problem of strong lens modeling, based on earlier works by some of the same coauthors of this work \cite{poh2022stronglensingparameterestimation,agarwal2025neuralnetworkpredictionstrong,swierc2024domainadaptiveneuralposteriorestimation}. 
MAFs differ from the mixture density networks used in previous studies as they are not restricted to modeling probability densities using mixtures of Gaussians. 

Finally, we compare an NPE model \cite[]{Legin2021, Legin2023} to a BNN trained on the same data. 
BNNs have already been successfully used in astrophysics and cosmology --- e.g., for the classification of blazars~\cite{BF2022}, calibration of gamma-ray bursts as cosmological distance indicators using supernovae type Ia~\cite{ER2022}, and galaxy morphology measurements~\cite{WL2022, TC2022}, among others. 
In the context of strong lens parameter estimation, they are the most commonly used architecture in the literature \cite{Gentile_2023, levasseur2017uncertainties, pearson2021strong, wagner2021hierarchical, park2021large, bom2019deep}. 
As such, they provide a suitable method for comparison. 

Another reason we use BNNs for comparison is that both methods have computational costs that are amortized, enabling us to evaluate the performance of both models across a large ensemble of one thousand lenses on a reasonable timescale. 
We do not rely on the models' performances on any single lens image, as that can be prone to selection bias and may not be indicative of a model's typical performance. 
Due to the prohibitive computational expense of MCMC modeling, it cannot be performed for a large number of lenses, so we did not compare NPE to MCMC.

We evaluate the performance of the MAF-driven NPE model in this work with extensive diagnostics. 
It is difficult to achieve statistical guarantees from any model without making post hoc calibrations, which is beyond the scope of this work. 
Due to this difficulty, we emphasize the importance of using multiple diagnostics to characterize the stability and sensitivity of our models, such as posterior/empirical coverage plots, Simulation-Based Calibration \cite[SBC;][]{Talts2018}, and weight initialization and generalization.
No single diagnostic can fully characterize the performance of a model:  it is important to deploy a suite of diagnostics that complement each other. 

This paper is organized as follows. 
In \mysection{}~\ref{sec:sgl}, we give an overview of strong lens system modeling. 
Details about how NPE is implemented, as well as BNN that are used for comparison, are given in \mysection{}~\ref{sec:methods}. 
We describe the simulator we use and discuss how we create our datasets in \mysection{}~\ref{sec:simulations}. 
Metrics for evaluating model performance are discussed in Section~\ref{sec:modeldiagnostics}. 
Our results are presented in \mysection{}~\ref{sec:results}, with discussion in \mysection{}~\ref{sec:discussion} and conclusion in \mysection{}~\ref{sec:conclusion}.

\section{Strong Gravitational Lensing}
\label{sec:sgl}

General Relativity predicts that when light from a distant source passes near an intervening massive object (the lens), the mass causes light to deflect around the object. The deflection field is given by the lens equation, 
\begin{equation}
\label{eqn:lenseq}
    \hat{\beta} = \hat{\theta} - \hat{\alpha}, 
\end{equation}
where $\hat{\beta}$ is the 2D angular position in the source plane, $\hat{\theta}$ is the position in the image plane, and $\hat{\alpha}$ is the deflection angle caused by gravitational lensing.  
The mapping from the source plane position $\hat{\beta}$ to the image plane position $\hat{\theta}$ is generally non-linear, and a single source location $\hat{\beta}$ may map to multiple image positions $\hat{\theta}$ in the image plane. 
The deflection angle $\hat{\alpha}$ depends on the mass distribution of the intervening lensing object(s) and on the impact parameter of a given ray in the lens plane relative to the lens mass distribution. 
Solving this lens equation is central to lens modeling problems, but analytic solutions exist for only the simplest parametric lens mass models. 
There is a wealth of literature describing the mathematical formalism of lensing \cite[][]{Narayan1996}, cosmological and astrophysical applications \cite[][]{2010ARA&A..48...87T}, and solutions for some of the most common lens mass profiles  \cite[][]{Kassiola1993, Kormann1994, Barkana1998}.

The population of strong lens systems predicted to be found in future surveys is dominated by galaxy-galaxy lensing systems, in which a single source galaxy is lensed by a single lensing galaxy, leading to a distorted (strongly sheared) image of the source.
The lens population is dominated by massive, early-type (elliptical) galaxies \cite{Oguri2010}, the mass distributions of which  
are well-approximated by Singular Isothermal Ellipsoid (SIE) profiles. 
For the SIE profile, the reduced surface mass density (i.e., convergence) is given by
\begin{align}
\label{eqn:lensconvergence}
\kappa(x',y') = \frac{1}{2} \left(\frac{\theta_{\rm E}}{\sqrt{q_{\mathrm{l}} x'^2 + y'^2/q_{\mathrm{l}}}}\right),
\end{align}
where ($x'$, $y'$) are 2D image plane coordinates oriented such that $x'$ is aligned with the semi-major axis of the mass distribution, $q_{\mathrm{l}}$ is the ratio of semi-minor to semi-major axis of the mass profile (the lens ellipticity), and $\theta_{\mathrm{E}}$ is the Einstein radius, defined as the effective radius within which the average convergence $\langle\kappa\rangle=1$, and which characterizes the scale length of deflections of the source light by the lensing galaxy. 

The solution for the deflection angle $\hat{\alpha}$ in \myequation{}~\ref{eqn:lenseq} in the same coordinate system as \myequation{}~\ref{eqn:lensconvergence} is given analytically for SIE lenses by \cite{Kochanek2004}:

\begin{align}
\label{eqn:lensdeflection}
   \alpha_{x'} &= \theta_E \left(\frac{\sqrt{q_{\mathrm{l}}}}{\sqrt{1-q_{\mathrm{l}}^2}}\right)\arctan \left(\frac{x'\sqrt{1-q_{\mathrm{l}}^2}}{\sqrt{q_{\mathrm{l}}^2x'^2 + y'^2}}\right), \\
   \alpha_{y'} &= \theta_E \left(\frac{\sqrt{q_{\mathrm{l}}}}{\sqrt{1-q_{\mathrm{l}}^2}}\right)\arctanh \left(\frac{y'\sqrt{1-q_{\mathrm{l}}^2}}{\sqrt{q_{\mathrm{l}}^2x'^2 + y'^2}}\right), 
\end{align}
where $\alpha_{x'}$ and $\alpha_{y'}$ are the $x'$- and $y'$- components of the deflection angle. 
To fully describe the SIE profile, we must also specify the position of the center of the lens mass ($x$, $y$) in an observer's 2D Cartesian coordinate system ($\hat{x}$, $\hat{y}$) in the image plane and the position angle of the lens $\phi_{\mathrm{l}}$ in that system; $\phi_{\mathrm{l}}$ is thus the rotation angle between the coordinates  ($x'$, $y'$) and ($\hat{x}$, $\hat{y}$). 
Because $\phi_{\mathrm{l}}$ is a cyclic parameter, which can complicate the numerical analysis, it is often convenient to perform a parameter transformation from the lens ellipticity and position angle to the eccentricity moduli:
\begin{align}
\label{eqn:lenseccentricity1}
\eccentricity{l}{1} &= \frac{1-q_{\mathrm{l}}}{1+q_{\mathrm{l}}}\cos(2\phi_{\mathrm{l}}), \\
\label{eqn:lenseccentricity2}
\eccentricity{l}{2} &= \frac{1-q_{\mathrm{l}}}{1+q_{\mathrm{l}}}\sin(2\phi_{\mathrm{l}}). 
\end{align}
The SIE model profile is therefore specified by five parameters: the Einstein radius (\einsteinrad), the eccentricity moduli ($\eccentricity{l}{1}$, $\eccentricity{l}{2}$), and the 2D position of the center of mass ($x$, $y$). 

In addition to the mass profile parameters, we include an external shear component to emulate the effect of external mass perturbers near the line of sight. 
The external shear is parameterized by a shear modulus $\gamma$ and position angle $\phi_{\gamma}$ in the $(x,y)$ coordinate system \cite{Keeton1997}.
However, as with the lens ellipticity and position angle, we avoid working in the cyclic coordinate $\phi_{\gamma}$ by changing variables to two external shear components: 
\begin{align}
\gamma_1 &= \gamma\cos(2\phi_{\gamma}),\\
\gamma_2 &= \gamma\sin(2\phi_{\gamma}). 
\end{align}

Next, we model the source galaxy light profile as a 2D Sersic profile \cite{Sersic1968}:
\begin{equation}
\label{eqn:Sersic}
    I(r) = I_e\exp\left(-b_n\left(\left(\frac{r}{R}\right)^\frac{1}{n}-1\right)\right),
\end{equation}
where
\begin{equation}
r = \sqrt{q_{\mathrm{s}}x''^2 + y''^2/q_{\mathrm{s}}}
\end{equation}
is the projected radial distance from the center of the light profile, $q_{\mathrm{s}}$ is the ratio of semi-minor to semi-major axis (the ellipticity) of the source light profile, $x''$ and $y''$ are 2D Cartesian coordinates oriented such that $x''$ is aligned with the semi-major axis of the light profile, and the source-light position angle in the $(x,y)$ coordinate system is given by $\phi_{\rm src}$.
The coefficient $b_n$ in \myequation{}~\ref{eqn:Sersic} is a function of the Sersic index $n$ and is well-approximated by
\begin{equation}
    b_n \approx 2n - \frac{1}{3} + \frac{4}{405n} + \frac{46}{25515n^2},
\end{equation}
in the range $0.5 < n < 10$ \cite{Ciotti1999}. 
The parameter $R$ is the half-light radius (i.e., the radius which contains half the emitted flux of the galaxy), and $I_e$ is the intensity at the half-light radius $R$, such that $I(R)=I_e$. 

In this work, we use the source galaxy's apparent magnitude $m_s$ in some passband in place of the source intensity $I_e$ in that band, as the former is the quantity directly measured in astronomical observations. 
Converting between the two quantities requires the magnitude zero point, defined as the magnitude that produces 1 count per second on the detector in that band, which varies from telescope to telescope. 
We define the zero point using DES observing conditions in the $g$-band filter, which we describe in \mysection{}~\ref{sec:simulations}.

As with the lens ellipticity (\myequation{}'s~\ref{eqn:lenseccentricity1}-\ref{eqn:lenseccentricity2}), we replace the corresponding source parameters ($q_{\mathrm{s}}$, $\phi_{\mathrm{s}}$) with the eccentricity moduli:
\begin{align}
\eccentricity{s}{1} &= \frac{1-q_{\mathrm{s}}}{1+q_{\mathrm{s}}}\cos(2\phi_{\mathrm{s}}), \\
\eccentricity{s}{2} &= \frac{1-q_{\mathrm{s}}}{1+q_{\mathrm{s}}}\sin(2\phi_{\mathrm{s}}). 
\end{align}

The lensing morphology depends only on the relative difference between the 2D positions of the centers of the lens mass and source light profiles.
We fix the center of the source light profile to the origin of the $(x,y)$ image coordinate system without any loss of generality; we instead infer the relative positions as ($x_{\mathrm{c}}$, $y_{\mathrm{c}}$).
This reduces the total number of parameters needed to describe the source light profile to five: the source apparent magnitude $m_s$, the effective radius $R$, the Sersic index $n$, and the eccentricity moduli ($s_{e1}$, $s_{e2}$). 

In this study, we make a few additional simplifying assumptions in our modeling setup. 
First, we assume that the lens galaxy's light has been perfectly subtracted from the image.
The lens light is often subtracted before lens modeling, and there has been promising work done to automate the process of lens light subtraction as a pre-processing step in an automated pipeline \cite{Hezaveh2017}. 
The alternative would be to include the lens light in the model with additional parameters.
Second, we set the lens redshift to $z_l=0.5$ and the source redshift to $z_s=2.0$ because those are the peaks of the projected redshift distributions for galaxy-galaxy strong lenses in forthcoming surveys \cite{Collett2015}. 
Finally, we assume a flat $\Lambda$CDM cosmology with $H_0=70$ kms$^{-1}$Mpc$^{-1}$ and $\Omega_M=0.3$. 

In summary, each strong lens is characterized by 12 parameters: five to describe the lens mass (\einsteinrad, $l_{e1}$, $l_{e2}$, $x_c$, $y_c$), two to describe the external shear ($\gamma_1, \gamma_2$), and five to describe the source light ($m_s$, $R$, $n$, $s_{e1}$, $s_{e2}$). 
Table~\ref{table:params} contains further details on the distributions of these parameters.

\section{Methodology: Posterior Estimation}
\label{sec:methods}

\subsection{Density Modeling and Posterior Estimation }

In this section, we describe the two density modeling techniques we use to model the posterior probability for a lens model. 
The first method, NPE, is an SBI approach that leverages recent advancements in deep learning techniques to train a neural density estimator to directly learn the posterior distribution of strong lens model parameters. 
NPE requires three elements --- a mechanistic model (simulator), priors on parameters of the model being inferred, and data (or corresponding summary statistics), which are outputs of the simulator. 
The priors are utilized to sample the parameters and simulate synthetic data that is passed to a neural network (density estimator).
The network is trained to learn the relation between simulated data and the underlying parameters and is later deployed on empirical data to obtain the posterior parameter distribution corresponding to that data. 
NPE avoids explicit likelihood calculations and uses simulations to train the network and to get the relevant parameters.

The second approach, BNNs, involves using stochastic neural networks trained using a Bayesian approach \cite{BNN2022}. 
Traditional artificial neural networks produce the equivalent of maximum likelihood (MLE) or maximum a posteriori (MAP) point estimates for the labels when given an observed data point. 
Although these so-called deterministic neural network models are easy to deploy, they do not include uncertainty quantification, which is crucial in science applications. 
BNNs address this flaw in traditional neural networks by replacing the deterministic network weights with probability distributions. 
By introducing stochasticity into the network, the BNN model produces a probability density estimate instead of a point estimate as its output, thus enabling posterior density estimation and uncertainty quantification.

\subsection{Neural Posterior Estimation}

\subsubsection{Overview}
\label{sec:npe_overview}

There are a few subcategories of approaches for this method: the original method, NPE-A \cite{NPE-A} and two variants, NPE-B \cite{NPE-B} and NPE-C \cite{greenberg2019automatic}.
In this work, we follow the nomenclature of \cite{mancini2022bayesian}, which is consistent with the taxonomy of other SBI literature \cite{durkan2020contrastive}). 
These methods differ in their handling of active learning, also known as sequential estimation. 
The sequential aspect is optional, and we opt not to use it in this work for reasons we discuss in \mysection{}~\ref{sec:SNPE_discussion}. 
In the case where no sequential estimation is performed, all NPE methods reduce to the same approach described in NPE-A. 

In NPE, $q_{\phi}(\theta|X)$ is parameterized by a vector $\phi$ that represents the neural network weights.
It is trained on $N$ pairs of training data $\{\theta_i, X_i\}^N_{i=1}$ from the simulator to approximate the posterior density $p(\theta|X)$, the posterior probability distribution of model parameters $\theta$ given some observed data $X$. 

\begin{align}
   &\theta_i \sim p(\theta), & X_i \sim p(X|\theta_i),
\end{align}
where $p(\theta)$ is the prior and $p(X|\theta_i)$ is a statistical model representing the simulator's data-generating process.
The model's goodness-of-fit is evaluated by maximizing the average log-likelihood

\begin{equation}
\label{eq:gen_likelihood}
    \mathcal{L}(\phi) = \frac{1}{N}\sum^N_i\log q_{\phi}(\theta_i|X_i)
\end{equation}
with respect to $\phi$. 
By the strong law of large numbers, in the limit $N \rightarrow \infty$, \myequation{}~\ref{eq:gen_likelihood} converges to the expectation value of the log-likelihood $\log q_{\phi}(\theta|X)$ over the true underlying joint probability distribution of the training data $\{\theta_i, X_i\} \sim p(X|\theta)p(\theta)$:

\begin{align}
    \mathcal{L}(\phi) &= \lim_{N\to\infty} \frac{1}{N}\sum^N_i\log q_{\phi}(\theta_i|X_i)  \notag \\
    &= \int\int p(X|\theta)p(\theta) \log q_{\phi}(\theta_i|X_i) dXd\theta \notag \\
    &\equiv \mathbb{E}_{p(X|\theta)p(\theta)}[\log q_{\phi}(\theta|X)].
\end{align}
This is equivalent to minimizing the Kullback-Leibler (KL) divergence~\cite{KL1951} between the true and model probability distributions
\begin{align}
    \mbox{KL}(p(\theta|X) || q_{\phi}(\theta|X)) &\equiv \int \int p(\theta|X) \log \frac{p(\theta|X)}{q_{\phi}(\theta|X)} dX d\theta \notag \\
    & = -\mathbb{E}_{p(X|\theta)p(\theta)}[\log q_{\phi}(\theta|X)] + \text{Const.}
\end{align}
with respect to $\phi$. 
In other words, maximizing \myequation{}~\ref{eq:gen_likelihood} for $q_{\phi}(\theta|X)$ decreases its statistical distance from the true posterior underlying $p(\theta|X)$, enabling it to be a better approximation of $p(\theta|X)$.

Therefore, if $q_{\phi}(\theta|X)$ is a sufficiently flexible model, then as $N \rightarrow \infty$, $q_{\phi}(\theta|X)$ will converge to the true posterior $p(\theta|X)$ when the average log-likelihood in \myequation{}~\ref{eq:gen_likelihood} is maximized. 
By using the negative average log-likelihood as a loss function, we can train $q_{\phi}(\theta|X)$ on the training data $\{\theta_i, X_i\}^N_{i=1}$ to approximate the true posterior $p(\theta|X)$.

\subsubsection{Masked Autoregressive Flow Density Estimators}
\label{sec:maf}

In this work, we used Masked Autoregressive Flows \cite[MAF;][]{papamakarios2017masked} as the neural density estimator for their state-of-the-art flexibility in modeling complex density distributions. 
MAFs are a fusion of two state-of-the-art families of neural density estimators --- autoregressive models \cite{uria2016neural} and normalizing flows \cite{rezende2015variational}. 
Autoregressive density estimators use the property that any $n$-dimensional probability density $p(\textbf{v})$, where $\textbf{v} \equiv [v_i,...,v_n]$, can be factorized by the chain rule into a product of $1$-dimensional conditional probability distributions: 

\begin{equation}
    \label{eqn:auto}
    p(\textbf{v}) = \prod_{i=1}^n p(v_i | v_1,...,v_{i-1}) =\prod_{i=1}^n p(v_i | \textbf{v}_{1:i-1}),
\end{equation}
where $\textbf{v}_{1:i-1}$ is the subset of the data vector containing data points with indices less than $i$. 
These conditional probabilities, where $p(v_i)$ is only conditional on data points with indices less than $i$, are {autoregressive}. 
The family of autoregressive density estimation methods uses neural networks to model $p(v_i)$ for all $n$, which can then be combined using \myequation{}~\ref{eqn:auto} to get the target density $p(\textbf{v})$.

By contrast, normalizing flows work by transforming samples from a simple base density (e.g., typically a standard normal distribution), $\textbf{u} \sim p(\textbf{u})$ into samples from the target density $\textbf{v} \sim p(\textbf{v})$ through an invertible transformation or flow $T$:

\begin{equation}
    p(\textbf{v}) = p(\textbf{u}) \left| \frac{\partial T}{\partial \textbf{u}} \right|^{-1},
\end{equation}
where $\textbf{u} = T^{-1}(\textbf{v})$. 
$p(\textbf{v})$ is well-defined under the transformation $T$ if and only if $T$ is invertible and both $T$ and $T^{-1}$ are differentiable (i.e., $T$ has a tractable Jacobian). 
Enforcing this property ensures that the resulting transformation is both {normalized} (i.e., $\int p(\textbf{v}) d\textbf{v} = \int p(\textbf{u}) d\textbf{u} = 1$) and {composable} (i.e., the composition of two transformations $T = T_2\cdot T_1$ also has a tractable Jacobian). 
The base density $p(\textbf{u})$ is chosen such that it is easily evaluated for any $\textbf{u}$, with the standard normal distribution being a very common choice in the literature and one that we will continue to use in this work. 
The family of normalizing flow estimation methods uses neural networks to model the invertible transformations $T$. 

A particular family of autoregressive models --- autoregressive models that are parameterized as Gaussian conditional probabilities --- can also be re-interpreted as normalizing flows. 
In these models, the $i$-th autoregressive conditional can be written as:

\begin{equation}
\label{eq:autogauss}
p(v_i | \mathbf{v}_{1:i-1}) = {\cal N} (v_i | \mu_i, (\mathrm{exp}\, \sigma_i)^2),
\end{equation}
where $\mu_i$ and $\sigma_i$ are the mean and log standard deviation of the Gaussian that $v_i$ is sampled from and are themselves scalar functions of previous data points $\mathbf{v}_{1:i-1}$ only:

\begin{align}
    \label{eq:autoscalar1}
     & \mu_i = f_{\mu_i}(\mathbf{v}_{1:i-1}), \\
     \label{eq:autoscalar2}
     & \sigma_i = f_{\sigma_i}(\mathbf{v}_{1:i-1}).
\end{align}

The probability density model in \myequation{}~\ref{eq:autogauss} can be used to generate data points recursively. 
To generate data, the model requires a source of randomness, typically that samples from some kind of base probability density, with the standard normal distribution being a common choice. 
Random numbers are first generated by that source of randomness and are then transformed into data points in data space.
For example, we generate a data point $v_i$ by first sampling a random number $u_i$ from a standard normal distribution $u_i \sim {\cal N} (0,1)$, and then transform $u_i$ by shifting and scaling $u_i$ by $\mu_i$ and $\sigma_i$ respectively to get the data point $v_i$:
\begin{equation}
\label{eq:autoscale}
 v_i = u_i \mathrm{exp}\, \sigma_i +\mu_i.   
\end{equation}
\myequation{}~\ref{eq:autoscale} can also be interpreted as a
transformation $T$ of a vector of random numbers $\mathbf{u}$ to a vector of data points $\mathbf{v}=T(\mathbf{u})$. 
By construction, $T$ is invertible and has a tractable Jacobian, which implies that the autoregressive model described in \myequation{}~\ref{eq:autogauss} is also definitionally a normalizing flow. 

Density models of the type described in \myequation{}~\ref{eq:autoscale} --- that are solely parameterized by Gaussian distributions --- have limited flexibility. 
The insight that \myequation{}~\ref{eq:autoscale} also represents a normalizing flow leads to the realization that it is composable: instead of having a single autoregressive model transform the initial simple base density $\mathbf{u}$ into the target density $\mathbf{v}$, we can compose stacks of autoregressive model blocks, such that samples of the output density of the last model is used as the input of the next model. 
For example, consider a model comprising $k$ autoregressive model blocks $M_1,..., M_k$ of the kind described in \myequation{}'s~\ref{eq:autogauss}-\ref{eq:autoscale}. 
The block $M_k$ takes as input samples from the density $\mathbf{u_k}$, which are the output samples from $M_{k-1}$. 
The input samples of $M_{k-1}$, $\mathbf{u_{k-1}}$, is, in turn, the output samples of $M_{k-2}$, and so on, until we recurse to $\mathbf{u_0}$, which samples from the base density, which for this example we will take to be the standard normal distribution $\mathbf{u_0} \sim {\cal N} (0,1)$.
To maintain consistency with the rest of this paper, the indexing is the inverse of that used in \cite{papamakarios2017masked}. 

This composition greatly increases the flexibility of the density model, as samples from the initial input density --- the standard normal distribution $\mathbf{u_k} \sim {\cal N} (0,1)$ --- can be progressively transformed into samples from increasingly more complex distributions $\mathbf{u_{<k}}$, which are in turn used as input densities for intermediate model blocks to produce even more complex output densities for the next block in the stack, and so on, until we obtain the final target data $\mathbf{v}$. 
This stack of models can be trained jointly in an end-to-end fashion to produce a normalizing flow that is far more flexible than its individual model blocks. 
The stacking of autoregressive model blocks in this manner is the essential structure of an MAF, and we refer to \cite{papamakarios2017masked} for a full treatment of the remaining technical details.

MAFs use a Masked Autoencoder for Distribution Estimation~\cite[MADE;][]{GG2015} to encode the autoregressive model blocks $M_j$ described above. 
For disambiguation, we will use $j$ to index the $j$-th model block $M_j$ and their associated input data vector $\mathbf{u_j}$, and $i$ to index the $i$-th data point in that data vector. 
For a given  block $M_j$, MADEs are neural networks that model the functions $\{f_{\mu_{j,i}}, f_{\sigma_{j,i}}\}$ in \myequation{}'s~\ref{eq:autoscalar1}-\ref{eq:autoscalar2} for all $i$. 
More specifically, it takes as input the data of the previous block, $\mathbf{u_{j-1}} = [u_{j-1,1},...,u_{j-1,i},...,u_{j-1,n}]$ and outputs the mean $\mu_{j, i}$ and log standard deviation $\sigma_{j, i}$, which is then used to generate $u_{j, i}$ for all $i$ using \myequation{}~\ref{eq:autoscale}. 
The advantage of MADE over other autoregressive model architectures is that it avoids the need to compute the recursion in \myequation{}'s~\ref{eq:autogauss}-\ref{eq:autoscale}. 
By judiciously applying binary masks to some of the weight matrices in MADE, such that the autoregressive property in \myequation{}'s~\ref{eq:autogauss}-\ref{eq:autoscale} are enforced, the function $\{f_{\mu_{j,i}}, f_{\sigma_{j,i}}\}$ for all $i$ can be calculated efficiently in a single forward pass. 
We refer to \cite{GG2015} for a full treatment of the technical details.

\subsubsection{Data Compression with Summary Statistics}
\label{sec:embed_discussion}

For NPE models in which the number of dimensions of the data space $X$ is relatively small ($\lesssim100$), the data vectors can be directly used as inputs to the NPE model itself. 
However, when the dimensionality of the data is large ($\gtrsim1000$) as with imaging data, it can become expensive computationally to learn the sampling distribution of the data as a function of the parameters. 
Traditional inference methods address this problem by constructing summary statistics that reduce the dimensionality of the data to a smaller number of summary statistics. 
The inference process, which may involve comparing the summary statistics of the observed data to the simulated data or performing density estimation of the model parameters over the summary statistics, becomes far more computationally tractable. 
This approach has also been extended to NPE methods and has become a common practice when performing inference on high-dimensional data  \cite[e.g.,][]{CB2019, AC2019, Charnock_2018, rodrigues2021hnpeleveragingglobalparameters}. 

The choice of summary statistics can have a big impact on the inference process, as a well-chosen set of summary statistics will preserve the largest amount of information from the original data, while an ill-chosen set of summary statistics can result in significant information loss that impacts the quality of the inference. 
The choice of summary statistics depends on the dataset itself. 
As such, summary statistics are usually prescribed by domain experts, guided by their intuition and knowledge of the subject. 
In recent years, however, there have been efforts to systematize the construction and evaluation of summary statistics, especially in relation to its use in NPE~\cite[e.g.,][]{Blum_2013, prangle2015summary}. 
In machine learning, {features} or {representations} are often synonymous with summary statistics, and advancements have been made in developing methods to learn or extract summary statistics/features in an automated fashion in recent years \cite{bengio2014representation}. 
Convolutional neural networks (CNN) are a common type of neural network architecture used for this purpose, and it remains state-of-the-art in numerous machine learning applications, including, saliently, computer vision problems \cite{li2022cnn}.  
In this work, we leverage these machine learning advancements by using a CNN to compress the output of the simulators (lens imaging data) into a lower-dimensional vector of summary statistics, which are then passed on as inputs to the density estimator. 
The parameters of both the summary CNN and the density estimator are then learned together during training.  

\subsubsection{NPE implementation} 
\label{sec:npe_implementation}

In this section, we describe the model architectures we use in this work and the hyper-parameters we chose for those models. 
We used NPE (\mysection{}~\ref{sec:npe_overview}) as implemented in Macke Lab \sbi{}~\cite{tejero-cantero2020sbi} to train an MAF-based neural density estimator (\mysection{}~\ref{sec:maf}) to infer the lens parameter posteriors from simulated single-band lens images. 
The simulated lens images are first passed through a custom CNN which compresses the higher dimensional image data into informative summaries statistics (\mysection{}~\ref{sec:embed_discussion}). 
This CNN, also known as an embedding network, comprises six convolutional layers (Conv2d), with each convolutional layer followed by a batch normalization layer (BatchNorm2d). 
We perform a max pooling operation (Maxpool2d) after alternating convolutional layers --- specifically after the 2nd, 4th, and 6th convolutional layers. 
The output from the convolutional layers is flattened before being passed to the fully connected (Linear) layer.
Table~\ref{table:embedding_net} provides a detailed summary of the architecture.

In our experiments, we output $4P$ summaries, where $P$ is the number of model parameters we aim to infer. 
Through trial-and-error experimentation, we found that the performance of the model did not demonstrate improvement when the number of summaries per inferred parameter was increased beyond four summaries per parameter for the data used in this work. 
In later sections, we apply the method to models with one, five, and 12 parameters, for which we used four, 20, and 48 summaries, respectively. 
The summaries from the embedding network are inputs to the MAF density estimator. 
The embedding network is simultaneously trained with the density estimator and acts as an automatic summary statistic/feature extractor. 

The MAF implementation in \sbi\ allows for a variety of hyperparameters to be adjusted, with the most important two being the number of hidden units (nodes) in each hidden layer of each MADE block in the MAF, and the number of flow transformations in the MAF, as these hyperparameters change the structure of the MAF architecture. 
We tuned the model by testing hyperparameters until a stable and performant model was identified --- increasing the number of hidden units up to 1000 and the number of transformations up to 50. 
Each model was trained on the same data set and evaluated on the same test set. 
We evaluated each model on its posterior coverage calibration until we noticed no further improvement in its coverage (\mysection{}~\ref{sec:eval_poscoverage}). 
We settled on the following hyperparameter choices for each model: for the 1-parameter lens model, we used 100 hidden units and 10 transformations. 
For the 5-parameter model, we used 120 hidden units and 20 transformations. For the 12-parameter model, we used 400 hidden units and 20 transformations. 

For the remaining hyperparameters, we opted to leave them at their default values in the \sbi{} library, as modifying them did not appear to result in any significant gains. 
The adaptive learning rate optimizer \texttt{Adam} \cite{kingma2017adam} was used, with the initial learning rate and batch size set to 0.0005 and 50, respectively.
The validation fraction was set to 0.1 of the training set. 
The training was automatically terminated if there was no improvement in training loss on the validation set after 20 epochs.
In contrast with cosmology, in deep learning, an epoch is defined as a single pass of the training dataset through the training process.

\tableembeddingnetnpe

\subsubsection{Why not use Sequential NPE (SNPE)?}
\label{sec:SNPE_discussion}

As mentioned in \mysection{}~\ref{sec:npe_overview}, NPE methods have an optional active learning component, {sequential} estimation, which we did not use. 
In this section, we briefly describe sequential estimation, the trade-offs involved when using it, and why we ultimately chose not to use it in this work. 
Sequential inference involves training a density estimator iteratively over several rounds to learn a proposal prior distribution $\Tilde{p}(\theta)$. 
This proposal prior distribution is then used to simulate the $N$ pairs of training data $\{\theta_i, X_i\}^N_{i=1}$ used to train the target posterior density estimator. 

The motivation behind this sequential step is as follows: if we are interested in the posterior distribution corresponding to a given observation $X_0$, $p(\theta|X=X_0)$, the training data that is most informative for training $q_{\phi}(\theta|X=X_0)$ to accurately approximate $p(\theta|X_0)$ should be drawn from a distribution that closely matches $p(\theta|X_0)$. 
The prior $p(\theta)$ is typically much wider than $p(\theta|X_0)$, so the vast majority of the simulated training data generated from the prior $p(\theta)$ is uninformative for training $q_{\phi}(\theta|X_0)$ to approximate $p(\theta|X_0)$. 
Therefore, by first learning a proposal prior $\Tilde{p}(\theta)$ that closely matches $p(\theta|X_0)$, we require fewer simulations sampled from $\Tilde{p}(\theta)$ to accurately train $q_{\phi}(\theta|X_0)$ to approximate $p(\theta|X_0)$. 

While this sequential step improves the sampling efficiency in principle, it has two major drawbacks.
The method is no longer amortized. 
For every new data point $X_i$ for which we want to infer the posterior $p(\theta|X_i)$, a new proposal prior has to be first learned for each datum. 
This additional training cost for each new datum may offset any improvements in sampling efficiency in cases where the objective is to rapidly infer the posteriors for a great number of data points.
This breaks amortization.
Sequential methods have additional failure modes and complications because there is no guarantee that sequential rounds of training will result in a proposal prior that is appropriate for training a target posterior. 
Examples of such failure modes include posterior leakage, which can cause inference to be prohibitively expensive over multiple rounds \cite{deistler2022truncated}, as well as overconfident uncertainty predictions \cite{hermans2022trust}.
SNPE appears to work best in cases where the number of evaluated data points are evaluated {separately} is relatively small. 
As such, we do not think SNPE is suitable for our use case, as our objective in this work is to study a pipeline to model a large number of lens systems as efficiently as possible. 
This is consistent with the recommendations by \cite{lueckmann2021benchmarking} in their benchmarking work.

\tableembeddingnetbnnoned
\tableembeddingnetbnn

\subsection{Bayesian Neural Networks}
\label{sec:bnn}

To compare the results of our NPE method to another deep learning-based method, we train a BNN~\cite{NR1996} on the same simulated datasets. 
With BNNs, the deterministic weights of a standard CNN network are replaced by probability distributions, which can be used subsequently to provide a measure of the uncertainty. 

The objective of training a BNN is to find the posterior distribution $p(\phi|\cal{D})$ of the weights $\phi$, given the training datasets ${\cal D} =(X,\theta)$, where $X$ are the inputs and $\theta$ are the corresponding labels. 
Since inference of the BNN posterior is a very difficult task that includes calculation of an integral over all model weights, approximate methods are utilized. 
Here we use variational inference to model the posterior using a simple variational distribution $q(\phi|\zeta)$, which we take to be a Gaussian (i.e., $\zeta$ consists of the mean and standard deviation of the Gaussian distribution), and fit the distribution’s parameters $\zeta$ to be as close as possible to the true posterior $p(\phi|{\cal D})$~\cite{GR2011, SL2019}. 
This is done by minimizing the Kullback-Leibler (KL) divergence~\cite{KL1951} between the true and variational posterior probability distributions:
\begin{equation}
    \mbox{KL}(q(\phi|\zeta) || p(\phi|{\cal{D}})) \equiv \int q(\phi|\zeta) \log \frac{q(\phi|\zeta)}{p(\phi|{\cal{D}})} d\phi.
\end{equation}
\noindent This produces the objective of BNN training, which is the minimization of the Evidence Lower Bound (ELBO) loss that comprises the negative log-likelihood loss and the KL divergence:
\begin{equation}
\label{eq:loss}
{\cal{L}}({\cal{D}},\zeta) =  \mathbb{E}_{q(\phi|\zeta)}\left[\log q(\phi|\zeta) - \log p(\phi)p({\cal{D}}|\phi)\right].
\end{equation}

The outputs of the BNN are multivariate normal distributions --- i.e., BNN outputs both mean and variance for each of the output parameters. 
However, the utility of BNNs is generally limited by the distributions used to model the neural network weights, which in practice are typically Gaussian or Gaussian mixture models. 
If the true posterior distribution follows a more complicated distribution, BNNs may not be flexible enough to approximate the correct distribution.

The BNN architecture details for the 1-parameter model are provided in Table~\ref{table:BNN1d}, and the architecture for the 5- and 12-parameter models is provided in Table~\ref{table:BNN}. 
To avoid overspecification (e.g., overparameterization), the 1-parameter model is less complex than the 5- and 12-parameter model.
Conv2dFlipout denotes the 2D convolution layer with Flipout gradient estimator~\cite{wen2018flipout}, and DenseFlipout denotes the Dense layer with Flipout gradient estimator, which is used for decorrelating the gradients within a mini-batch by implicitly sampling pseudo-independent weight perturbations for each example. 
MNTriL denotes the output multivariate normal distribution. 

We trained the network for $500$ epochs as we found that increasing the number of training epochs did not significantly improve the training loss, and we used the Adadelta optimizer~\cite{Z2012} with an initial learning rate of 0.08. 
Like \texttt{Adam}, the Adadelta optimizer has an adaptive learning rate, and the results are insensitive to the choice of initial learning rate. 

\subsection{Computing Environment}
We employed an Nvidia RTX A5000 GPU for training purposes. 
The 1-, 5- and 12-parameter models took about 1, 2.5, and 6 hours respectively for both NPE and BNN models.

In comparison, a conventional MCMC fit would take around the order of an hour per lens system.

\section{Data: Simulations of Strong Lenses}
\label{sec:simulations}

To develop the NPE and BNN models, we consider three levels of model complexity: 1) a baseline $1$-parameter model, where we only vary the Einstein radius, \einsteinrad; 2) a $5$-parameter lens-only model, where the external shear and source galaxy light parameters are fixed and only the lens mass parameters are varied; 3) a $12$-parameter model, where lens mass, source light, and environmental shear parameters are varied. 
As we will describe in \mysection{}~\ref{sec:results} and \mysection{}~\ref{sec:discussion}, we have found that for more model parameters, a larger training set is needed to reach good model performance. 
We use the same training and test data for the NPE and BNN models. 
For the BNN models, we generate an additional validation dataset for model training. 
For the NPE model, $10\%$ of the training data is used as a validation dataset, as internally set by the \sbi{} package. 
Table~\ref{table:dataset_size} shows the sizes of datasets used in our experiments. 

For the 1- and 5-parameter models, we model the light from the source galaxy as a fixed circular Sersic profile 
with $n=4$, $R=0.35''$, and $m_s=21$, which are typical values found in cosmic surveys \cite{SLACS4, Collett2015}. 
In these two experiments, we do not include external shear. 
Furthermore, for the simple 1-parameter model, we vary only the Einstein radii, keeping the lens eccentricity fixed at zero and with a fixed lens-source position offset $(x_c,y_c)=(0.2'',0.0'')$. 

\tabledatasetsizes

We use \deeplenstronomy~\cite{MN2021}, which is built on \lenstronomy ~\cite{BA2018, BS2021}, to simulate strong lensing data~\cite{BA2018, BS2021}.
These packages together provide state-of-the-art simulations of lensing systems, including high-fidelity ray-tracing of light from source objects around lenses, emulation of cosmic survey observations, and the generation of sample distributions. 

To mimic data from cosmic survey experiments, we emulate empirical DES Data Release 1 (DR1) survey conditions following \cite{DESdatarelease1}. 
The Dark Energy Camera (DECam) has a pixel scale of 0.263 arcsec/pixel, CCD gain of 6.083 e$^-$/count, and read noise of 7.0 e$^-$. 
We draw on empirical distributions of sky brightness, atmospheric seeing, and the effective number of exposures from Figures 4, 5, and 6 of \cite{DESdatarelease1} respectively\footnote{After the paper was finalized, we encountered an error in the magnitude calculations in \deeplenstronomy{}. 
This does not affect the main conclusions of the work, so we did not re-run the experiments.
The error is discussed in Appendix~\ref{AppMagnitudes}}.

In this work, we opted to restrict the scope of our simulation pipeline to single-band imaging in order to focus more on demonstrating the utility of these deep learning methods. For the single-band,
we simulate $g$-band images with a magnitude zero point of 26.58. 
The choice to emulate g-band imaging is because it is a fairly representative photometric band in which optical surveys have observed a wide variety of lensed galaxies and lensed features in the past (see, e.g. \cite{Knabel_2023,Collett_2017}). We plan to explore how to incorporate multiple-band imaging in a deep learning pipeline in a future work.

In generating the simulation samples, we draw from uniform prior distributions of each parameter, which are based on the ranges we expect to encounter in real data (Table~\ref{table:params}).
We restrict the test set prior range to cover a portion of the training set range for each parameter.
This helps to mitigate biases in the training near the edge of the training range.

\begin{figure}
    \centering
    \includegraphics[width=0.99\linewidth]{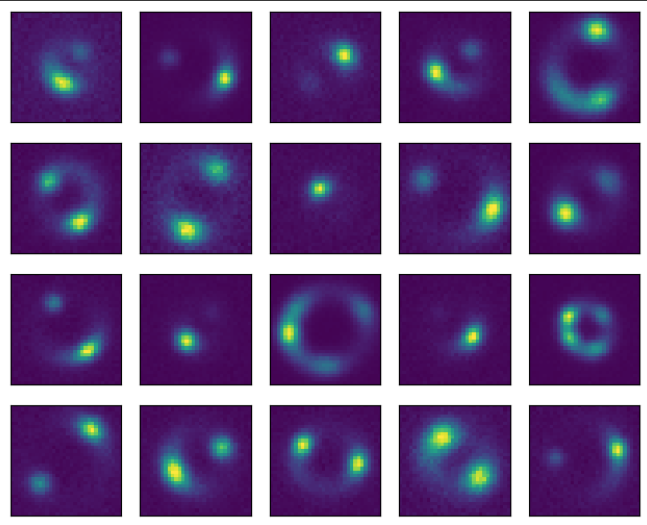}
    \caption{
    20 randomly selected examples of $g$-band images in the test set used in the 5-parameter model. 
    To mimic the lens light subtraction during traditional lens modeling, the lens light has not been added, so only the lensed source light is shown.
    The source magnitudes are $m_s = 21$, which makes them appear less noisy than some of the images in \myfigure{}~\ref{fig:12paramsingleparam_examples}, which have a larger variation in magnitude.
    }
    \label{fig:5paramsingleparam_examples}
\end{figure}

\begin{figure}
    \centering
    \includegraphics[width=0.99\linewidth]{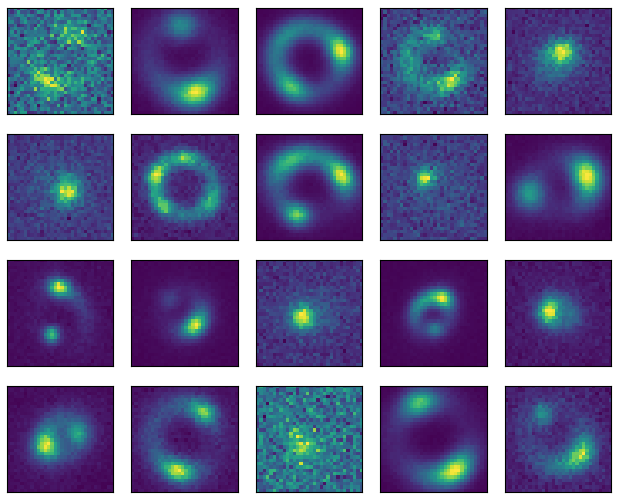}
    \caption{
    20 randomly selected examples of $g$-band images in the test set used in the 12-parameter model. 
    To mimic the lens light subtraction during traditional lens modeling, the lens light has not been added during the simulation phase, so only the lensed source light is shown.
    The source magnitudes in these images have a wide range: $ 18 < m_s < 25$. 
    Due to the magnitude range, some fainter objects appear noisier than the images in \myfigure{}~\ref{fig:5paramsingleparam_examples}.
    }
    \label{fig:12paramsingleparam_examples}
\end{figure}

\myfigures{}~\ref{fig:5paramsingleparam_examples} and \ref{fig:12paramsingleparam_examples} show $20$ randomly selected lenses from the 5- and 12-parameter test sets, respectively. 
All the images used for training and testing are single-band images that are $32 \times 32$ pixels in size. 

The data we used for this work are published on \href{https://doi.org/10.5281/zenodo.13961234}{Zenodo}\footnote{10.5281/zenodo.13961234}. 
We include code for generating data, training models, and evaluating models in a \href{https://github.com/deepskies/DeepLensSBI}{GitHub Repository}\footnote{\url{https://github.com/deepskies/DeepLensSBI}}.

\tableparameterdistributionmain

\section{Model Evaluation Diagnostics}
\label{sec:modeldiagnostics}

For each model in this work, we performed multiple statistical and stability tests.
The statistical checks include simulation-based calibration (SBC), which operates in parameter space \cite{Talts2018, stanhyunjimoon}, and posterior predictive checks (PPCs), which operate in the observable/data space. 
We tested the model's stability with respect to the random initialization of the network parameter weights and its generalization capacity.

\subsection{Simulation-based Calibration}

\subsubsection{Posterior Coverage}
\label{sec:eval_poscoverage}

For a model to be trustworthy, it must produce uncertainty estimates that are credible and well-calibrated. 
The posterior coverage measures how well-calibrated the predicted uncertainties of the posterior models are. 
For a posterior distribution to be well-calibrated, it must contain the true value $p\%$ of the time in $p\%$ of the posterior probability volume. 
For example, a posterior distribution with 68\% (1$\sigma$) confidence intervals should contain the true value within those confidence intervals 68\% of the time. 
Moreover, these uncertainties will have to be averaged across a large ensemble of lenses in a test set, as the estimated posteriors of a single lens image alone may not be indicative of the average performance of the model.

We present the model posterior coverage, which is calculated using the following steps: 
\begin{enumerate}
    \item Create a test set of 1000 lens images as described in \mysection{}~\ref{sec:simulations}.
    \item For each lens in the test set, use the trained posterior estimator to obtain an estimate of the posterior distribution of the model parameters for that particular lens by sampling the posterior estimator model.
    \item From each posterior distribution, calculate parameter confidence intervals for that lens. 
    \item For each confidence interval, calculate the fraction of the lenses in the test set whose true parameter values fall within that confidence interval range. 
    \item For each parameter, plot the fraction of the lenses whose true parameter values fall within the confidence interval as a function of the confidence interval. 
\end{enumerate}

If the model's posterior uncertainties are perfectly calibrated for the entire test ensemble, we expect the posterior coverage to follow the 1-1 line (\myfigure{}~\ref{fig:1_param_main_PC}). 
If the posterior coverage falls above the 1-1 line, it is underconfident because it overestimates the uncertainties, and a larger fraction of the test set's true values fall within the given confidence interval than predicted by the model. 
Conversely, if it falls below the 1-1 line, it is overconfident and underestimates the true uncertainties.  

\subsubsection{Combined Distance Metric}
\label{sec:eval_distance_metric}
In addition to calculating the posterior coverage for individual parameters, we calculate the distance metric from \cite{wagner2021hierarchical}, defined as:

\begin{align}
\label{eqn:distance_metric}
D(\bm{\theta}_i) = (\bm{\theta_i}-\bm{\mu_{\theta_i}})\Sigma_{\textrm{data}}(\bm{\theta_i}-\bm{\mu_{\theta_i}})^T ,
\end{align}
where $\bm{\mu_{\theta_i}}$ is the vector of the means of the $i$-th lens' posterior parameters and $\Sigma_{\textrm{data}}$ is the empirical covariance matrix for the training set parameters. 
$\bm{\theta_i}$ is a vector of parameter values of a posterior sample for the $i$-th lens. 
The distance metric $D(\bm{\theta}_i)$ combines a vector of parameter values (e.g., from a posterior sample) into a single objective function that takes into account the covariance between different parameters. 

\subsubsection{Residuals}
\label{sec:eval_scatterplot}

To characterize the statistical performance of the best-fit values over an ensemble of test images, we use the residuals. 
We calculate it as follows: 
\begin{enumerate}
    \item For each lens in the test set, we use our model/s to infer the lens parameter posteriors. 
    \item We then subtract the best-fit (maximum likelihood) parameter values from the true value of the lens parameters for each lens to obtain the residual.
    \item  For the case of a single-parameter fit, the plot shows the true posterior as a function of the residual. 
    For the case of a multi-parameter fit, we plot the residuals for all lenses in the test set as a multidimensional (sometimes called a ``corner'' or ``triangle'') plot. 
\end{enumerate}
We show 68\% and 95\% contours of the residuals (\myfigure{}~\ref{fig:5paramcorner}). 
The multi-parameter residual plots are similar to parameter covariance (``corner'' or ``triangle'') plots of an MCMC sampling analysis \cite[][]{Bocquet2016, corner, lewis2019getdist}, which displays covarying scatter for every pair of parameters and a probability density of each parameter in some parameter space.
The difference is that in an MCMC corner plot, each scatter point represents a posterior sample from the MCMC chain. 
However, in the plots described here, each scatter point represents the difference between the model's best-fit estimate and the ground truth values of the parameters of one image in the test set. 

The scatter characterizes the average best-fit point estimate of the model across the entire test ensemble and can be thought of as a multidimensional variant of the true-vs-predicted scatter plot visualization diagnostic that is common in machine learning literature. 
If the method is accurate (predicts the true value for every single lens), one expects a single point at (0,0). 
If the method produces systematically biased best-fit point estimates, we expect the scatter to shift away from the origin, and if the method is not precise, we expect a large scatter in the distribution. 

We present this diagnostic plot for two reasons.
First, if the models are systematically biased in a way that is correlated between any pair of parameters, we will be able to identify highly skewed (degenerate) contour plots for those pairs of parameters. 
It would not be possible to visualize correlated biases in true-vs-predicted scatter plots. 
Second, when plotting true-vs-predicted scatter plots for a large number of test images with their associated uncertainties, the visualization becomes cluttered, and it is difficult to discern structure in the inferences. 

\subsubsection{Rank Histogram}
\label{sec:eval_sbc}

Simulation-based calibration \cite[SBC;][]{Talts2018} tests are a general method of diagnosing and validating Bayesian inference methods based \cite{Cook2006}. 
The SBC procedure is:  

\begin{enumerate}
    \item From the prior, we randomly generate $N$ independent parameter samples $\theta_i$. 
    \item We simulate observations of each lens image $X_i$. 
    \item From the trained NPE model, we obtain $M$ independent randomly chosen posterior samples $p_m(\theta_i|X_i)$ from each of the $N$ posterior models.
    \item For each NPE model, we count the number of samples of $p_m(\theta_i|X_i)$ that fall below the corresponding true parameter value $\theta_i$. 
    This number is the rank.
\end{enumerate}

The main principle behind SBC is that if the trained posterior model is well-calibrated, the SBC ranks should be uniformly distributed between 0 and $N$. 
The reason for this is that samples from each of the randomly chosen posteriors should be distributed according to the prior. 
This corresponds to the claim that the $p\%$ confidence interval of this posterior model should contain the simulated value in $p\%$ of simulations for all values of $X$. 
Histograms of the SBC rank distributions serve as a diagnostic for checking if the posterior model is correctly computing the posterior. 
Deviations from uniformity in SBC rank histograms indicate how the computed posteriors are systemically biased. 
Specifically, if the histogram is:

\begin{itemize}
    \item left-skewed: The posterior model is biased high compared to the true posterior and systematically overestimates the posterior mean. 
    \item right-skewed: the posterior model is biased low compared to the true posterior and systematically underestimates the posterior mean.
    \item $\cup$-shaped: On average, the posterior model is narrower than the true posterior and systematically underestimates the uncertainty. 
    \item $\cap$-shaped: On average, the model posterior is wider than the posterior model and systematically overestimates the uncertainty.  
\end{itemize}
\noindent In addition to the visual indications of bias listed above, we plot the 99\%-confidence interval of a uniform distribution given the number of samples provided. 
For a uniform distribution, we expect 1 out of 100 histogram bars to lie outside the confidence interval. 
In the SBC plots, this is indicated by a gray region (\myfigure{}\ref{fig:1_paramSBC}, top row).

\subsubsection{Rank Cumulative Distribution Function (CDF)}
\label{sec:eval_cdf}

The rank CDF is another method of presenting the rank information from \mysection{}~\ref{sec:eval_sbc}.
If the posterior is perfectly calibrated, one expects a perfectly diagonal CDF, which is equivalent to a uniformly distributed rank in the histogram. 
We plot the 99\%-confidence interval of a uniform distribution given the number of samples provided, indicated by a gray region (\myfigure{}\ref{fig:1_paramSBC}, bottom row).  
If the posterior model is systematically biased high (low), the rank CDF will deviate above (below) the diagonal line. 
If the posterior model systematically overestimates (underestimates) the uncertainty, the CDF will resemble a diagonal 'S' (inverted 'S') shape.  
The choice of visualization of ranks involves some trade-offs.
CDFs do not depend on the choice of histogram binning size \cite{stanhyunjimoon}. 
The plots also allow for multiple parameter rank CDFs to be overlaid on the same plot, resulting in more compact visuals compared to the histogram. 
However, it is often easier to visually identify model pathologies from the rank histograms. 
Due to these trade-offs, we opt to show both rank histograms and CDFs for our models.  

\subsection{Posterior Predictive Checks (PPCs)}
Posterior predictive checks \cite{PPC} are a commonly used diagnostic (\myfigure{~\ref{fig:1_param_main_single}}, top row). 
While they have been found lacking as a validation metric \cite{lueckmann2021benchmarking} for SBI, PPCs can be used as a starting point for diagnosing NPE performance by providing visual intuition about potential biases in the model's inference.  
These show how the best-fit parameters for a given model are reflected in the reconstructed data model of the lensing system.

\subsection{Stability Tests}
\label{sec:eval_seeds}

Network weights are randomly initialized, which allows variation in results amongst models trained on the same data \cite{Picard2021}. 
While the effects of initialization don't generally produce a large variance in the performance \cite{Picard2021}, it is not difficult to fortuitously pick random seeds that perform significantly differently from the average. 
While the use of those seeds does not mean the results of those models are invalid, their performance may not be representative of the typical performance of that particular deep learning architecture, making it challenging to determine a model's true performance. 
Therefore, it is important to diagnose the stability across multiple seeds. 
We train NPE models with three different seeds. 
We do not perform this diagnostic for the BNN models because the computation is prohibitively expensive, and BNNs are not the primary focus of this study.

\section{Results}
\label{sec:results}

We present the results of NPE and BNN model inference with three levels of complexity: a baseline 1-parameter model where one lens parameter is varied and inferred; a 5-parameter model where five lens parameters are varied; and a 12-parameter model that varies the seven lens-plane parameters and five source-plane parameters. 

\subsection{1-parameter Model: Varying One Lens Parameter}
\label{sec:results_theta}

\begin{figure}
    \centering
    \includegraphics[width=0.5\linewidth]{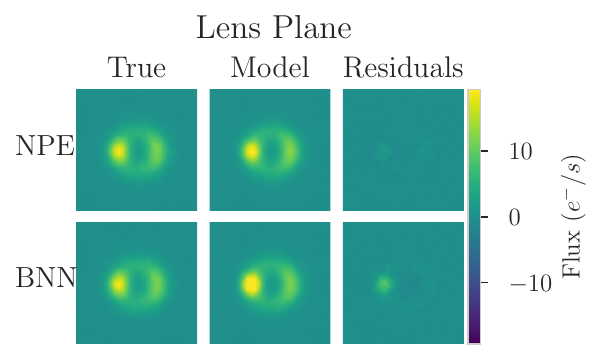}
    \includegraphics[width=0.5\linewidth]{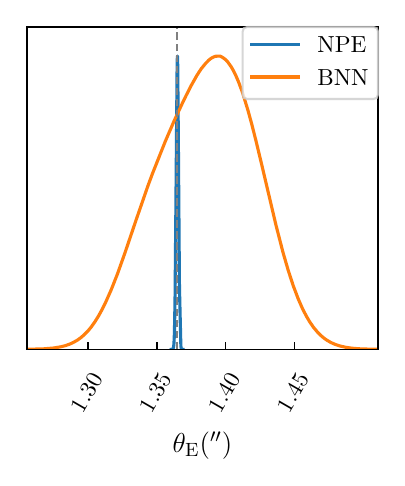}
    \caption{
    Performance for a single image in the 1-parameter NPE and BNN models. 
    Top: the posterior predictive check for the lens plane image for the NPE (upper row) and the BNN (lower row) models, which includes the true image, the image predicted by each method, and the residual between the true and the predicted image for this test object.
    Bottom: inferred posterior of the Einstein radius $\theta_E$ for the NPE (blue) and BNN (orange) models.
    We provide more examples in Appendix~\ref{AppA}.
    }
    \label{fig:1_param_main_single}
\end{figure}

We use the 1-parameter model as a starting point for studying training regimes and as a baseline for comparison with models of greater complexity. 
In this problem, as described in \mysection{}~\ref{sec:methods} and \mysection{}~\ref{sec:simulations}, we vary the Einstein radius while leaving other parameters fixed. 
We trained the models on 200,000 images in which the Einstein radius ranged from 0.3 to 4.0 arcsec. 
We used 100 hidden features and 10 transformations for the NPE model. 
\myfigure{}~\ref{fig:1_param_main_single} shows the posterior distribution of $\theta_E$ inferred by both the NPE and BNN model for a single lensed image in the test set. 
This figure also shows a comparison of the posterior predictive check for the NPE and BNN models: the true image, the image reconstructed with the best-fit parameters, and the residuals between the true and reconstructed image. 
In this simple problem, both NPE and BNN models are extremely accurate and precise, inferring the true $\theta_E$ value with $\sim10^{-3}$ arcsec and $\sim10^{-2}$ arcsec $68\%$ uncertainties, respectively. 
However, the NPE method is both more accurate and $\sim10$ times more precise than the BNN model.

\begin{figure*}[ht]
\begin{center}
\includegraphics[width=0.4\linewidth]
{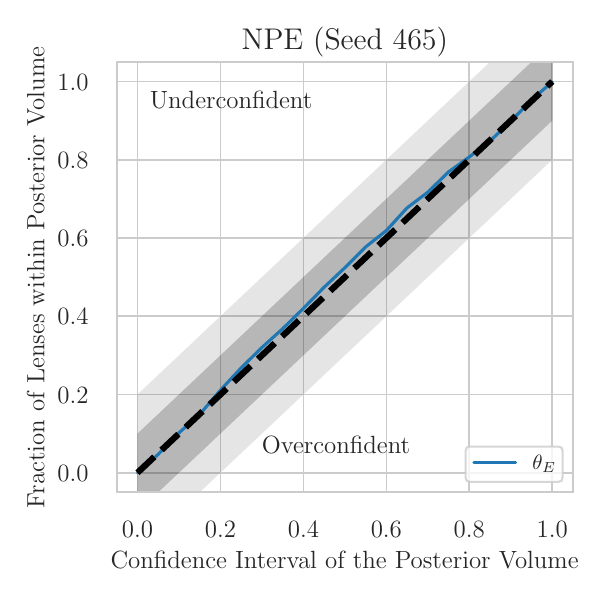}
\includegraphics[width=0.4\linewidth]
{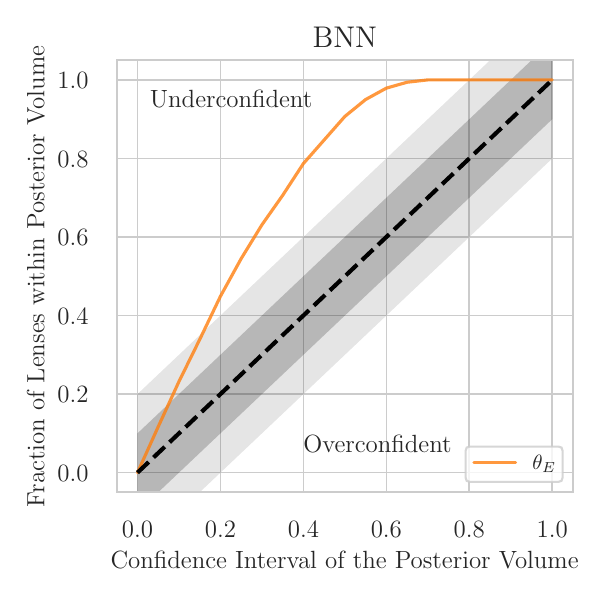}
\caption{
Posterior coverage of the 1-parameter NPE (left, blue) and BNN (right, orange) models computed on the test set (solid).
The 1-1 line (black, dashed) indicates perfect uncertainty calibration. 
The gray regions indicate thresholds of 10\% (dark gray) and 20\% (light gray) uncertainty miscalibration. 
}
\label{fig:1_param_main_PC}
\end{center}
\end{figure*}

\begin{figure*}
    \centering
    \includegraphics[width=0.49\linewidth]{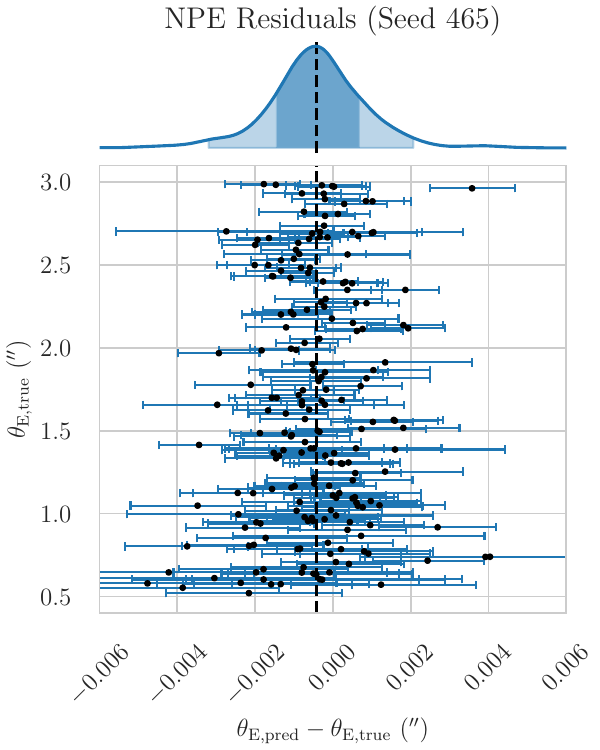}
    \includegraphics[width=0.49\linewidth]{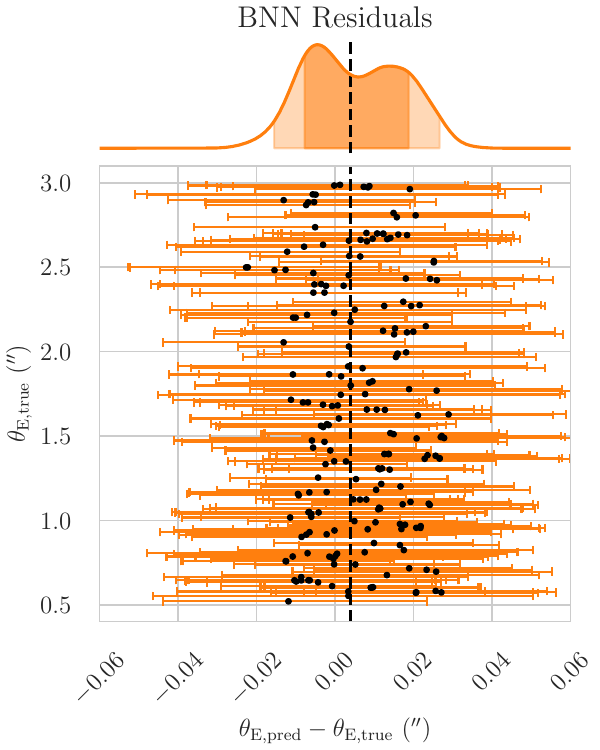}
    \caption{
    Residuals of the inferred posterior of the Einstein radius \einsteinrad{} in the 1-parameter NPE (blue, left) and BNN (orange, right) models for 200 images randomly selected from the test set.
    The error bar represents the 68\% uncertainty in \einsteinrad{} for each image. 
    Above the scatter plot of the residuals, the marginal distribution of best-fit values for the full test set of 1000 is shown.
    The 68\% and 95\% scatter in best-fit values are indicated by the dark and lighter-shaded regions, respectively.
    The NPE model random seed for weight initialization is set to $465$.
    \label{fig:1paramcorner}
    }
\end{figure*}

We quantify model performance primarily in two ways. 
First, we study the posterior coverage (\mysection{}~\ref{sec:eval_poscoverage}). 
\myfigure{}~\ref{fig:1_param_main_PC} shows the posterior coverage for the NPE and BNN models. 
The NPE method is slightly underconfident (within 5\% of perfect calibration), while the BNN model is extremely underconfident. 
Second, the residuals show the deviations of posteriors from the true parameter for 1000 test images. 
The lower panels of \myfigure{}~\ref{fig:1paramcorner} show the scatter of residuals in point values with error bars (68\% confidence intervals) for 200 randomly selected test images.
We also show the distribution of residuals in the upper panel of that figure. 
We perform kernel density estimation on the posterior distributions with a bandwidth calculated using Scott's rule \cite{scot1992multivariate}, with the 68\% and 95\% confidence intervals in the distribution of the residuals. 
The 68\% confidence intervals in the residuals are $\sim10^{-3}$ arcsec for the NPE model and $\sim1-2 \times 10^{-2}$ arcsec for the BNN model, with the predicted model uncertainties in the same regime. 
This is in agreement with the single-image example in \myfigure{}~ \ref{fig:1_param_main_single} and further validates that specific example as representative of the typical performance of both models. 
The distribution of residuals in the BNN model (\myfigure{}~\ref{fig:1paramcorner}, right) appears to exhibit bimodality, pointing to non-Gaussian systematic uncertainty in the BNN model point estimates.

To study model stability, we change the random seed initialization on the NPE model. 
We present the main results of those diagnostics in \myappendix{}~\ref{1paramappendix}.

\subsection{5-parameter Model: Varying Five Lens Parameters}
\label{sec:results_five_param}

\begin{figure}
    \centering
    \includegraphics[width=0.6\linewidth]{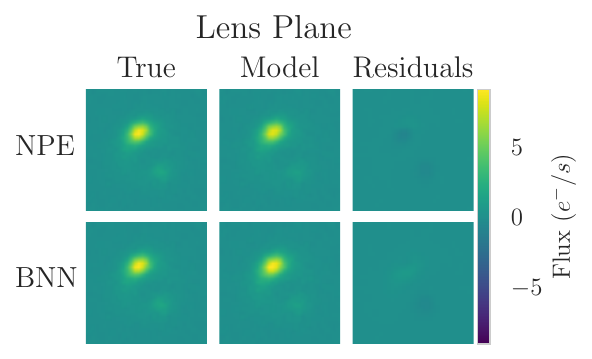}
    \includegraphics[width=0.6\linewidth]{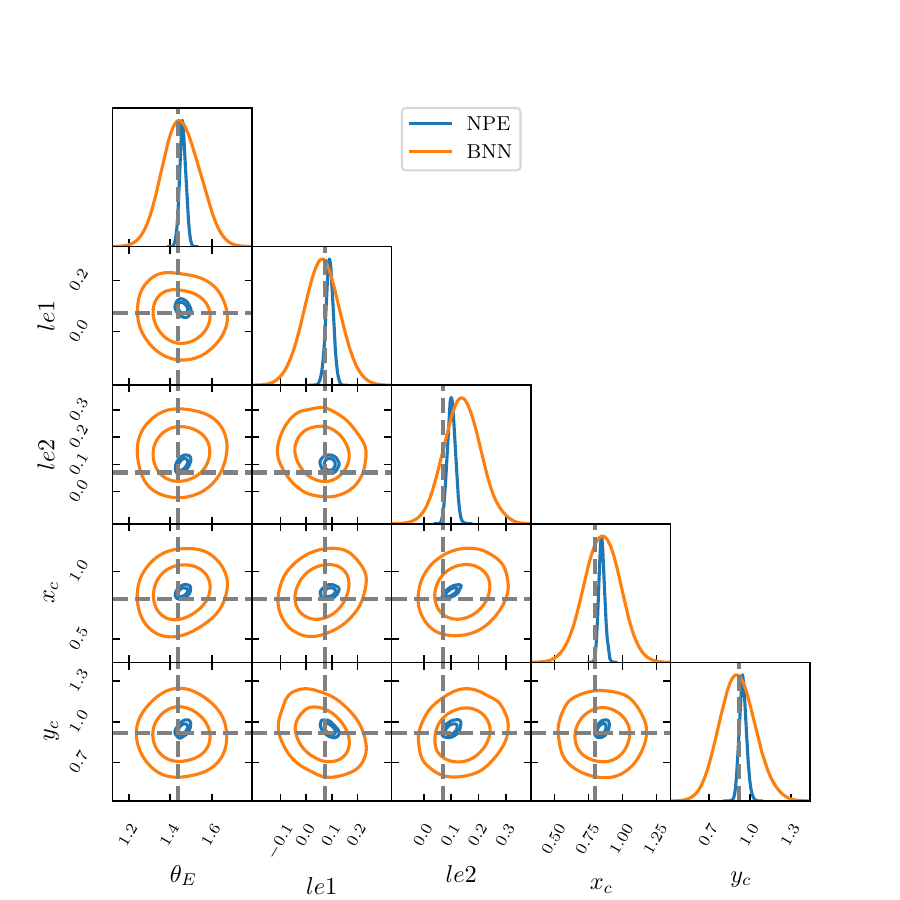}
    \caption{
    Performance of the 5-parameter NPE and BNN models for a single lensing system from the test set. 
    Top: the posterior predictive check for the lens plane images for the NPE (upper row) and the BNN (lower row) models, which includes the true image, the image predicted, and the residual between the true and the predicted image.   
    Bottom: inferred posterior distributions of five parameters for the NPE (blue) and BNN (orange) models. 
    The value and 68\% uncertainty listed at the top of each column are for the NPE model. 
    The gray dashed line indicates the true value.   
    We provide more examples in Appendix~\ref{AppA}.}
    \label{fig:5paramsingle9}
\end{figure}

\begin{figure*}
\begin{center}
\includegraphics[width=0.45\linewidth]{images/5_param_npe_pos_465}
\includegraphics[width=0.45\linewidth]{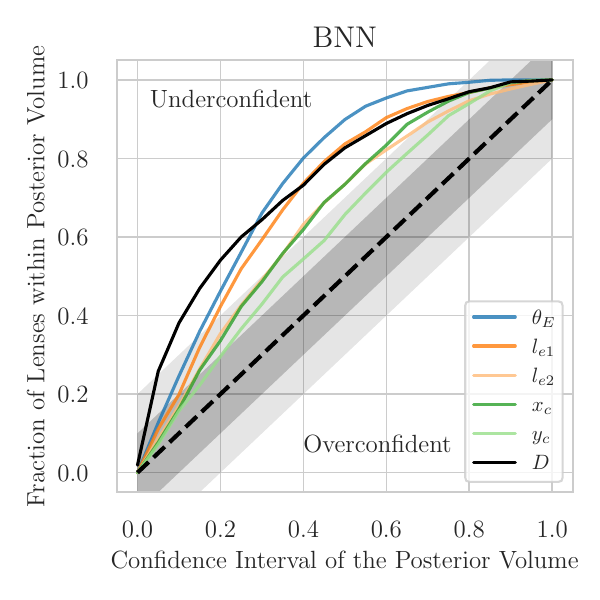}
\caption{
    Posterior coverage for the five-parameter NPE (left) and BNN (right) models for the test set. 
    The 1-to-1 line (dashed, black) indicates perfect uncertainty calibration. 
    The gray regions indicate thresholds of 10\% (dark gray) and 20\% (light gray) uncertainty miscalibration. 
    The solid black line represents the distance metric described in \myequation{}~\ref{eqn:distance_metric}.
    The NPE model random seed for weight initialization is set to $465$.
    }
    \label{fig:5param_poscov}
\end{center}
\end{figure*}

\begin{figure*}
    \centering
    \includegraphics[width=0.6\linewidth]{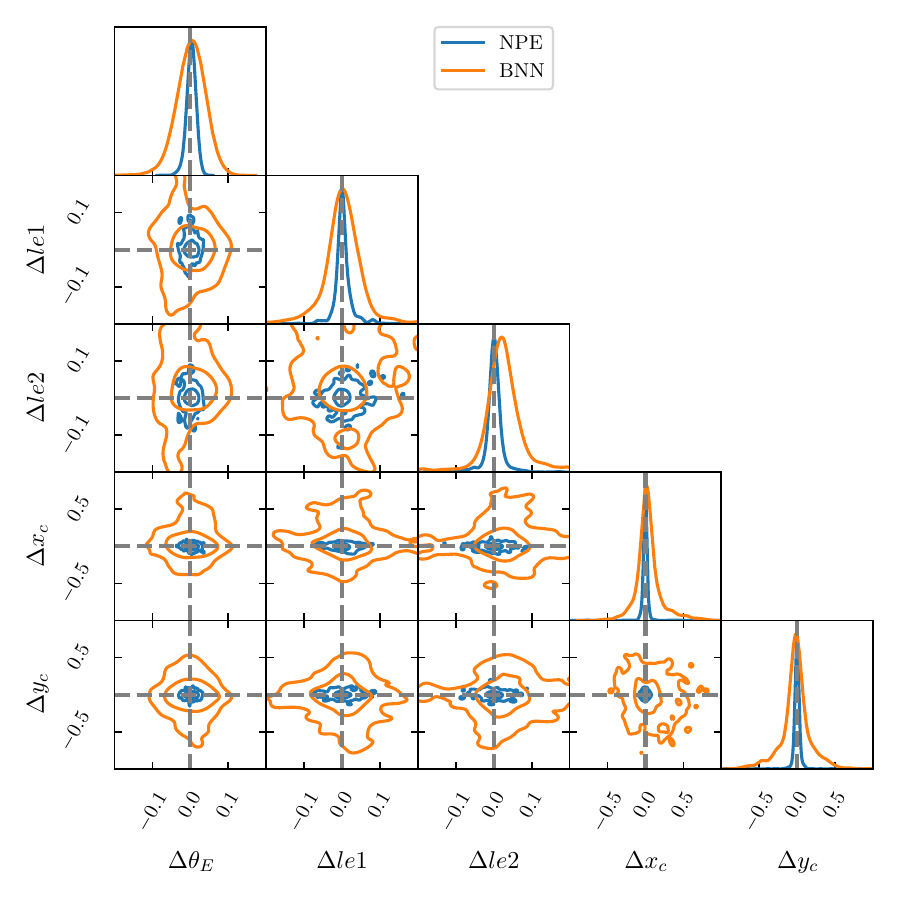}
    \caption{
    Residuals of the inferred posteriors for the 5-parameter NPE (blue) and BNN (orange) models for 1000 images from the test set.
    The average zero residual is represented by the gray dashed line.
    The same sample is used in the NPE model and the BNN model. 
    The contours represent approximate $68$th and $95$th percentile uncertainties in the scatter.
    The NPE model shown here is for the case in which the random seed for neural network initialization has been set to $465$.
    }
    \label{fig:5paramcorner}
\end{figure*}

The 5-parameter model has more complexity than the 1-parameter model: aside from \einsteinrad,  there are two additional parameters for lens eccentricity moduli ($\eccentricity{l}{1}$, $\eccentricity{l}{2}$) and two additional parameters for the positional offset between the lens and the source ($x_{\mathrm{c}}$, $y_{\mathrm{c}}$).  
We trained the models on 400,000 images with parameter distributions shown in Table~\ref{table:params}. 
During training, the neural density estimator was instantiated with 120 hidden features and 20 transforms for the MAF. 
We evaluated the performance on a test set of 1000 lenses generated using the priors recorded in Table~\ref{table:params}.
\myfigure{}~\ref{fig:5paramsingle9} shows the results of inference on a single image drawn randomly from the test set --- a corner plot of posteriors and the posterior predictive check for the model-based lens reconstruction. 
Table~\ref{table:12paramuncertainties} summarizes the results for the 1000-sample ensemble.

The performance of the NPE and BNN models on the 5-parameter model is similar to that on the 1-parameter model. 
First, the NPE model produces significantly smaller uncertainties than does the BNN model.
Compared to the 1-parameter model, the average uncertainty in \einsteinrad{} is about $\sim0.02$ arcsec for the NPE, and it is $\sim0.08$ arcsec for the BNN model, compared to $\sim10^{-3}$ arcsec and $\sim10^{-2}$ arcsec $68\%$, respectively, in the 1-parameter model. 
Unlike the 1-parameter model, the BNN uncertainties are $\sim1-5$ times larger than the NPE model, rather than 10 times larger, as in the 1-parameter model. 

We hypothesize that the decrease in accuracy going from a 1-parameter model to the 5-parameter model is related to the training set size and the parameter space volume.
Despite increasing the training set size for the 5-parameter model from 200,000 to 400,000 images, the parameter space volume has increased by a power of 5, exponentially outpacing the increase in training set size. 
Hence, the average sampling distance between points in the 5-parameter space is exponentially larger than that of the 1-parameter space, resulting in a less accurate model fit for the 5-parameter model.

The residuals in \myfigure{}~\ref{fig:5paramcorner} show that the NPE and BNN models are both unbiased.
The median values are consistent with zero within the 68\% scatter intervals 
(Table~\ref{table:12paramuncertainties}).
However, the 68\% scatter is two to three times wider in the BNN model than in the NPE model, showing that the BNN model is less precise than the NPE. 
There appear to be no degeneracies or correlations amongst the parameter residuals.
The posterior coverage in \myfigure{}~\ref{fig:5param_poscov} exhibits similar patterns as for the 1-parameter model.
The NPE model uncertainties are well calibrated, whereas the BNN uncertainties are underconfident for every parameter in the model --- albeit slightly less so for the position parameters. 

We also explore model stability for the 5-parameter model.
We find that the model is stable with respect to different random seed initializations of the network weight parameters. 
Details of that exploration can be found in \myappendix{}~\ref{5paramappendix}.

\subsection{12-parameter model: Varying Five Lens and Seven Source Parameters}
\label{sec:results_full_model}

\begin{figure*}
    \centering
    \includegraphics[width=0.9\linewidth]{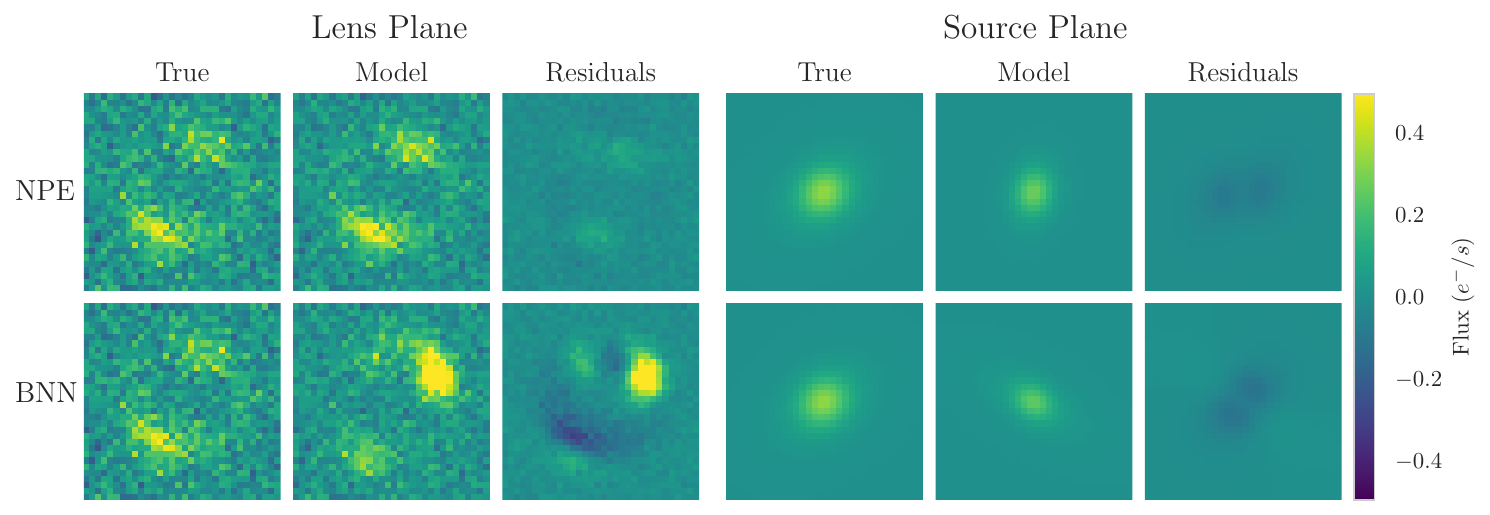}
    \vspace{-0.2cm}
    \includegraphics[width=0.9\linewidth]{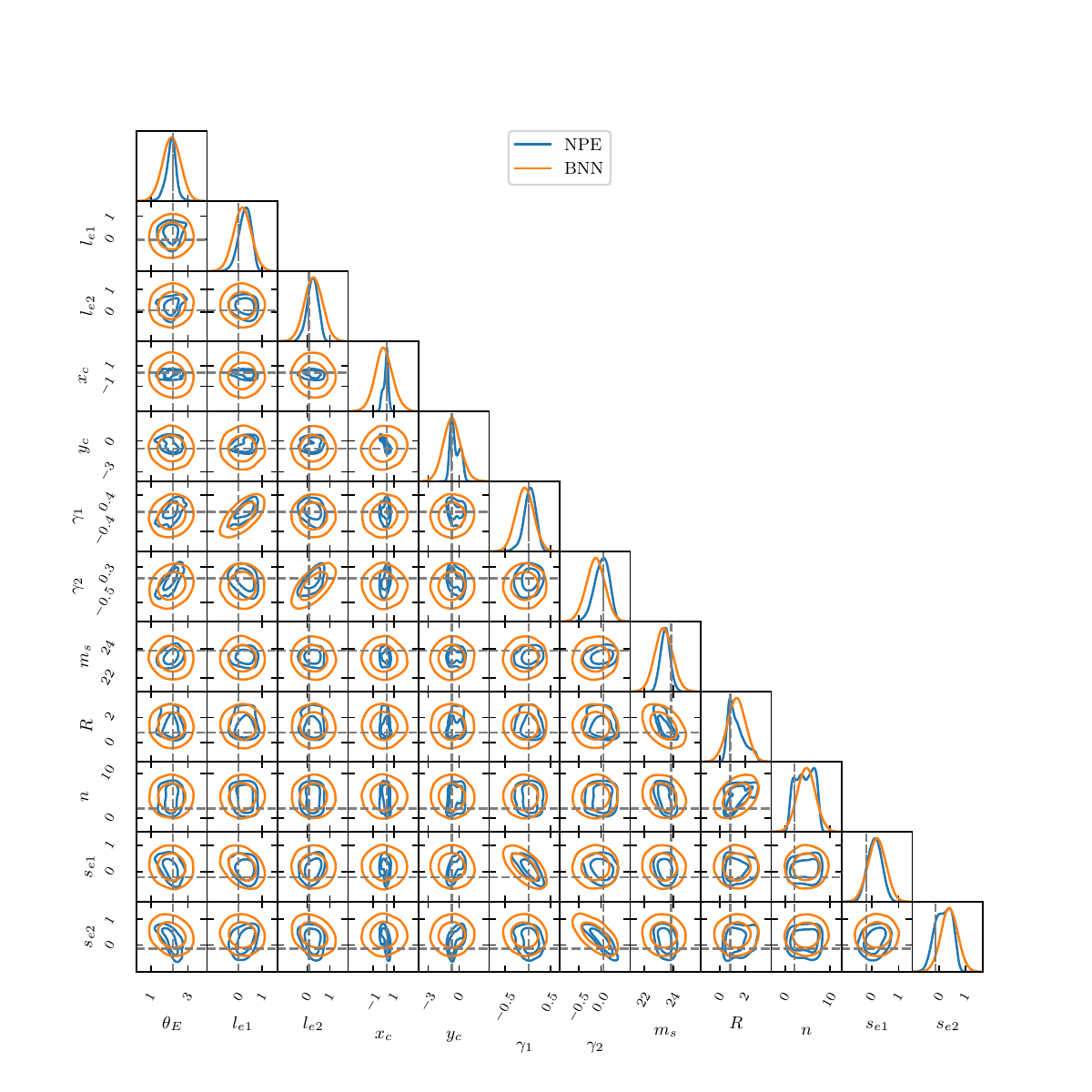}
    \caption{
    Performance of the 12-parameter NPE and BNN models for a single lensing system from the test set. 
    Top: the posterior predictive check for the lens plane and the source plane, which includes the true image, the image predicted by NPE (top) and the BNN (bottom), and the residual between the true and the predicted image.     
    Bottom: inferred posterior distributions of 12 parameters for the NPE (blue) and BNN (orange) models. 
    The true value is indicated by the gray dashed line. 
    We provide more examples in Appendix~\ref{AppA}.
    }
    \label{fig:12_paramsinglecorner}
\end{figure*}

\begin{figure*}
\begin{center}
\includegraphics[width=0.49\linewidth]
{images/12_param_npe_pos_465}
\includegraphics[width=0.49\linewidth]
{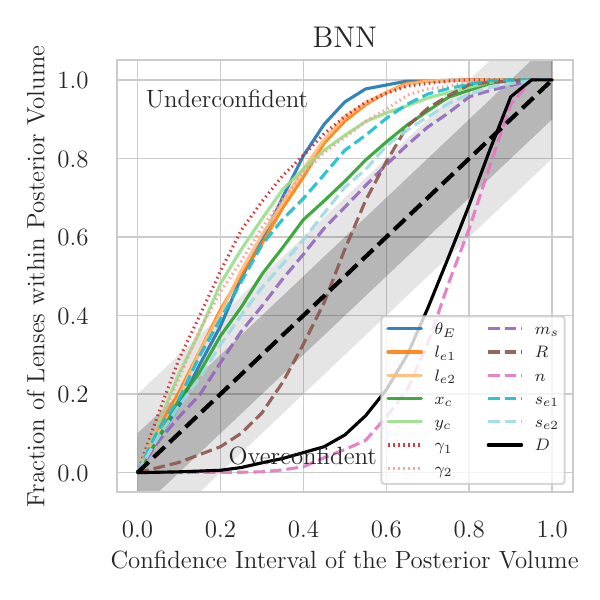}
\caption{
    Posterior coverage of 12-parameter NPE (left) and BNN (right) models for the test set. 
    The 1-to-1 line indicates perfect uncertainty calibration and is indicated by a dashed black line. 
    The gray regions indicate thresholds of 10\% (dark gray) and 20\% (light gray) uncertainty miscalibration. 
    The solid black line shows the combined distance metric $D$ (\myequation{}~\ref{eqn:distance_metric}).
    The lens parameter coverage, source parameter coverage, and shear parameter coverage are indicated by solid, dashed, and dotted lines, respectively.
    The NPE model random seed for weight initialization is set to $465$.
}
\label{fig:poscoverage12param}
\end{center}
\end{figure*}

\begin{figure*}
\begin{center}
\centerline{\includegraphics[width=1\linewidth]
{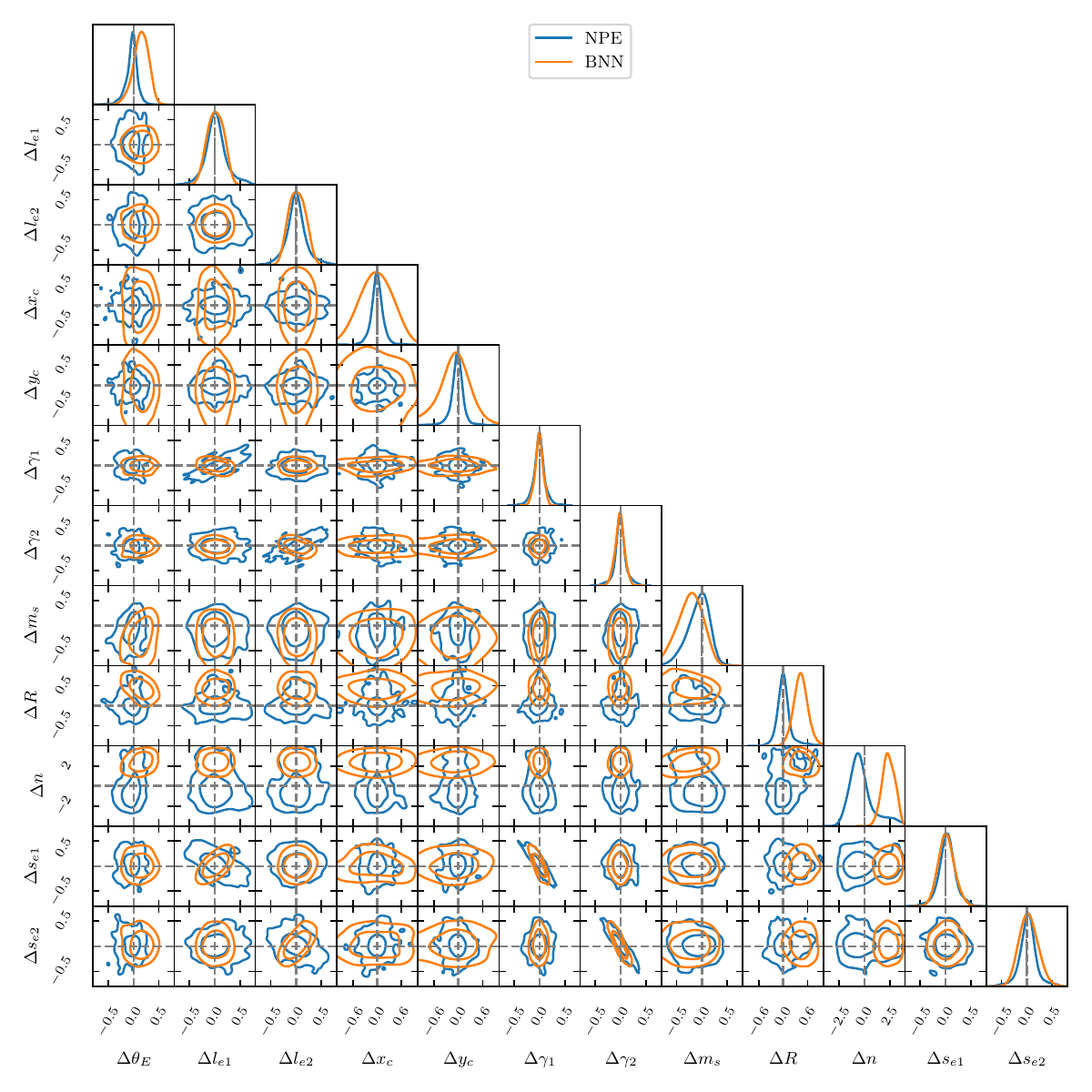}}
\caption{
    Residuals of the inferred posteriors of the parameters for the 12-parameter NPE (blue) and BNN (orange) models for 1000 images from the test set.
    Zero residuals are indicated by gray dashed lines.
    The contours represent 68th- and 95th- percentile uncertainties in the scatter.
    The NPE model is for the random seed set to $465$.
}
\label{fig:1000uncertainties}
\end{center}
\end{figure*}

The 12-parameter model is the most complex model that we consider. 
In addition to the five lens parameters discussed in \mysection{}~\ref{sec:results_five_param}, we model two parameters for the external shear ($\gamma_1, \gamma_2$) and five parameters for the source light distribution ($m_s$, $R$, $n$, $\eccentricity{s}{1}$, $\eccentricity{s}{2}$).
We trained the models on 800,000 images with parameter distributions shown in Table~\ref{table:params}. 
For training, the neural density estimator was instantiated with 400 hidden features and 20 transforms for the MAF. 
We primarily evaluated the performance on a test set of 1000 lenses.
\myfigure{}~\ref{fig:12_paramsinglecorner} shows the inferred posterior distributions and model reconstruction for a single image from the test set. 

The NPE model was more accurate and precise than the BNN model, inferring smaller uncertainties on the same test image by a factor of $\sim$1-3.
This is similar to the trends exhibited in the 1- and 5-parameter models.
The posterior coverage of the models (\myfigure{}~\ref{fig:poscoverage12param}) shows that the NPE outperforms the BNN model with respect to uncertainty calibration: for the NPE model, the coverage for all parameters is within 10\% of perfect calibration.
The BNN model is underconfident for every parameter except $R$ and $n$, for which it is overconfident. 
The BNN's posterior coverages for $R$ and $n$ exhibit 'S'-shaped coverage, which is usually an indication that the inferred posteriors are failing to accurately capture the ground truth values. 
The parameters $R$ and $n$ are source light parameters that produce extended lensing observables that map from the source plane to the lens plane in less straightforward ways. 
Inferring these parameters may be more challenging as a result.  
We hypothesize that the NPE model's better performance is due to MAF's inherently more flexible density estimators than the Gaussian densities of the BNN, as MAFs are proven to be universal approximators, capable of representing any probability distribution given certain reasonable conditions \cite{papamakarios2021}. 
Given enough training data, MAFs can more accurately estimate the true posterior distribution of the lens parameters. 

The residuals in \myfigure{}~\ref{fig:1000uncertainties} show that the NPE model outperforms the BNN model in accuracy for all parameters.
The NPE model is not highly accurate for $n$, but it is more accurate than the BNN model.
For  ($s_{e_1}$, $s_{e_2}$), ($l_{e_1}$, $l_{e_2}$), and ($\gamma_{1}$, $\gamma_{2}$), both NPE and BNN models are similarly precise and produce residuals of similar size.
For those parameters, the NPE model residuals also have a longer tail than the BNN model, with slightly wider 95\% contours than the BNN.

As mentioned in \S 6.2, we hypothesize that the difference in accuracy between the 5-parameter model and the 12-parameter model is related to the training set size and the parameter space volume.
Despite increasing the training set size for the 12-parameter model to 800,000 images, the parameter space volume increases exponentially with the number of dimensions, outpacing the increase in training set size. 
Hence, the average sampling distance between points in the 12-parameter space is exponentially larger than that of the 1- and 5-parameter spaces, resulting in a less accurate model fit for the 12-parameter model. 
Despite the NPE method being more flexible than the BNN, it is still constrained by the size of the training set. 
Due to computational constraints, we were unable to increase the training set size above 800k images. 
This prohibits increasing the training set size to that required to maintain the same sampling distance as the 1- or even 5-parameter model; this is not the same distance metric as \myequation{}~\ref{eqn:distance_metric}. 
 
For the position parameters ($x_{\mathrm{c}}$, $y_{\mathrm{c}}$), both NPE and BNN models were accurate and unbiased.
However, the BNN residuals exhibit significantly larger scatter (lower precision) than the NPE residuals, with the BNN's 68\% uncertainties being similar in size to the NPE model's 95\% uncertainties. 
The NPE model is unbiased for all the parameters except for $n$, which it underestimates by $\sim0.2$, although it is within the 68\% uncertainties.
The BNN, however, is systematically biased for four parameters  \einsteinrad, $m_s$, $R$, and $n$. 
The BNN overestimates \einsteinrad~and underestimates $m_s$, as evidenced by the peaks of the residuals shifted away from zero. 
The BNN performed worst on $R$ and $n$, overestimating the true values for more than 95\% of the test set. 
The BNN also overestimates $R$ by 
$>0.4$ arcsec and $n$ by 2.4.
This poor performance is also reflected in the 'S'-shaped posterior coverage for these parameters in \myfigure{}~\ref{fig:poscoverage12param}. 
As hypothesized earlier, these are source light parameters, which have non-linear mappings to the observed image plane. 
The BNN model relies on normal distributions for its parameters and may not have enough flexibility to capture the true posterior distribution accurately. 
In comparison, the NPE was more accurate and precise in inferring the best-fit parameter values for all these parameters. 

Some parameters demonstrate correlations in their residuals. 
For both the NPE and BNN models, the residuals of \einsteinrad~ are slightly positively correlated with residuals of $m_s$. 
For both models, the residuals of $l_{e_1}$ and $l_{e_2}$ and $s_{e_1}$ and $s_{e_2}$ are positively correlated with residuals of $\gamma_{1}$ and $\gamma_{2}$.
However, the BNN (NPE) residuals for the lens eccentricity parameters are positively (negatively) correlated with the source eccentricity parameters.
This parameter degeneracy is well-documented in strong lens modeling literature \cite{birrer2017, Unruh2017}.
However, the uncertainties and residuals of the eccentricities and shear may be correlated because we are jointly inferring them: offsets or errors in one parameter could be accounted for by offsets in another.

\tableresultssummary

The NPE model exhibits slight variations in uncertainty calibration for three different weight initializations. 
This may be explained by the increase in average sampling distance between training data points in the 12-parameter space compared to the 1- and 5-parameter spaces, which results in greater uncertainty in model fitting. 
$m_s$ appears to be biased high, and $n$ appears to be biased low for all three seeds.

\section{Discussion}
\label{sec:discussion}

In this section, we discuss challenges in model development and evaluation for NPE models. 

\subsection{Adopting Sequential SBI for Follow-up Studies}

The goal of this work is to efficiently infer astrophysical parameters from a large sample of lensing images. 
This procedure requires amortization --- the fast training and optimization of a model. 
Sequential methods, however, require a model to be trained each time new results are obtained (\mysection{}~\ref{sec:SNPE_discussion}).
However, we can imagine using sequential methods to follow up on a small sub-sample of lensing images to refine the NPE model. 
For example, after an initial analysis using NPE and ancillary observations (e.g., lenses where we have access to spectroscopic data), one could identify a smaller sub-sample of information-rich lenses --- e.g., quadruply-imaged lenses or time-delay lenses --- that we would like to refine the posterior estimates of, we can perform SNPE on those lenses by using the posteriors obtained from NPE in this work as a proposal prior for subsequent rounds of inference, thereby ensuring that no computation is wasted. 

\subsection{Increasing Modeling Complexity}

In this study, the most complex model has 12 physics parameters. 
This level of complexity is appropriate for the majority of lenses we expect to discover in ground-based imaging data --- i.e., single-lens, single-source galaxy-galaxy lenses with $\sim$1-arcsec resolution \cite[e.g.,][]{Agnello_2017, Knabel_2023}. 
However, more complicated lens system models, such as cluster-scale lenses, multi-source-plane lenses, and lens subhaloes, could require $\sim$30-50 parameters \cite[e.g.,][]{mahler2023precision}. 
Moreover, even with high-resolution, lower-noise space-based imaging (e.g., Roman, Euclid, JWST), the dimensionality of the model remains a challenge for SBI methods. 

Generally, SNPE appears to scale well to high-dimensional observations, but its scaling to parameter space dimensions greater than 30 is more challenging \cite{goncalves2020}.
Given that the difficulty of estimating full posteriors scales exponentially with dimensionality, this is an inherent challenge for all approaches that aim at full inference. 
The dimensionality of the observation refers to the dimensionality of the dataset: in our work, the dimensionality of our observations is 1024, corresponding to the number of pixels in each image.   
Advances in both SBI and density estimation techniques have overcome the limitations of the curse of dimensionality suffered by classical ABC techniques; although SBI doesn't yet have statistical guarantees, which are especially crucial for trustworthiness when these models are applied to real data.
However, further advances must be made for SBI techniques to scale for models in very high-dimensional spaces. 
In addition, the validation techniques used to ensure the credibility of the model also need to scale up to that number of parameters.

Advancements in recent years may enable NPE methods to continue scaling to higher-dimensionality problems in the future.
First, many simulators provide only pairs of input data and parameters $(X, \theta)$. 
When simulators provide more information --- e.g., the score $\nabla_{\theta}p(X|\theta)$, which is the gradient of the likelihood with respect to the parameters of interest --- models can be more performant and more efficiently trained. 
This has been applied to strong lensing analyses in the form of a proof-of-concept analysis of dark matter subhalo population properties in strong lenses \cite{Brehmer_2019}. 
 
Second, combining SBI with hierarchical modeling may efficiently provide greater constraining power when population-level (i.e., cosmological) parameters like the cosmic matter density $\Omega_m$ and the dark energy equation of state $w$ are the goal of the inference. 
Hierarchicality in SBI has been studied in a few strong lensing applications \cite{wagner2021hierarchical, WagnerCarena2022, jarugula2024}, and it has been directly implemented with neural posterior estimation \cite{rodrigues2021hnpeleveragingglobalparameters, heinrich2024hierarchicalneuralsimulationbasedinference}.

\subsection{Model Calibration}

Post-training calibration in machine learning models is an actively studied field, with much of the literature focused on calibrating neural network predictions in a classification or regression task \cite{guo2017calibration}. 
In this work, we did not apply post-training model calibration to evaluate or enhance performance. 
Calibration of SBI models in an astrophysical context is largely unexplored, although beta calibration \cite{pmlr-v54-kull17a} has been used to calibrate BNN models to predict cosmological parameters from Cosmic Microwave Background maps \cite{Hort_a_2020}. 
Conformal methods \cite{angelopoulos2022gentle,vovk1999} offer a promising calibration technique because they provide guarantees regarding the coverage, provided that the test set is independent and identically distributed to the validation set used for the calibration. 
However, the calibrated model, while valid, predicts the same uncertainties for all test set data and is unable to predict larger uncertainties for data within the test set that have higher risk --- such as noisy images. 
While there has been recent work to make conformal prediction both valid and discriminative \cite{lin2021locally}, it is unclear how this may be combined with existing SBI methods of uncertainty prediction.

\section{Conclusion}
\label{sec:conclusion}

In this work, we explore the capacity of NPE methods to model large numbers of strong lensing systems at various degrees of complexity, which is marked by the number of parameters in the model --- one, five, and 12.
We compare NPE to BNN because they are one of the few approaches that can quickly predict model parameters from images.
Traditional lens modeling algorithms are too inefficient to infer parameters from large numbers of lenses.
We use SBC's and PPC's to study the validity of the NPE and BNN models.
We also study the stability of these models with respect to network weight initialization.
The comprehensiveness of these experiments and diagnostics is a contribution to SBI applications in astrophysics contexts.

Our main results are:

\begin{enumerate}
    \item Both NPEs and BNNs provide informative inferred posterior distributions of lens parameters in experiments with noisy ground-based galaxy-galaxy lenses.  
    \item Due to their amortized nature, the NPE and BNN models infer the posterior densities of thousands of lens systems in minutes. 
    This meets the scaling requirement for tens to hundreds of thousands of lens systems that are projected to be discovered in future survey data. 
    For the 12-parameter model, both NPE and BNN models took about six hours to train. 
    In comparison, a conventional MCMC fit would take around the order of an hour per lens system. 
    \item This work demonstrates the use of Masked Autoregressive Flows (MAF) for lens parameter posterior density estimation. 
    Previous work used mixture density models \cite{Legin2023}.
    \item We found that the NPE model uncertainties are generally more well-calibrated than the BNN model uncertainties. 
    In this work, the BNN models tend to be underconfident --- overpredicting the uncertainties of the test images. 
    One potential method to improve the uncertainty calibration of the BNN model is to increase the number of Gaussian components in the model weights.  
    However, this comes with the trade-off of a steep computational cost increase because each additional Gaussian component requires two additional free parameters to be trained for every weight in the network. 
    \item The effect of random seed initialization in NPEs is subdominant to the model uncertainties in the 5- and 12-parameter models, implying that the choice of model will not significantly impact the results (see \myappendix{}~\ref{AppA}). 
    In the 1-parameter model, the choice of random seed results in a $\sim0.1\%$ uncertainty, which is on the same order of magnitude as the uncertainties from the posterior itself. 
    However, this is unlikely to impact the modeling of real lenses in practice, as achieving a $\sim0.1\%$ uncertainty estimate is only possible in the 1-parameter model, which is too simple to accurately model real lenses. 
    In practice, real lenses have uncertainties on the order of 5\% \cite{Ruff_2011}.
    \item NPE methods do not yet come with statistical guarantees. 
    Therefore, it's critical that applications are accompanied by diagnostics to assess model calibration and stability.  
    This includes assessing the following: (1) both maximum-likelihood point estimates and uncertainty estimates are unbiased and well-calibrated when averaged across a large ensemble of data; (2) model performance does not exhibit large variation with the choice of random seeds for weight initialization. 
    These diagnostics are essential for showing that a model produces reliable and reproducible results.
\end{enumerate}

This work presents another step in the study of SBI methods, which have advanced in recent years due to developments in deep learning, as well as exponential gains in computational efficiency.
There are multiple avenues for further exploration and development --- both in fundamental algorithm challenges and in the application to strong gravitational lensing and other astrophysical probes.
First, for algorithm development, as discussed in \mysection{}~\ref{sec:discussion}, SBI methods that leverage not only the training data but additional information that characterizes the latent process of the simulators --- e.g., the score --- may increase model performance. 
Second, the potential performance improvements and computational trade-offs from ensembling large numbers of SBI models should also be investigated in more detail. 
Third, the lack of statistical guarantees and fully interpretable uncertainties hinders the trustworthiness of these methods.
Well-motivated diagnostics like SBC present the best tools for evaluating SBI model fitness. 
Aside from diagnostics, post-calibration techniques may provide a near-term partial substitute for formal theoretical guarantees.

\appendix

\section{Correction to Magnitude Calculation}
\label{AppMagnitudes}

While finalizing this manuscript, we discovered a bug in the simulation code.
When the background noise was simulated, the sky brightness was not correctly converted from magnitude to counts per second. 
The consequence of this bug is that the average root mean square noise per pixel in the simulation is lower than it should have been by a factor of $\sim1.6$: this factor leads to $\sim0.08$ counts per pixel instead of $\sim0.13$ counts per pixel. 
This does not significantly affect the results or conclusions of this work because the RMS noise is typically an order of magnitude lower or less compared to the average lensing signal in the simulated images on all the models.

\section{Experiments for Model Stability}
\label{AppA}

We study NPE model stability for the three levels of problem complexity in this work: 1-parameter, 5-parameter, and 12-parameter models. 
We initialized model weights with three different random seeds (42, 465, and 839) to study the variation in model performance on a single test data set with priors within the range of the training set.
Generally, the diagnostics indicate that the models are stable and reliable.


\subsection{1-parameter Model}
\label{1paramappendix}

First, regarding stability for the 1-parameter model, variations in model performance on $\theta_E$ were $\sim10^{-3}$ arcsec ($\sim0.1\%$) --- the same order of magnitude of the model uncertainties. 
\myfigure{}~\ref{fig:1param3seed} shows residuals for the different seed models on the same test images. 
It is undesirable for the choice of random seed to significantly impact the model's posterior predictions. 
However, this is an edge case because a $\sim10^{-3}$-arcsec modeling uncertainty is only achievable for extremely constrained and unrealistic models, such as this 1-parameter model. 
In practice, modeling uncertainties for $\theta_E$ in the most optimal real-world cases --- i.e., high-resolution space-based images of simple lens systems \cite{Ruff_2011} --- are on the order of 5\%. 
The residuals for the 1-parameter model indicate that there may be a stochastic noise floor caused by weight initialization and that the uncertainty of this noise floor is on the order of $\sim$0.1\%. 
The results show that given a sufficiently constrained model, it is possible to attain the precision and accuracy level up to the level of this noise floor. 

We find that the NPE model is well-calibrated for each of the random seeds: the rank histograms and the accompanying cumulative distribution functions (\myfigure{}~\ref{fig:1_paramSBC}), along with the posterior coverages (\myfigure{}~\ref{fig:1param3seed}), show that one model (seed 465) is almost perfectly calibrated and that two models (seeds 42 and 839) are slightly underconfident. 
All three models in \myfigure{}~\ref{fig:1_paramSBC} exhibit slight $\cap$-shaped distributions in the rank histograms, indicating that the posterior models slightly overestimate the true posterior uncertainty. 
The rank histogram for the seed 42 model skews slightly to the left, indicating a slight bias towards lower values compared to the true posterior (underestimation). 
In comparison, the seed 465 model skews slightly to the right, indicating a slight systematic bias towards higher values (overestimation). 
These variations are not readily perceptible in the residual plots (\myfigure{}~\ref{fig:1paramcornerseed}) or the posterior coverage plot (\myfigure{}~\ref{fig:1param3seed}).

\vspace{8mm}
\noindent\begin{minipage}{\linewidth}
\centering
    \includegraphics[width=0.32\linewidth]{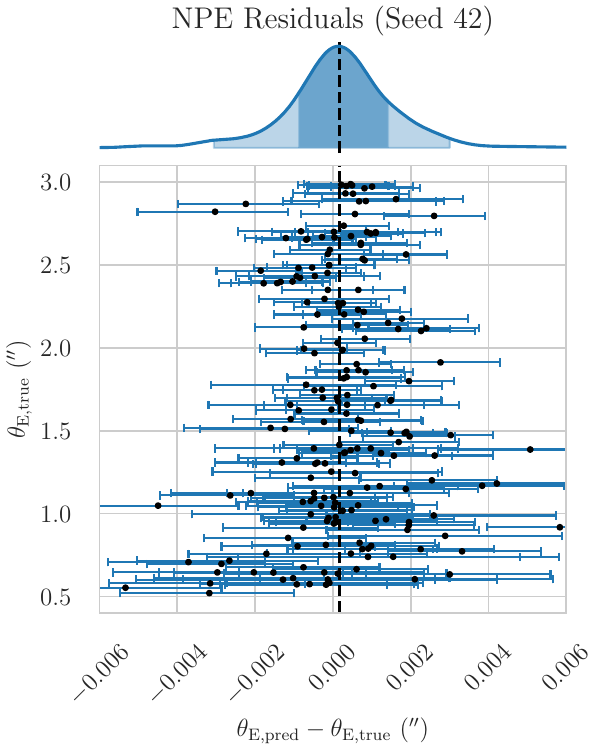}
    \includegraphics[width=0.32\linewidth]{images/1_param_corner_465.pdf}
    \includegraphics[width=0.32\linewidth]{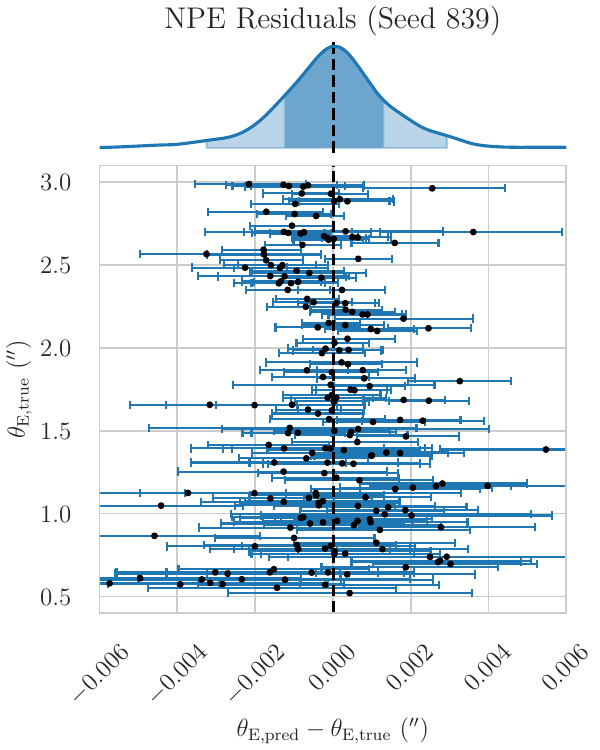}
    \captionof{figure}{
    Residuals of the inferred posterior of the Einstein radius \einsteinrad{} in the 1-parameter model for the NPE model, with three different seeds --- 42 (left), 465 (middle), and 839 (right) for the network weight initialization for 200 images randomly selected from the test set.
    The error bar represents the 68\% uncertainty in \einsteinrad{} for each image. 
    Above the scatter plot of the residuals, the marginal distribution of best-fit values for the full test set of 1000 is shown.
    The 68\% and 95\% scatter in best-fit values are indicated by the dark and lighter-shaded regions, respectively. 
    This is similar to the \myfigure{}~\ref{fig:1paramcorner}.
    \label{fig:1paramcornerseed}
    }
\end{minipage}

\vspace{8mm}
\noindent
\begin{minipage}{\linewidth}
\centering
    \includegraphics[width=0.32\linewidth]{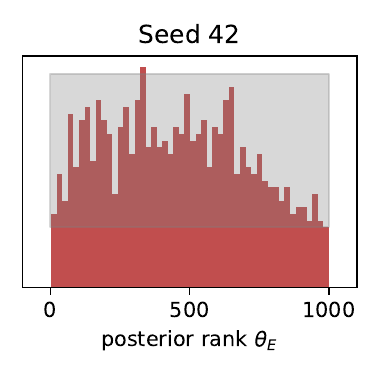}
    \includegraphics[width=0.32\linewidth]{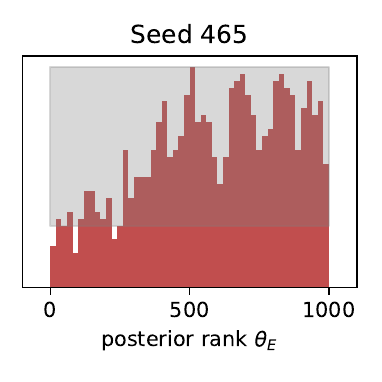}
    \includegraphics[width=0.32\linewidth]{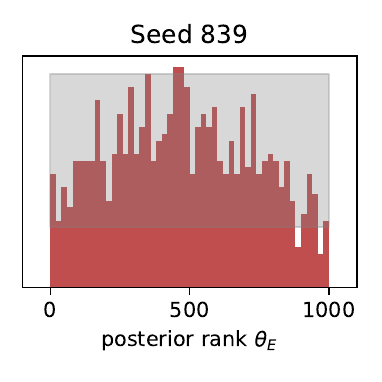}
    \includegraphics[width=0.32\linewidth]{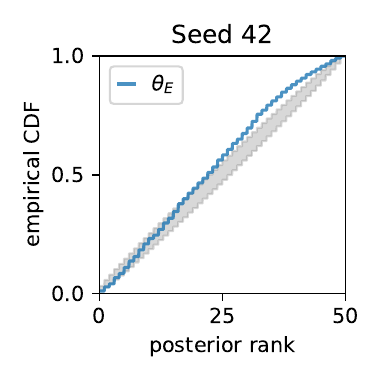}
    \includegraphics[width=0.32\linewidth]{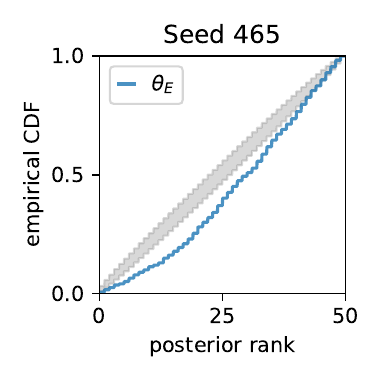}
    \includegraphics[width=0.32\linewidth]{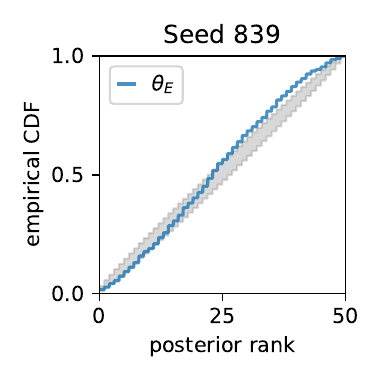}
    \captionof{figure}{
    Rank histograms (top) and cumulative density functions (bottom) for the 1-parameter model with seeds 42 (left), 465 (middle), and 839 (right). 
    We discuss the plots in \mysection{}~\ref{sec:results_theta}. 
    In the rank histograms, the gray region indicates uniformity.
    In the CDFs, the gray regions indicate the 99\% confidence interval of a uniform distribution given the number of samples provided.
    }
    \label{fig:1_paramSBC}. 
\end{minipage}

\vspace{8mm}
\noindent
\begin{minipage}{\linewidth}
\centering
    \includegraphics[width=0.45\linewidth]{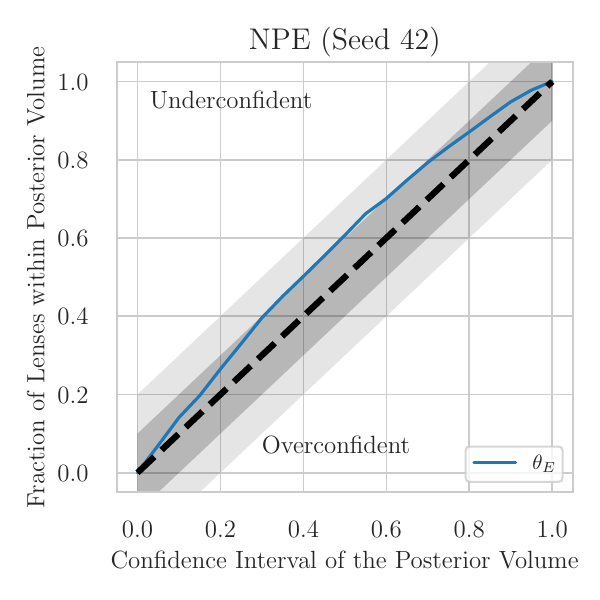}
    \includegraphics[width=0.45\linewidth]{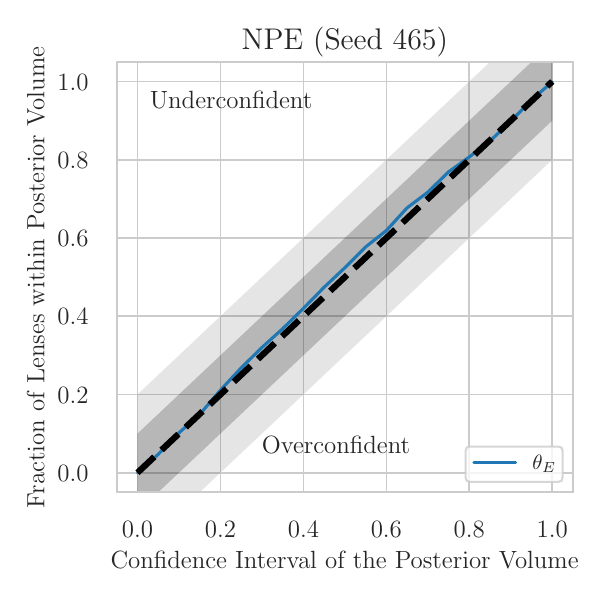}
    \includegraphics[width=0.45\linewidth]{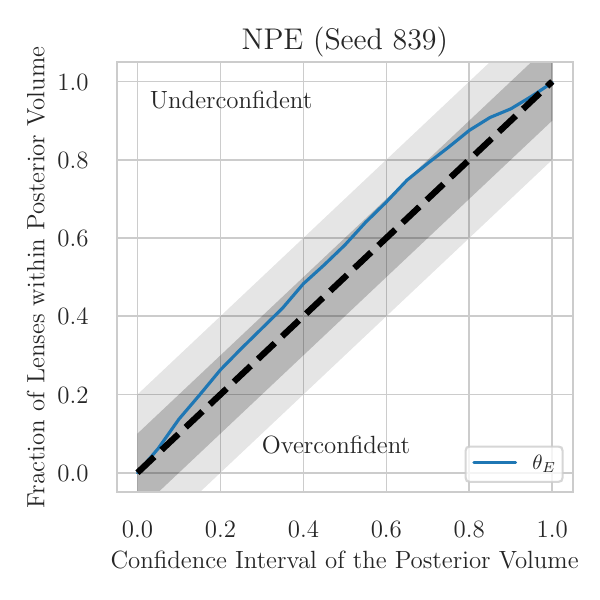}
    \includegraphics[width=0.45\linewidth]{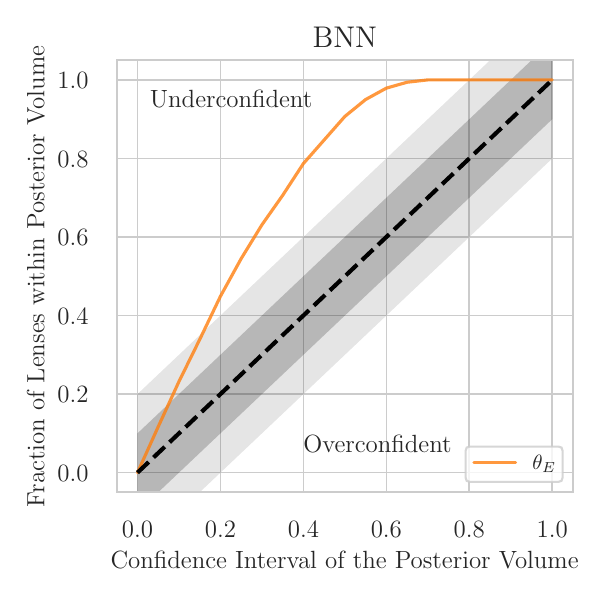}
    \captionof{figure}{
    Posterior coverage of the 1-parameter NPE and BNN models computed on the test set (solid). 
    The NPE models are initialised with three seeds --- 42, 465, and 839.
    The 1-1 line (black, dashed) indicates perfect uncertainty calibration. 
    The gray regions indicate thresholds of 10\% (dark gray) and 20\% (light gray) uncertainty miscalibration. 
    }
    \label{fig:1param3seed}
\end{minipage}

\subsection{5-parameter Model}
\label{5paramappendix}

The 5-parameter model is well-calibrated or nearly well-calibrated for most of the five parameters for all three seeds.
For seed 42, the model is slightly miscalibrated for \einsteinrad{}, $\eccentricity{l}{2}$, and $y_c$. 
For seeds 465 and 839, the models appears to be well-calibrated for all parameters.
This is reflected in the rank histograms (\myfigure{}~\ref{fig:5paramSBChist}) and CDFs (\myfigure{}~\ref{fig:5paramSBCcdf}): the rank bins are consistent with a uniform distribution, where only one or fewer bins lie outside the grey 99\% confidence region. 
Using the combined distance metric (\myequation{}~\ref{eqn:distance_metric}) as a proxy for the overall uncertainty calibration across all five parameters,
we find that the seed 42 model is just slightly overconfident, the seed 465 model is almost perfectly calibrated, and the seed 839 model is just slightly underconfident. 
Such slight calibration differences are unlikely to be the most significant source of uncertainty in real-life applications. 
Still, it highlights the importance of evaluating the effects of the random weight initialization on the performance and stability of deep learning models. 

\vspace{8mm}
\noindent
\begin{minipage}{\linewidth}
    \centering
    \includegraphics[width=0.99\linewidth]{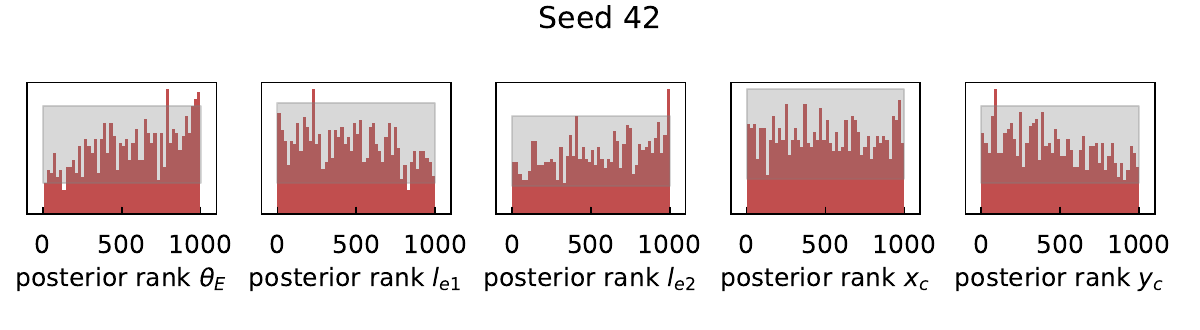}
    \includegraphics[width=0.99\linewidth]{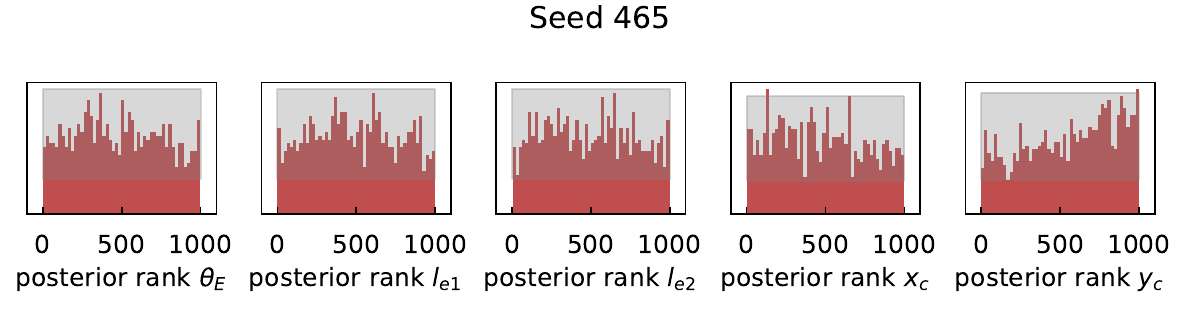}
    \includegraphics[width=0.99\linewidth]{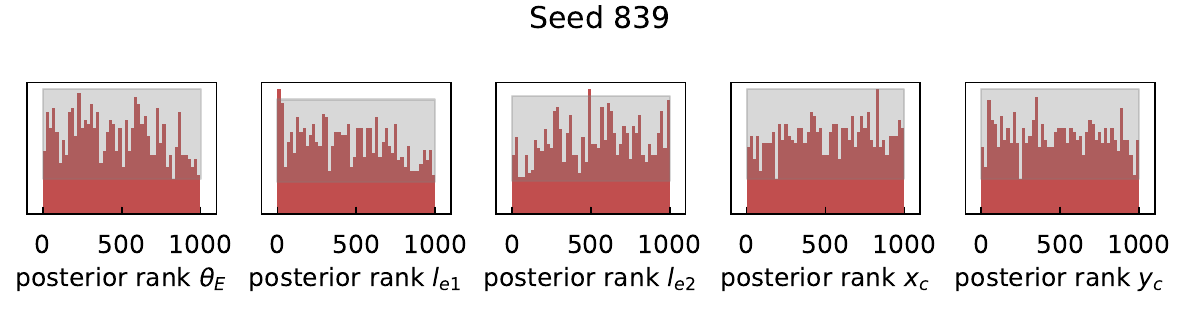}
    \captionof{figure}{
    Rank histograms for the 5-parameter NPE model with seeds 42, 465, and 839 (top, middle, bottom) for the weight initialization.
    The gray region indicates uniformity.
    }
    \label{fig:5paramSBChist}
\end{minipage}

\vspace{8mm}
\noindent
\begin{minipage}{\linewidth}
    \centering    
    \includegraphics[width=0.3\linewidth]{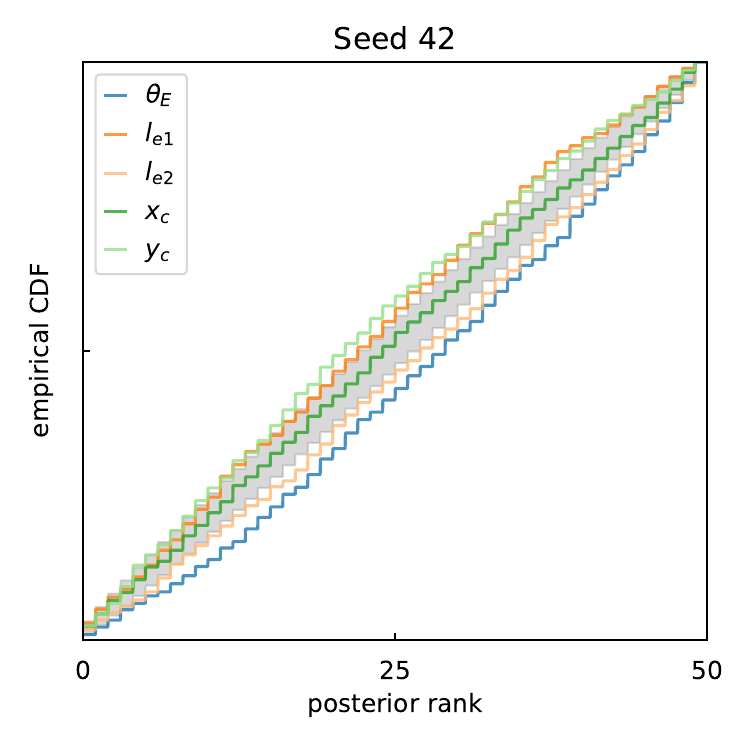}
    \includegraphics[width=0.3\linewidth]{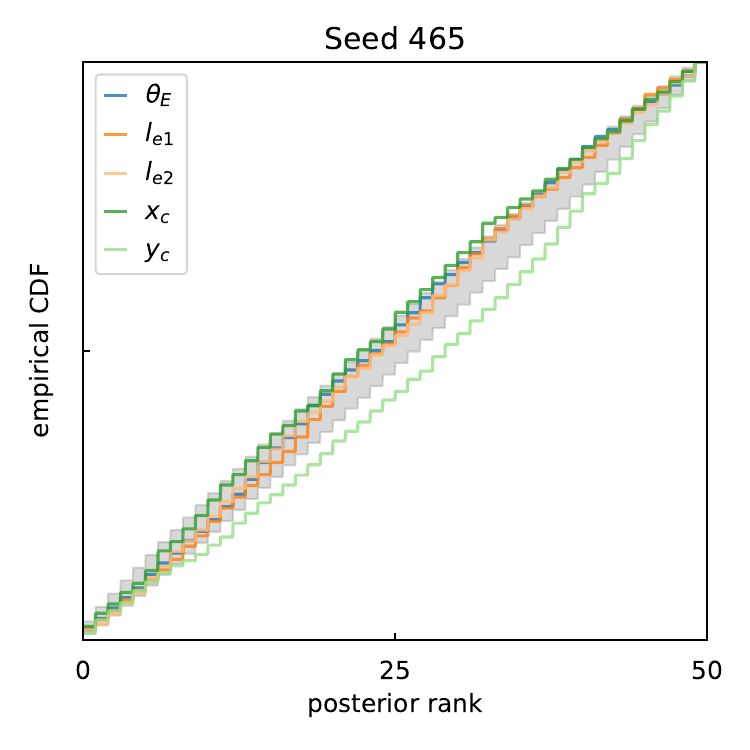}
    \includegraphics[width=0.3\linewidth]{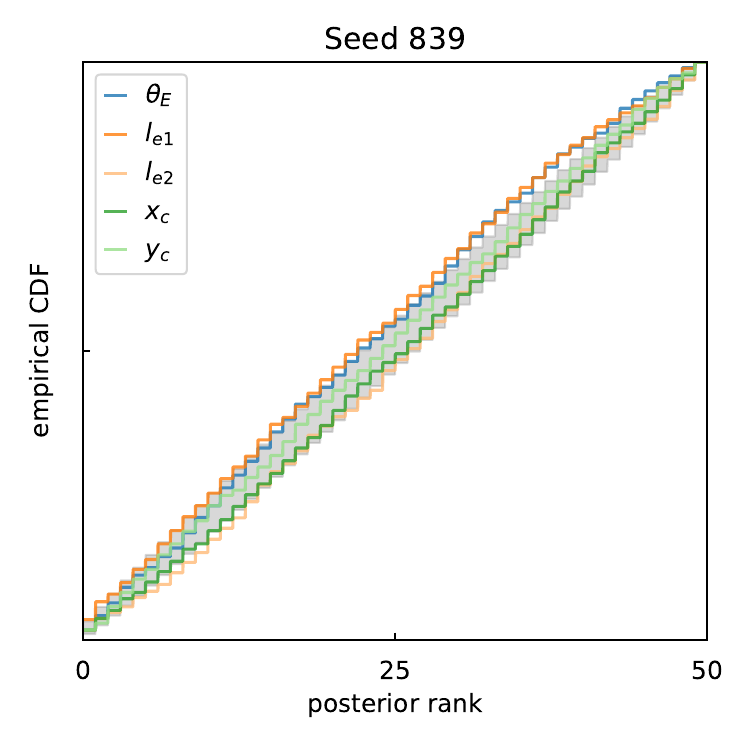}
    \captionof{figure}{
    CDFs for the 5-parameter NPE model for seeds 42, 465, and 839. 
    The gray regions indicate the 99\% confidence interval of a uniform distribution given the number of samples provided.
    }
    \label{fig:5paramSBCcdf}
\end{minipage}

\vspace{8mm}
\noindent
\begin{minipage}{\linewidth}
    \centering
    \includegraphics[width=0.32\linewidth]{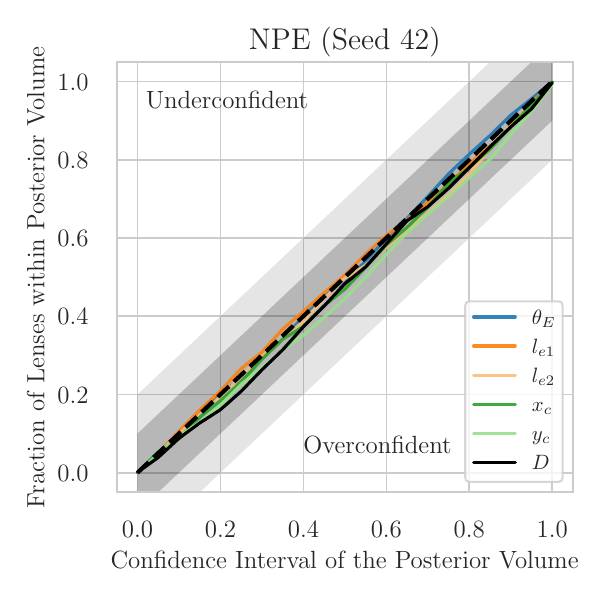}
    \includegraphics[width=0.32\linewidth]{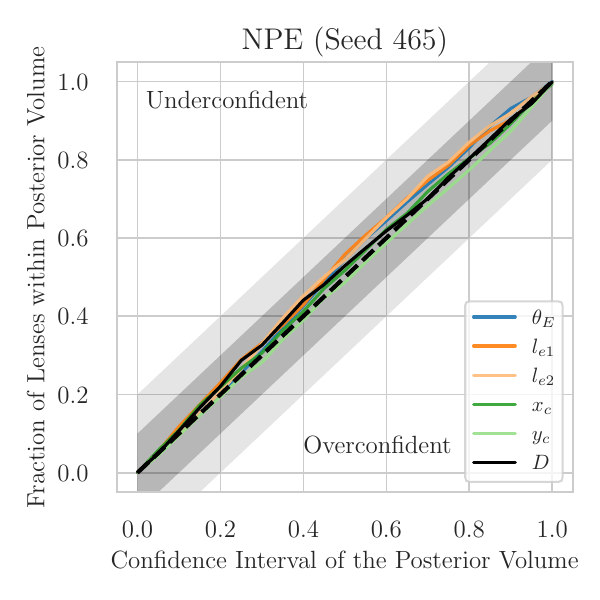}
    \includegraphics[width=0.32\linewidth]{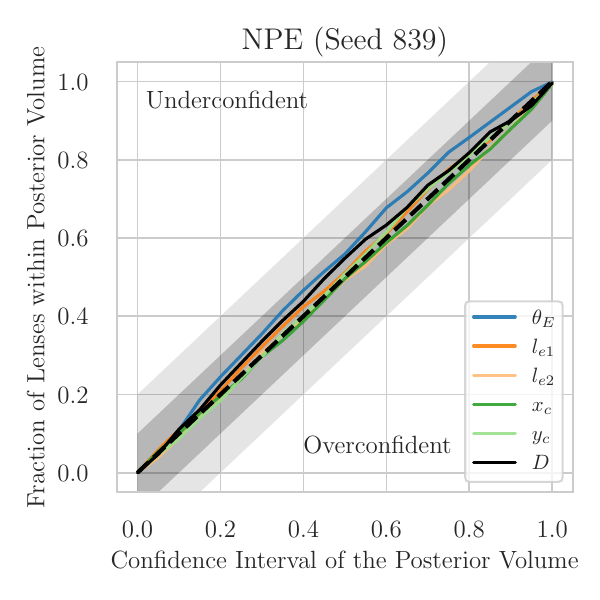}
    \captionof{figure}{
    Posterior coverage and the combined distance metric for the 5-parameter NPE models on the test set (first row) with three different seeds for the network weight initialization --- 42 (left), 465 (middle), and 839 (right).
    The 1-1 line (black, dashed) indicates perfect uncertainty calibration. 
    The gray contours indicate regions of 10\% (dark gray) and 20\% (light gray) uncertainty miscalibration. 
    }
    \label{fig:5param3seeds}
\end{minipage}

\subsection{12-parameter Model}
\label{12paramappendix}

For the 12-parameter model, according to the posterior coverage (\myfigure{}~\ref{fig:12param}), the rank histograms and the CDFs, our models exhibit slight variations in uncertainty calibration between seeds. 
These variations are larger than that of the 5-parameter model.
For example, the least well-calibrated parameter across the three seeds in the 5-parameter model is the  \einsteinrad\ for the seed 839 model (solid blue line on the rightmost plot in \myfigure{}~\ref{fig:5param3seeds}), which, at worst, exhibits about $\sim$5\% confidence interval difference from perfect calibration. 
In contrast, the 12-parameter seed 839 model is more miscalibrated for \einsteinrad{} (solid blue line on the rightmost plot in \myfigure{}~\ref{fig:12param}), with a 
$\sim$10\% difference from perfect calibration.
The effects of seed initialization can become increasingly significant when the dimensionality is increased.
The rank diagnostics for the model for the three different seeds (\myfigure{}~\ref{fig:12paramSBCcdf} and \myfigure{}~\ref{fig:12paramSBChist}) also provide insight on the uncertainty calibration of the model parameters: across the three seeds, most of the parameters produce rank diagnostics that are consistent with uniform distributions. 
However, there are a few exceptions.
$m_S$ appears to be biased high in all three seeds, and the CDF and rank histograms exhibit deviation from uniformity (\myfigure{}~\ref{fig:12paramSBCcdf} and \ref{fig:12paramSBChist}).

\vspace{8mm}
\noindent
\begin{minipage}{\linewidth}
    \centering
    \includegraphics[width=0.3\linewidth]{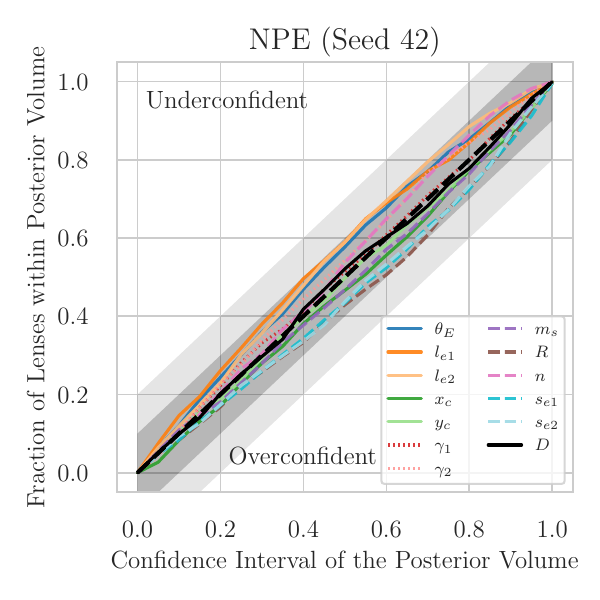}
    \includegraphics[width=0.3\linewidth]{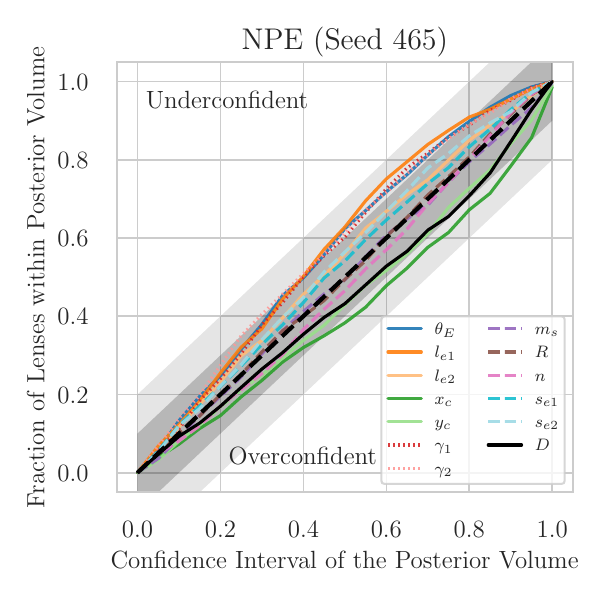}
    \includegraphics[width=0.3\linewidth]{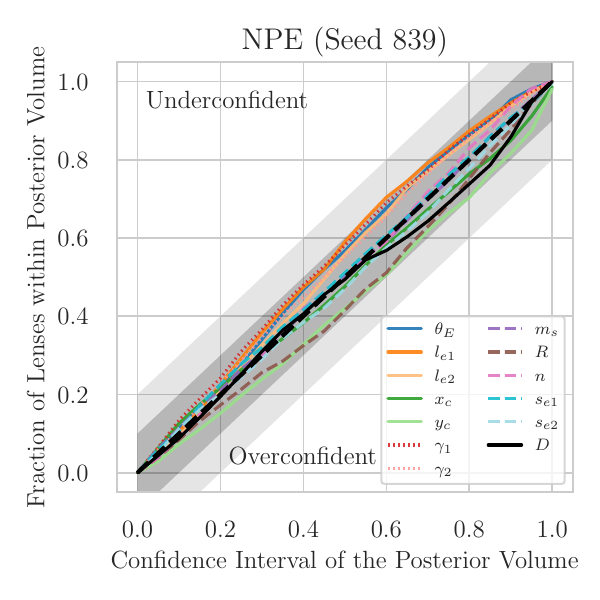}
    \captionof{figure}{
    Posterior coverage and the combined distance metric of the 12-parameter NPE models computed on the test set (first row)   for three seeds --- 42 (left), 465 (middle), 839 (right) --- for the network weight initialization.
    The 1-1 line (black, dashed) indicates perfect uncertainty calibration. 
    The gray regions indicate thresholds of 10\% (dark gray) and 20\% (light gray) uncertainty miscalibration. 
    }
    \label{fig:12param}
\end{minipage}

\vspace{8mm}
\noindent
\begin{minipage}{\linewidth}
    \centering
    \includegraphics[width=0.3\linewidth]{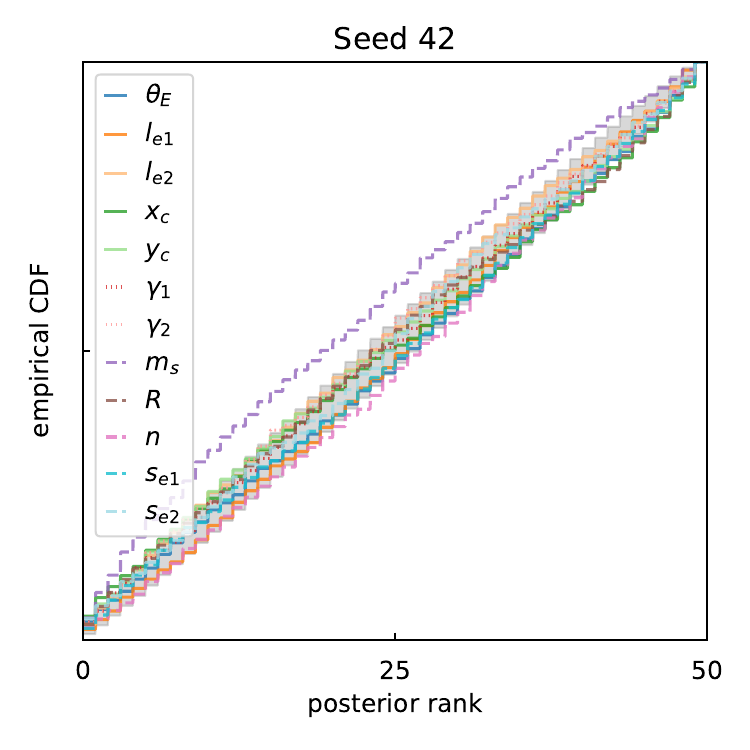}
    \includegraphics[width=0.3\linewidth]{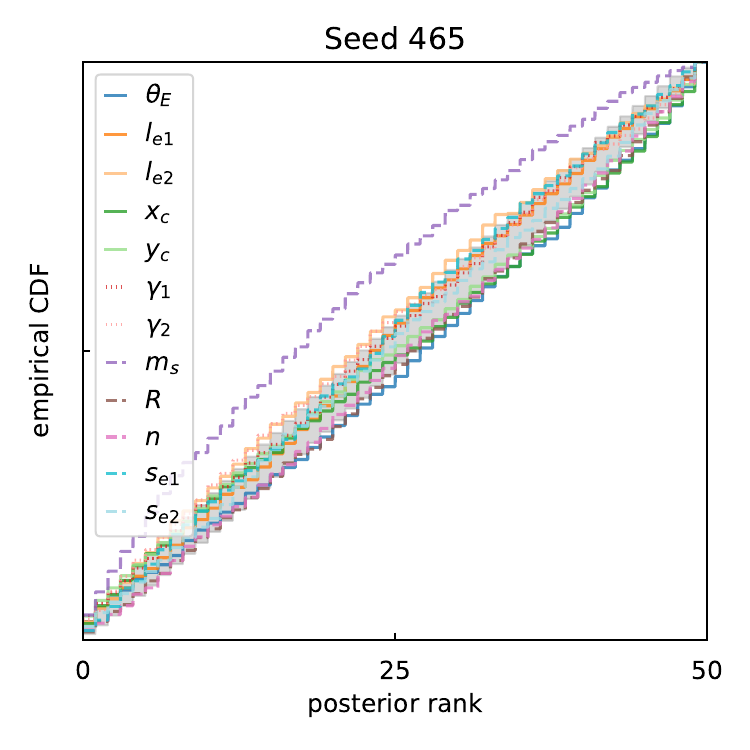}
    \includegraphics[width=0.3\linewidth]{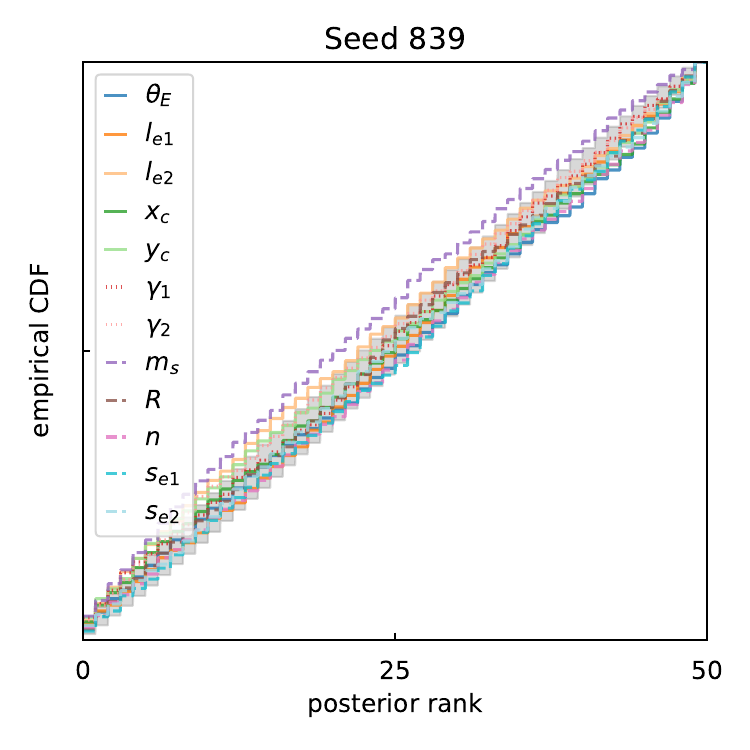}
    \captionof{figure}{
    CDFs for the 12-parameter NPE model for seeds 42 (left), 465 (middle), and 839 (right). 
    The gray regions indicate the 99\% confidence interval of a uniform distribution given the number of samples provided.
    }
    \label{fig:12paramSBCcdf}
\end{minipage}

\vspace{8mm}
\noindent
\begin{minipage}{\linewidth}
    \centering
    \includegraphics[width=0.6\linewidth]{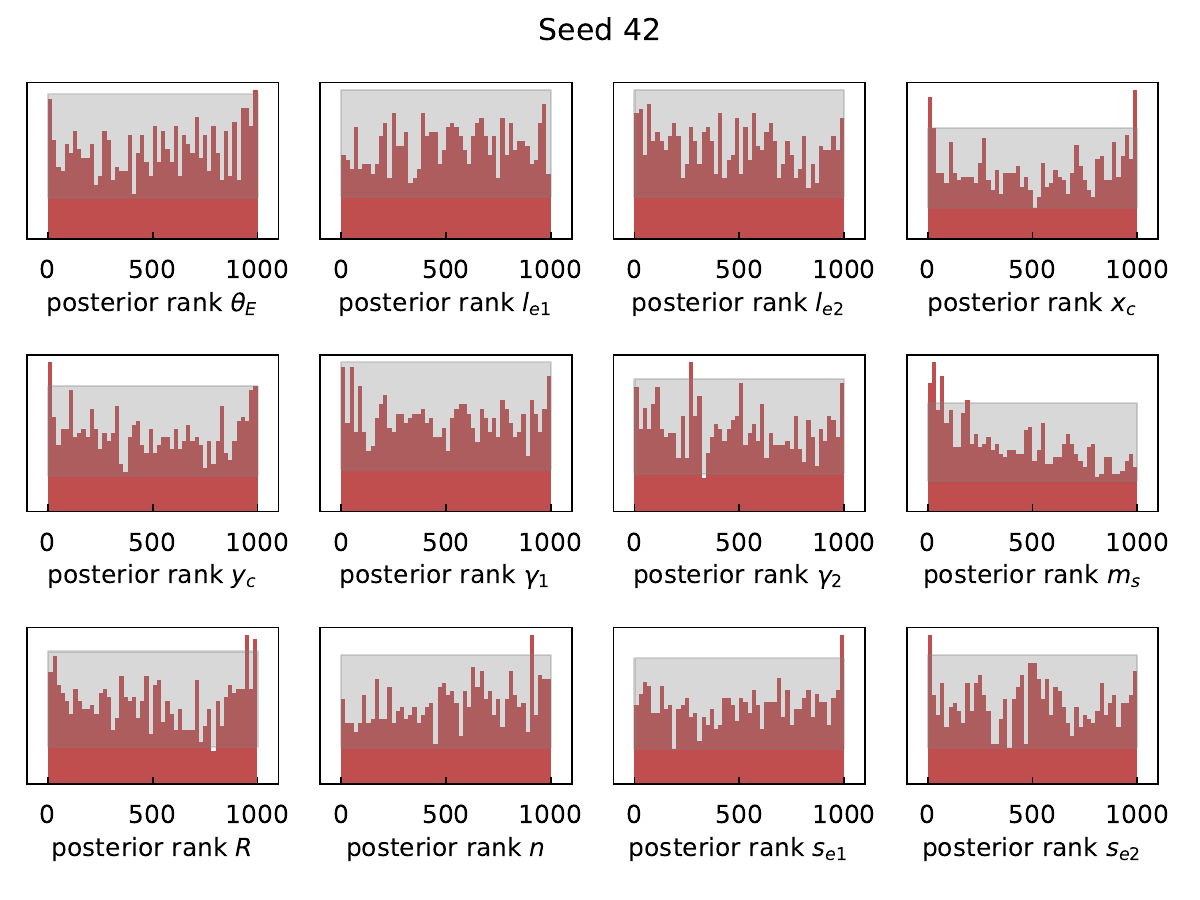}
    \includegraphics[width=0.6\linewidth]{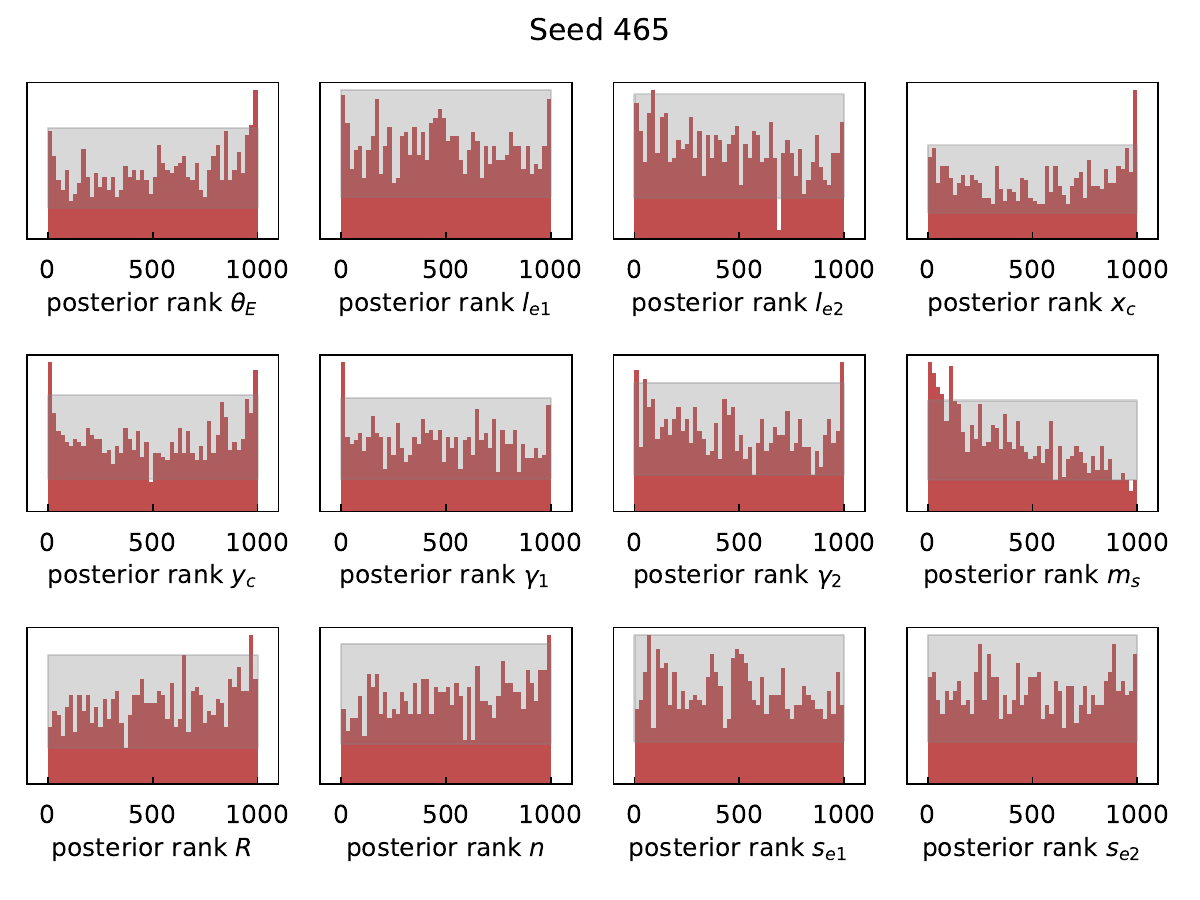}
    \includegraphics[width=0.6\linewidth]{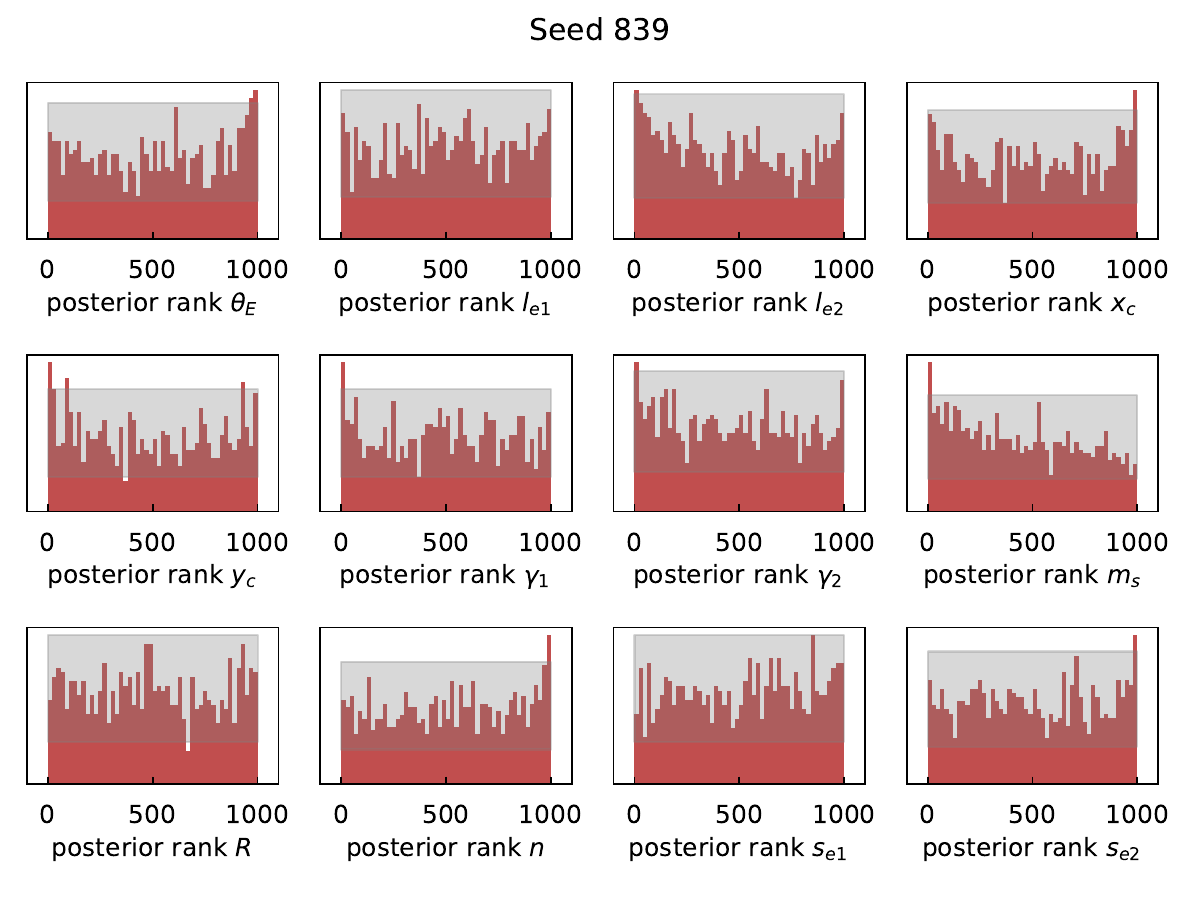}
    \captionof{figure}{
    Rank histograms for the 12-parameter NPE model with seeds 42, 465, and 839 (top, middle, bottom). 
    The gray region indicates uniformity.
    }
    \label{fig:12paramSBChist}
\end{minipage}

\section{Single-Image Inference Examples}

In addition to the single-lens image inference examples for the NPE and BNN models in the main text, here we include more examples for each of the 1-, 5-, and 12-parameter models. 
The examples here were also randomly selected from the test set.
We show the posterior-predictive check and the posterior itself for each model and lensing object.

\subsection{1-parameter model}

\begin{figure}
     \centering
     \begin{subfigure}[b]{\paramonewidthfactor\textwidth}
         \centering
         \includegraphics[width=\textwidth]{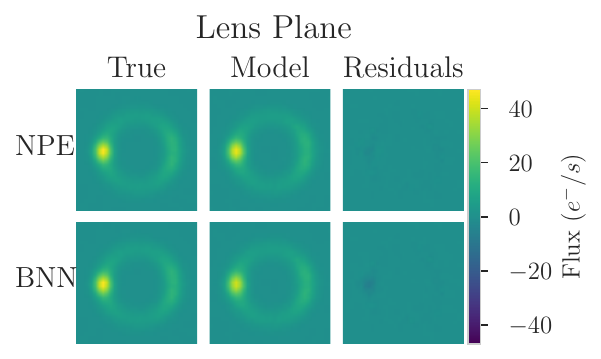}
     \end{subfigure}
     \hfill
     \begin{subfigure}[b]{\paramonewidthfactor\textwidth}
         \centering
         \includegraphics[width=\textwidth]{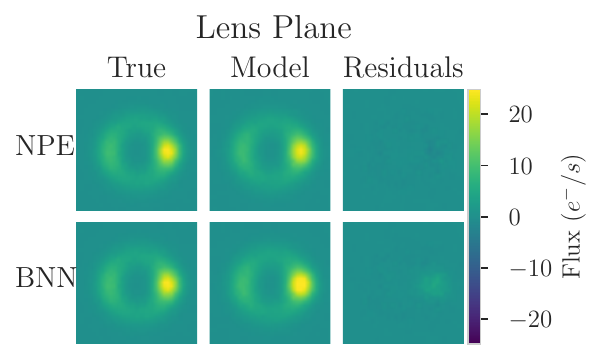}
     \end{subfigure}
     \hfill
     \begin{subfigure}[b]{\paramonewidthfactor\textwidth}
         \centering
         \includegraphics[width=\textwidth]{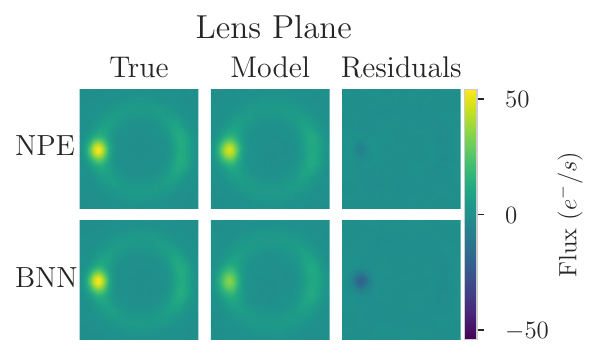}
     \end{subfigure}
     \hfill
     \begin{subfigure}[b]{\paramonewidthfactor\textwidth}
         \centering
         \includegraphics[width=\textwidth]{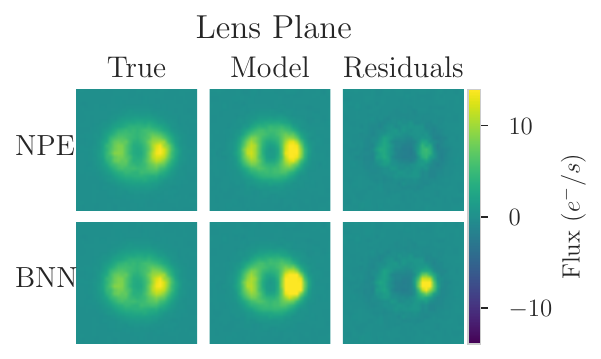}
     \end{subfigure}
     \hfill
     \begin{subfigure}[b]{\paramonewidthfactor\textwidth}
         \centering
         \includegraphics[width=\textwidth]{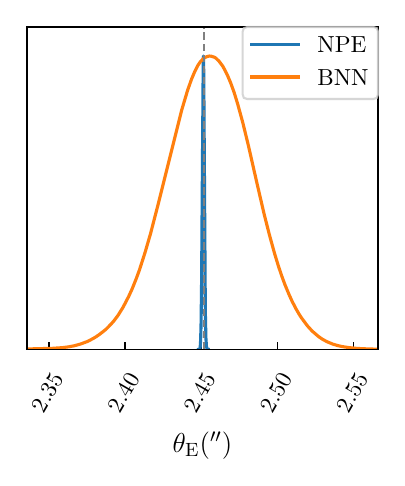}
         \caption{Lensing system A.}
     \end{subfigure}
     \hfill
     \begin{subfigure}[b]{\paramonewidthfactor\textwidth}
         \centering
         \includegraphics[width=\textwidth]{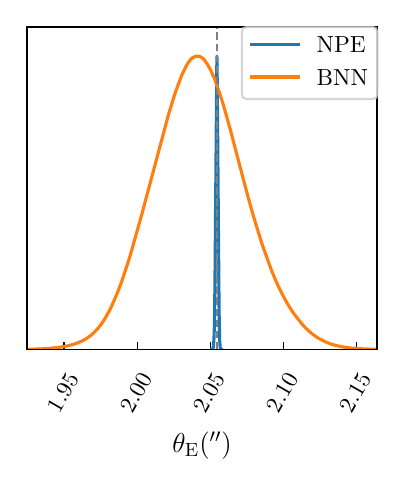}
         \caption{Lensing system B.}
     \end{subfigure}
     \hfill
     \begin{subfigure}[b]{\paramonewidthfactor\textwidth}
         \centering
         \includegraphics[width=\textwidth]{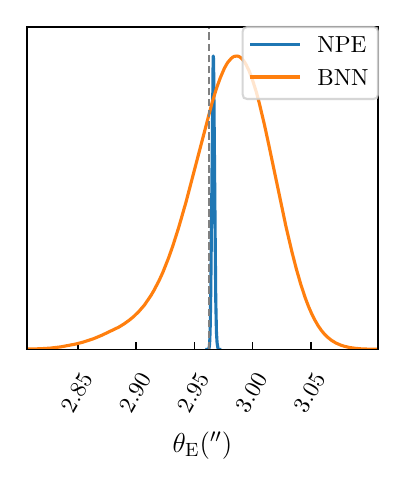}
         \caption{Lensing system C.}
     \end{subfigure}
     \hfill
     \begin{subfigure}[b]{\paramonewidthfactor\textwidth}
         \centering
         \includegraphics[width=\textwidth]{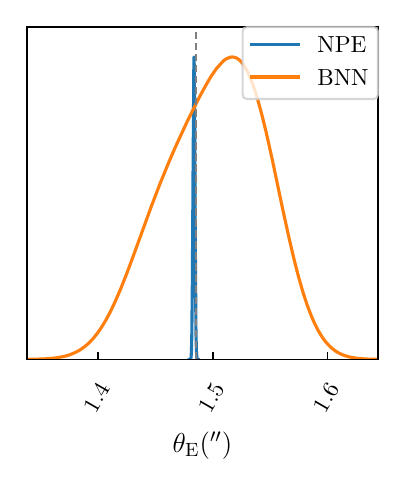}
         \caption{Lensing system D.}
     \end{subfigure}
     \hfill
     \caption{
        Performance for four lensing systems (A, B, C, and D) in the 1-parameter NPE and BNN models. 
        Top: the posterior predictive check for the lens plane image for the NPE (upper row) and the BNN (lower row) models, which includes the true image, the image predicted by each method, and the residual between the true and the predicted image for this test object.
        Bottom: inferred posterior of the Einstein radius $\theta_E$ for the NPE (blue) and BNN (orange) models.
      }
      \label{fig:1_paramsingle}
\end{figure}

\myfigure{}~\ref{fig:1_paramsingle} shows the posterior predictive image reconstruction (top) and the posterior inferences (bottom) of the NPE and BNN models for the Einstein radius (bottom) for individual lensing systems (A, B, C, and D) for the 1-parameter model.
The NPE model consistently provides significantly smaller uncertainties than the BNN model.
The NPE model is also consistently less biased than the BNN model for these examples; however, the NPE model is sometimes slightly biased with respect to the true posterior of the Einstein radius.

\begin{figure}[!ht]
     \centering
     \begin{subfigure}{\paramfivewidthfactor\textwidth}
         \centering
         \includegraphics[width=\textwidth]{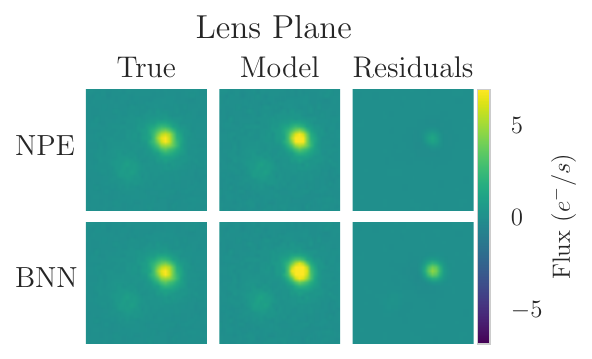}
     \end{subfigure}  
     \hfill
     \begin{subfigure}{\paramfivewidthfactor\textwidth}
         \centering
         \includegraphics[width=\textwidth]{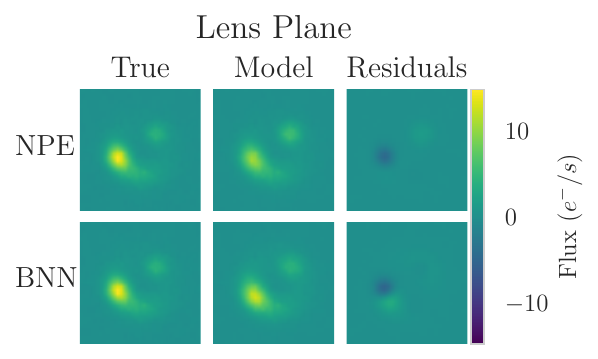}
     \end{subfigure}
     \hfill
     \begin{subfigure}[b]{\paramfivewidthfactor\textwidth}
         \centering
         \includegraphics[width=\textwidth]{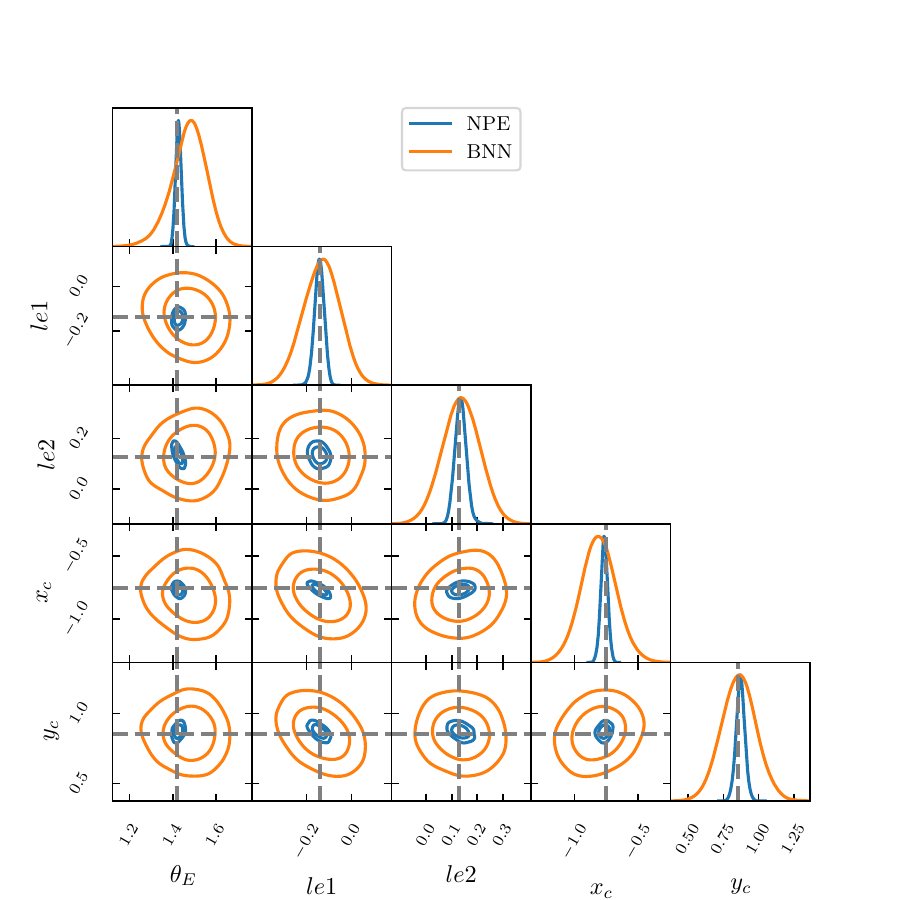}
         \caption{Lensing system E.}
     \end{subfigure}
     \hfill
     \begin{subfigure}[b]{\paramfivewidthfactor\textwidth}
         \centering
         \includegraphics[width=\textwidth]{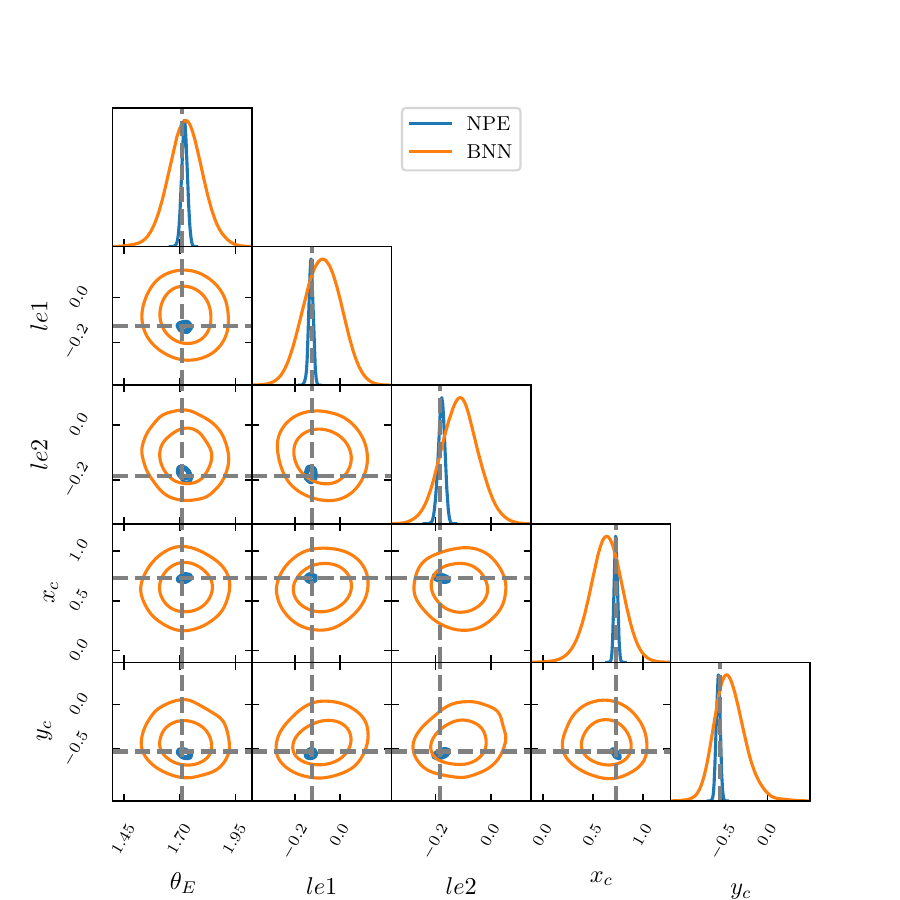}
         \caption{Lensing system F.}
     \end{subfigure}
     \hfill
     \begin{subfigure}{\paramfivewidthfactor\textwidth}
         \centering
         \includegraphics[width=\textwidth]{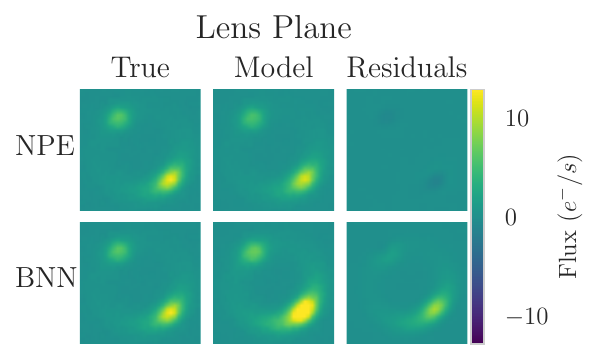}
     \end{subfigure}  
     \hfill
     \begin{subfigure}{\paramfivewidthfactor\textwidth}
         \centering
         \includegraphics[width=\textwidth]{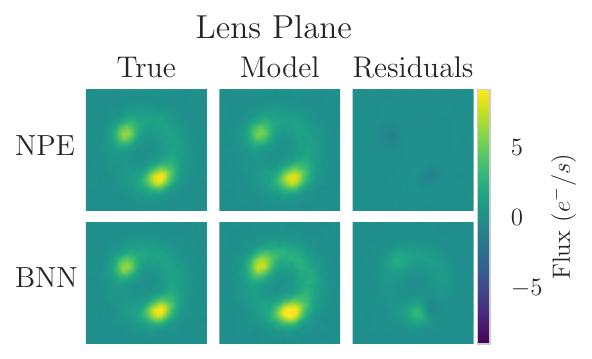}
     \end{subfigure}
     \hfill
     \begin{subfigure}[b]{\paramfivewidthfactor\textwidth}
         \centering
         \includegraphics[width=\textwidth]{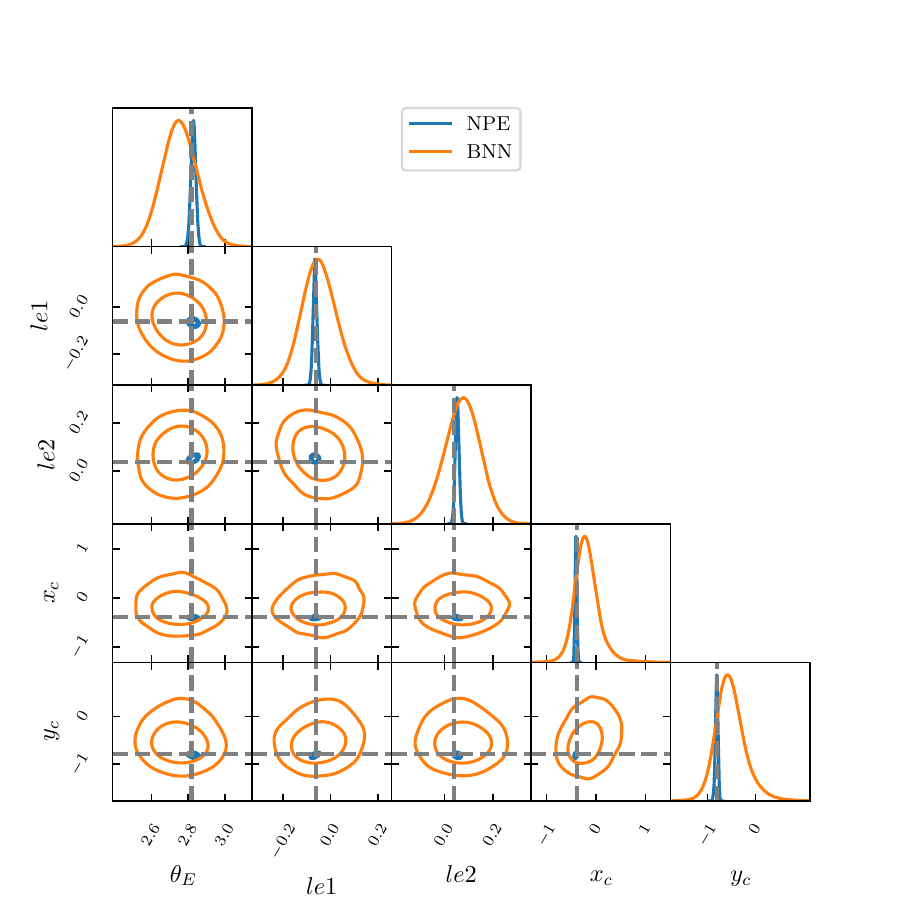}
         \caption{Lensing system G.}
     \end{subfigure}
     \hfill
     \begin{subfigure}[b]{\paramfivewidthfactor\textwidth}
         \centering
         \includegraphics[width=\textwidth]{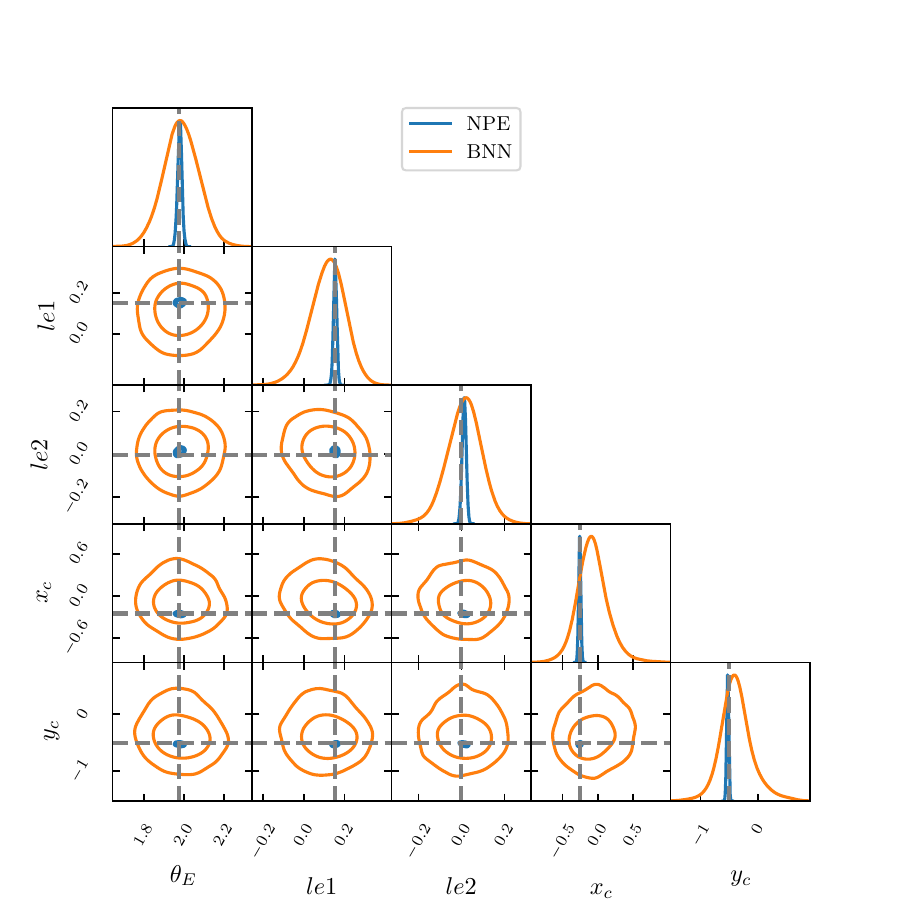}
         \caption{Lensing system H.}
     \end{subfigure}    
     \caption{
       Performance of the 5-parameter NPE and BNN models on four lensing objects drawn from the test set. 
       Top: the posterior predictive check for the lens plane image for the NPE (upper row) and the BNN (lower row) models, which includes the true image, the image predicted by each method, and the residual between the true and the predicted image for the object.       
       Bottom: inferred posteriors for the NPE (blue) and BNN (orange) models.
     }
     \label{fig:5_paramsingle}
\end{figure}

\subsection{5-parameter model}

\myfigure{}~\ref{fig:5_paramsingle} shows the posterior predictive image reconstruction (top) and the posterior inferences (bottom) of the NPE and BNN models for individual lensing systems (E, F, G, and H) for the 5-parameter model. 
The NPE model consistently provides smaller uncertainties than the BNN model for all parameters and examples.
The NPE model is also consistently less biased than the BNN model for these examples; however, the NPE model is sometimes slightly biased with respect to the true posteriors.
The NPE model uncertainties are about two orders of magnitude larger in the 5-parameter model than they are for the 1-parameter model.

\subsection{12-parameter model}

\myfigures{}~\ref{fig:12_paramsingle1}, ~\ref{fig:12_paramsingle2}, ~\ref{fig:12_paramsingle3}, and ~\ref{fig:12_paramsingle8} show the posterior predictive models (top) and the posterior inferences (bottom) from the NPE and BNN models for individual lensing systems  (I, J, K, and L) for the 12-parameter model. 
The NPE model consistently provides smaller uncertainties than the BNN model for all parameters and examples.
The NPE model is also consistently less biased than the BNN model for these examples.
However, the NPE model is sometimes slightly or moderately biased with respect to the true posteriors.
The biases for both the NPE and BNN models are larger for more parameters in the 12-parameter model than in the 5-parameter model.
The NPE model uncertainties are on the same order as or about one order of magnitude larger for the 12-parameter model compared to the 5-parameter model.

\begin{minipage}{\linewidth}
    \centering
    \includegraphics[width=\paramtwelvewidthfactor\linewidth]{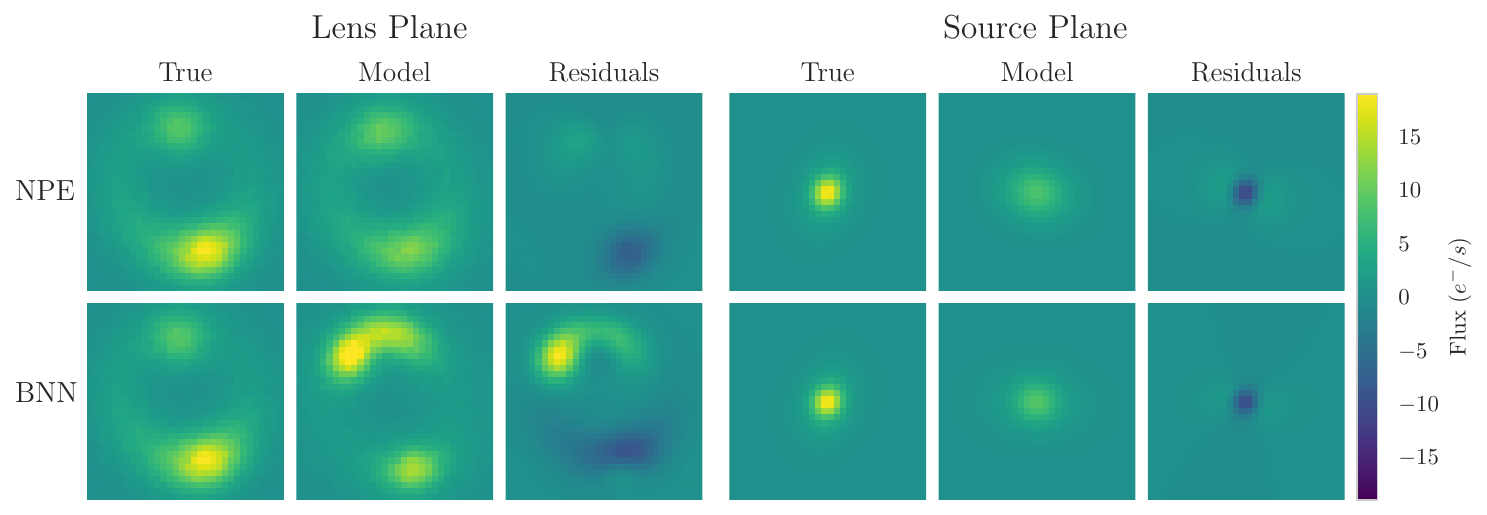}
    \includegraphics[width=1.0\linewidth]{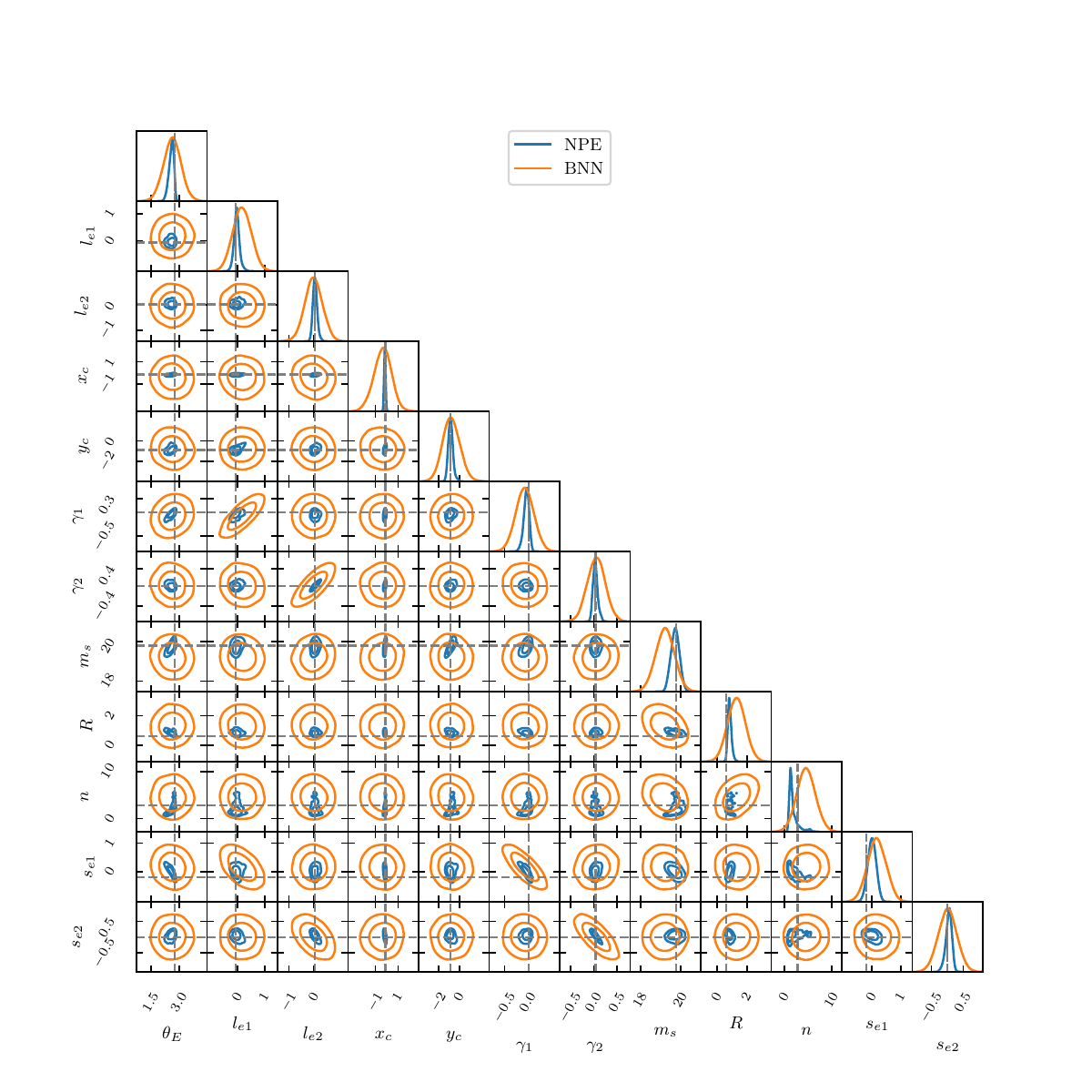}
    \captionof{figure}{
    Performance for the 12-parameter NPE and BNN models on system I (seed 465). 
    Top: the posterior predictive check for the lens plane image for the NPE (upper row) and the BNN (lower row) models, which includes the true image, the image predicted by each method, and the residual between the true and the predicted image for this test object.
    Bottom: inferred posteriors for the NPE (blue) and BNN (orange) models.
    The true values of parameters are indicated by dashed lines.
    }
    \label{fig:12_paramsingle1}
\end{minipage}

\begin{minipage}{\linewidth}
    \centering
    \includegraphics[width=\paramtwelvewidthfactor\linewidth]{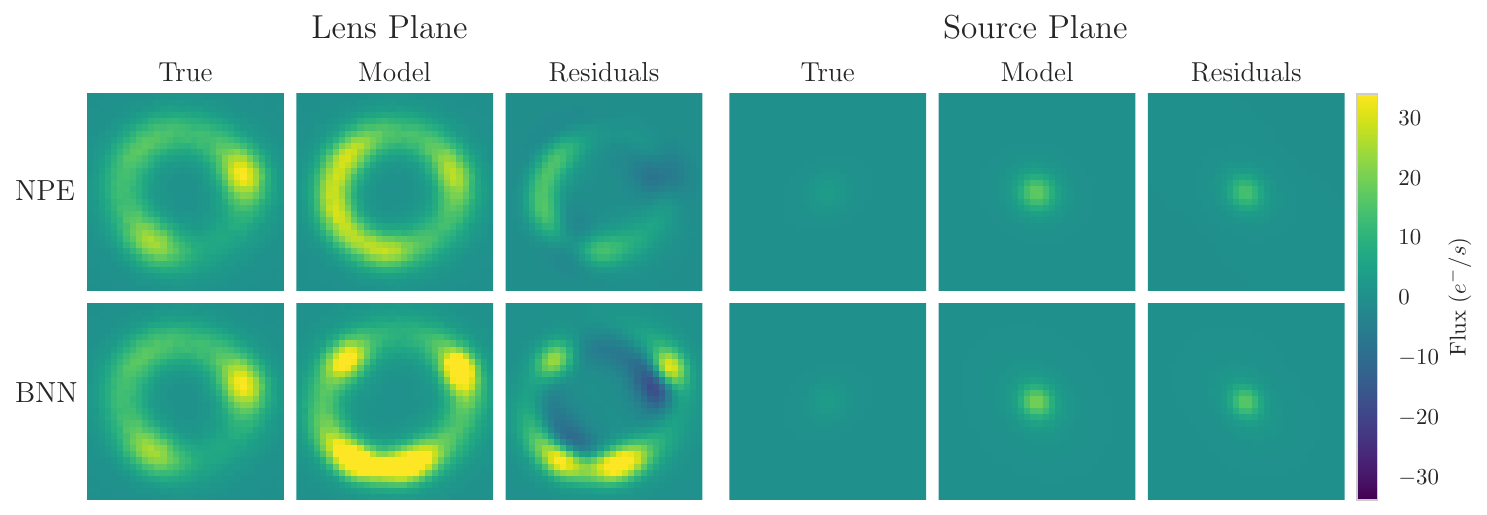}
    \includegraphics[width=1.0\linewidth]{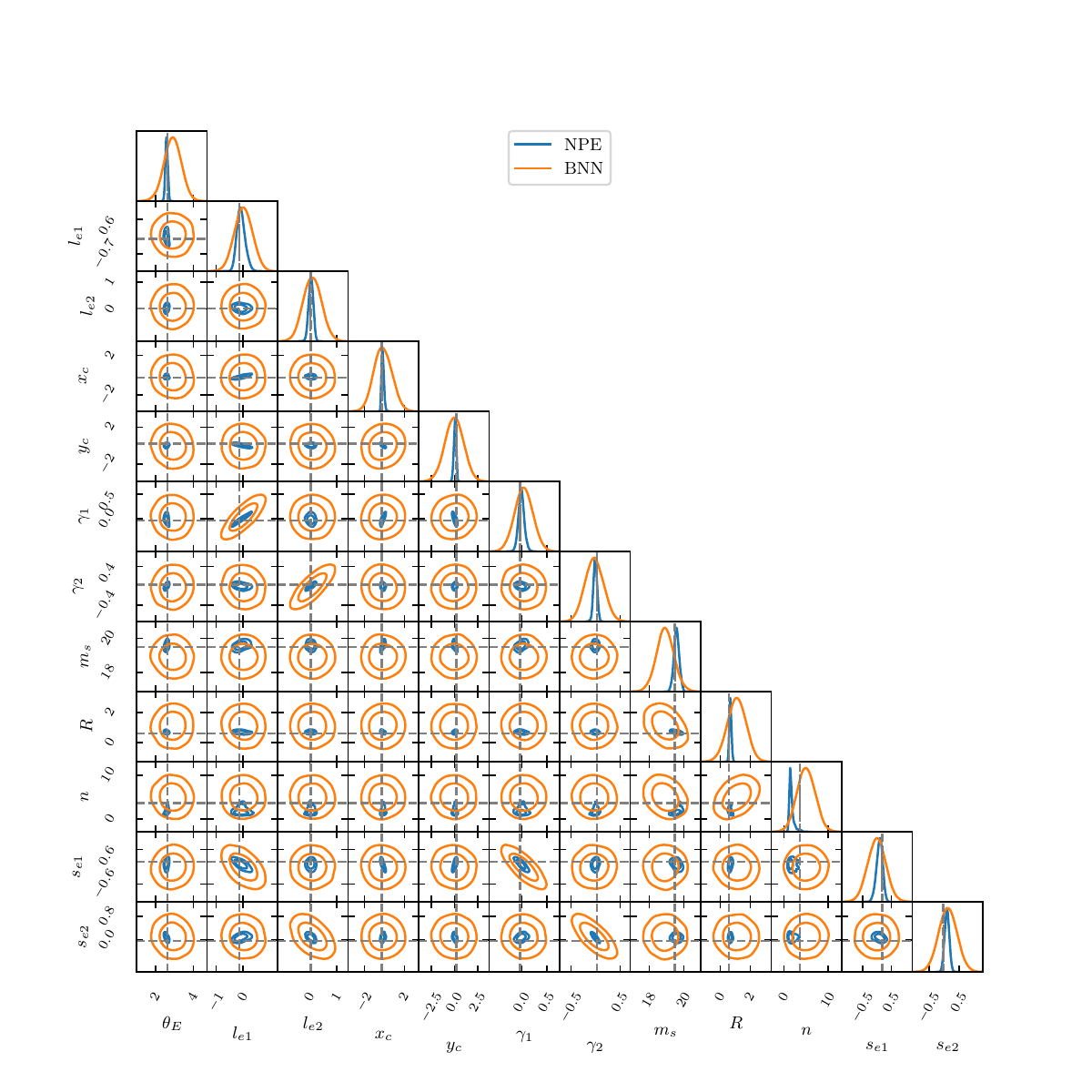}
    \captionof{figure}{ 
    Performance for the 12-parameter NPE and BNN models on system J  (seed 465). 
     Top: the posterior predictive check for the lens plane image for the NPE (upper row) and the BNN (lower row) models, which includes the true image, the image predicted by each method, and the residual between the true and the predicted image for this test object. 
    Bottom: inferred posteriors for the NPE (blue) and BNN (orange) models.
    The true values of parameters are indicated by dashed lines.
    }
    \label{fig:12_paramsingle2}
\end{minipage}

\begin{minipage}{\linewidth}
    \centering
    \includegraphics[width=\paramtwelvewidthfactor\linewidth]{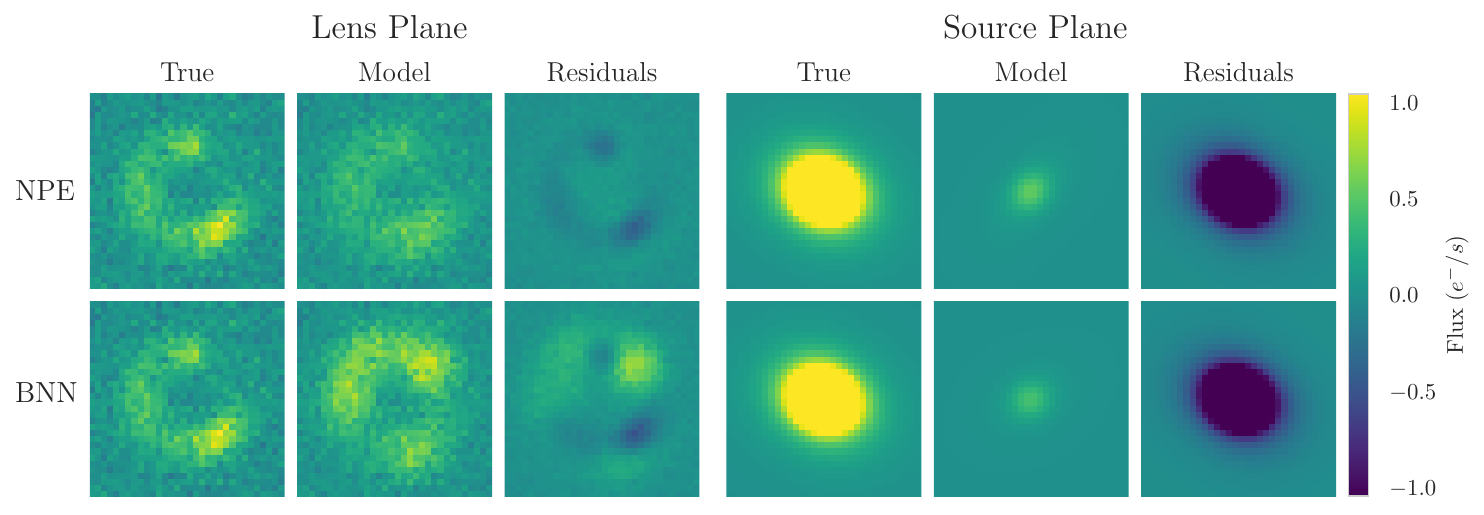}
    \includegraphics[width=1.0\linewidth]{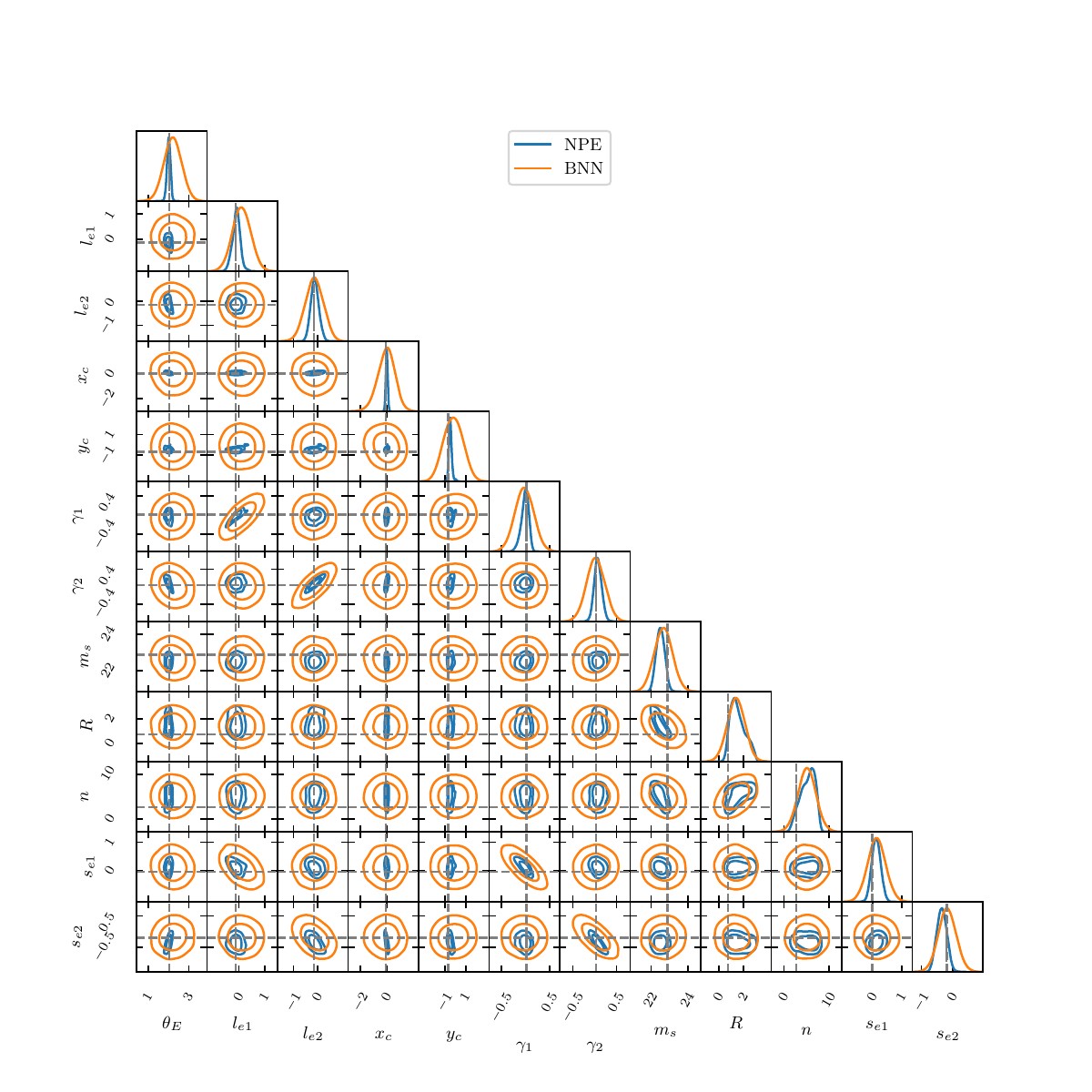}
    \captionof{figure}{
    Performance for the 12-parameter NPE and BNN models on system K  (seed 465). 
     Top: the posterior predictive check for the lens plane image for the NPE (upper row) and the BNN (lower row) models, which includes the true image, the image predicted by each method, and the residual between the true and the predicted image for this test object.
    Bottom: inferred posteriors for the NPE (blue) and BNN (orange) models.
    The true values of parameters are indicated by dashed lines.
    }
    \label{fig:12_paramsingle3}
\end{minipage}

\begin{minipage}{\linewidth}
\centering
    \includegraphics[width=\paramtwelvewidthfactor\linewidth]{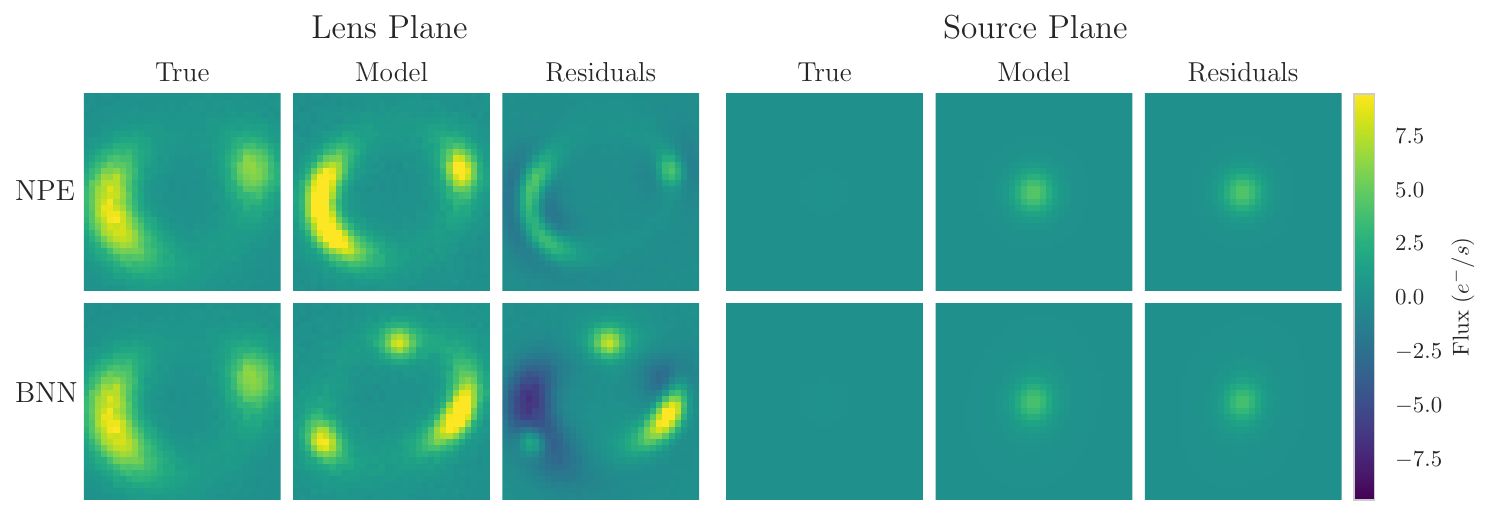}
    \includegraphics[width=1.0\linewidth]{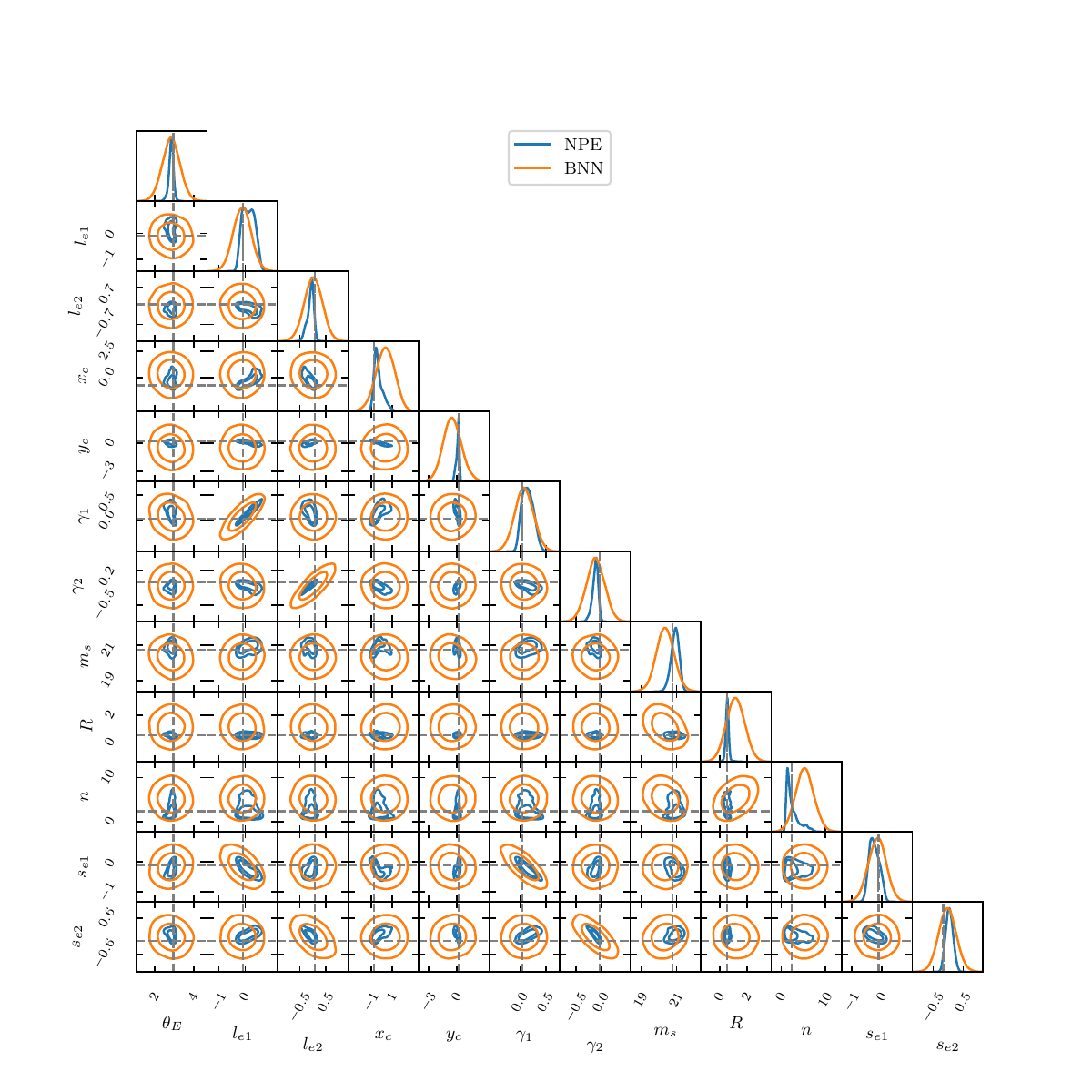}
    \captionof{figure}{
    Performance for the 12-parameter NPE and BNN models on system L  (seed 465). 
    Top: the posterior predictive check for the lens plane image for the NPE (upper row) and the BNN (lower row) models, which includes the true image, the image predicted by each method, and the residual between the true and the predicted image for this test object. 
    Bottom: inferred posteriors for the NPE (blue) and BNN (orange) models.
    The true values of parameters are indicated by dashed lines.
    }
    \label{fig:12_paramsingle8}
\end{minipage}



\newpage
\section*{Funding}

This work was produced by Fermi Forward Discovery Group, LLC under Contract No. 89243024CSC000002 with the U.S. Department of Energy, Office of Science, Office of High Energy Physics. The United States Government retains and the publisher, by accepting the work for publication, acknowledges that the United States Government retains a non-exclusive, paid-up, irrevocable, world-wide license to publish or reproduce the published form of this work, or allow others to do so, for United States Government purposes. The Department of Energy will provide public access to these results of federally sponsored research in accordance with the DOE Public Access Plan (\url{http://energy.gov/downloads/doe-public-access-plan}).

\section*{Acknowledgments}

The authors would like to thank the following colleagues for their insights and discussions during the development of this work: Sreevani Jarugula and Rebecca Nevin.

The authors of this paper have committed themselves to performing this work in an equitable, inclusive, and just environment, and we hold ourselves accountable, believing that the best science is contingent on a good research environment.
We acknowledge the Deep Skies Lab as a community of multi-domain experts and collaborators who have facilitated an environment of open discussion, idea generation, and collaboration. 
This community was important for the development of this project. 
We also would like to thank members of the University of Chicago Survey Science group for their feedback and comments. 

\section*{Author Contributions}

\noindent J.~Poh: Data curation, Formal analysis, Investigation, Methodology, Resources, Software, Visualization, Writing of original draft; \\

\noindent A.~Samudre: Conceptualization, Methodology, Software,  Writing  (review \& editing); \\

\noindent A.~\'Ciprijanovi\'c: Formal analysis, Investigation, Methodology, Software, Supervision, Writing of original draft; \\

\noindent G.~Khullar: Methodology, Software,  Writing  (review \& editing); \\

\noindent J.~Frieman: Supervision, Resources, Writing (review \& editing); \\

\noindent B.~D.~Nord: Conceptualization, Methodology, Resources, Supervision, Writing (review \& editing) \\

\bibliographystyle{plainnat}
\bibliography{main}

\begin{thebibliography}{144}
\providecommand{\natexlab}[1]{#1}
\providecommand{\url}[1]{\texttt{#1}}
\expandafter\ifx\csname urlstyle\endcsname\relax
  \providecommand{\doi}[1]{doi: #1}\else
  \providecommand{\doi}{doi: \begingroup \urlstyle{rm}\Url}\fi

\bibitem[{Abbott} et~al.(2018){Abbott}, {Abdalla}, {Allam}, {Amara}, {Annis},
  {Asorey}, {Avila}, {Ballester}, et~al., and {NOAO Data Lab}]{DESdatarelease1}
T.~M.~C. {Abbott}, F.~B. {Abdalla}, S.~{Allam}, A.~{Amara}, J.~{Annis},
  J.~{Asorey}, S.~{Avila}, O.~{Ballester}, et~al., and {NOAO Data Lab}.
\newblock {The Dark Energy Survey: Data Release 1}.
\newblock \emph{\apjs}, 239\penalty0 (2):\penalty0 18, December 2018.
\newblock \doi{10.3847/1538-4365/aae9f0}.

\bibitem[Abbott et~al.(2021)Abbott, Adamów, Aguena, Allam, Amon, Annis, Avila,
  Bacon, Banerji, Bechtol, Becker, Bernstein, Bertin, Bhargava, Bridle, Brooks,
  Burke, Carnero~Rosell, Carrasco~Kind, Carretero, Castander, Cawthon, Chang,
  Choi, Conselice, Costanzi, Crocce, da~Costa, Davis, De~Vicente, DeRose,
  Desai, Diehl, Dietrich, Drlica-Wagner, Eckert, Elvin-Poole, Everett, Evrard,
  Ferrero, Ferté, Flaugher, Fosalba, Friedel, Frieman, García-Bellido,
  Gaztanaga, Gelman, Gerdes, Giannantonio, Gill, Gruen, Gruendl, Gschwend,
  Gutierrez, Hartley, Hinton, Hollowood, Honscheid, Huterer, James, Jeltema,
  Johnson, Kent, Kron, Kuehn, Kuropatkin, Lahav, Li, Lidman, Lin, MacCrann,
  Maia, Manning, Maloney, March, Marshall, Martini, Melchior, Menanteau,
  Miquel, Morgan, Myles, Neilsen, Ogando, Palmese, Paz-Chinchón, Petravick,
  Pieres, Plazas, Pond, Rodriguez-Monroy, Romer, Roodman, Rykoff, Sako,
  Sanchez, Santiago, Scarpine, Serrano, Sevilla-Noarbe, Smith, Smith,
  Soares-Santos, Suchyta, Swanson, Tarle, Thomas, To, Tremblay, Troxel, Tucker,
  Turner, Varga, Walker, Wechsler, Weller, Wester, Wilkinson, Yanny, Zhang,
  Nikutta, Fitzpatrick, Jacques, Scott, Olsen, Huang, Herrera, Juneau, Nidever,
  Weaver, Adean, Correia, de~Freitas, Freitas, Singulani, and
  Vila-Verde]{Abbott_2021}
T.~M.~C. Abbott, M.~Adamów, M.~Aguena, S.~Allam, A.~Amon, J.~Annis, S.~Avila,
  D.~Bacon, M.~Banerji, K.~Bechtol, M.~R. Becker, G.~M. Bernstein, E.~Bertin,
  S.~Bhargava, S.~L. Bridle, D.~Brooks, D.~L. Burke, A.~Carnero~Rosell,
  M.~Carrasco~Kind, J.~Carretero, F.~J. Castander, R.~Cawthon, C.~Chang,
  A.~Choi, C.~Conselice, M.~Costanzi, M.~Crocce, L.~N. da~Costa, T.~M. Davis,
  J.~De~Vicente, J.~DeRose, S.~Desai, H.~T. Diehl, J.~P. Dietrich,
  A.~Drlica-Wagner, K.~Eckert, J.~Elvin-Poole, S.~Everett, A.~E. Evrard,
  I.~Ferrero, A.~Ferté, B.~Flaugher, P.~Fosalba, D.~Friedel, J.~Frieman,
  J.~García-Bellido, E.~Gaztanaga, L.~Gelman, D.~W. Gerdes, T.~Giannantonio,
  M.~S.~S. Gill, D.~Gruen, R.~A. Gruendl, J.~Gschwend, G.~Gutierrez, W.~G.
  Hartley, S.~R. Hinton, D.~L. Hollowood, K.~Honscheid, D.~Huterer, D.~J.
  James, T.~Jeltema, M.~D. Johnson, S.~Kent, R.~Kron, K.~Kuehn, N.~Kuropatkin,
  O.~Lahav, T.~S. Li, C.~Lidman, H.~Lin, N.~MacCrann, M.~A.~G. Maia, T.~A.
  Manning, J.~D. Maloney, M.~March, J.~L. Marshall, P.~Martini, P.~Melchior,
  F.~Menanteau, R.~Miquel, R.~Morgan, J.~Myles, E.~Neilsen, R.~L.~C. Ogando,
  A.~Palmese, F.~Paz-Chinchón, D.~Petravick, A.~Pieres, A.~A. Plazas, C.~Pond,
  M.~Rodriguez-Monroy, A.~K. Romer, A.~Roodman, E.~S. Rykoff, M.~Sako,
  E.~Sanchez, B.~Santiago, V.~Scarpine, S.~Serrano, I.~Sevilla-Noarbe, J.~Allyn
  Smith, M.~Smith, M.~Soares-Santos, E.~Suchyta, M.~E.~C. Swanson, G.~Tarle,
  D.~Thomas, C.~To, P.~E. Tremblay, M.~A. Troxel, D.~L. Tucker, D.~J. Turner,
  T.~N. Varga, A.~R. Walker, R.~H. Wechsler, J.~Weller, W.~Wester, R.~D.
  Wilkinson, B.~Yanny, Y.~Zhang, R.~Nikutta, M.~Fitzpatrick, A.~Jacques,
  A.~Scott, K.~Olsen, L.~Huang, D.~Herrera, S.~Juneau, D.~Nidever, B.~A.
  Weaver, C.~Adean, V.~Correia, M.~de~Freitas, F.~N. Freitas, C.~Singulani, and
  G.~Vila-Verde.
\newblock The dark energy survey data release 2.
\newblock \emph{\apjs}, 255\penalty0 (2):\penalty0 20, July 2021.
\newblock ISSN 1538-4365.
\newblock \doi{10.3847/1538-4365/ac00b3}.
\newblock URL \url{http://dx.doi.org/10.3847/1538-4365/ac00b3}.

\bibitem[Agarwal et~al.(2025)Agarwal, Ćiprijanović, and
  Nord]{agarwal2025neuralnetworkpredictionstrong}
Shrihan Agarwal, Aleksandra Ćiprijanović, and Brian~D. Nord.
\newblock Neural network prediction of strong lensing systems with domain
  adaptation and uncertainty quantification, 2025.
\newblock URL \url{https://arxiv.org/abs/2411.03334}.

\bibitem[Agnello et~al.(2017)Agnello, Lin, Buckley-Geer, Treu, Bonvin, Courbin,
  Lemon, Morishita, Amara, Auger, Birrer, Chan, Collett, More, Fassnacht,
  Frieman, Marshall, McMahon, Meylan, Suyu, Castander, Finley, Howell,
  Kochanek, Makler, Martini, Morgan, Nord, Ostrovski, Schechter, Tucker,
  Wechsler, Abbott, Abdalla, Allam, Benoit-L{\'{e} }vy, Bertin, Brooks, Burke,
  Rosell, Kind, Carretero, Crocce, Cunha, D'Andrea, da~Costa, Desai, Dietrich,
  Eifler, Flaugher, Fosalba, Garc{\'{\i}}a-Bellido, Gaztanaga, Gill, Goldstein,
  Gruen, Gruendl, Gschwend, Gutierrez, Honscheid, James, Kuehn, Kuropatkin, Li,
  Lima, Maia, March, Marshall, Melchior, Menanteau, Miquel, Ogando, Plazas,
  Romer, Sanchez, Schindler, Schubnell, Sevilla-Noarbe, Smith, Smith, Sobreira,
  Suchyta, Swanson, Tarle, Thomas, and Walker]{Agnello_2017}
A.~Agnello, H.~Lin, L.~Buckley-Geer, T.~Treu, V.~Bonvin, F.~Courbin, C.~Lemon,
  T.~Morishita, A.~Amara, M.~W. Auger, S.~Birrer, J.~Chan, T.~Collett, A.~More,
  C.~D. Fassnacht, J.~Frieman, P.~J. Marshall, R.~G. McMahon, G.~Meylan, S.~H.
  Suyu, F.~Castander, D.~Finley, A.~Howell, C.~Kochanek, M.~Makler, P.~Martini,
  N.~Morgan, B.~Nord, F.~Ostrovski, P.~Schechter, D.~Tucker, R.~Wechsler,
  T.~M.~C. Abbott, F.~B. Abdalla, S.~Allam, A.~Benoit-L{\'{e} }vy, E.~Bertin,
  D.~Brooks, D.~L. Burke, A.~Carnero Rosell, M.~Carrasco Kind, J.~Carretero,
  M.~Crocce, C.~E. Cunha, C.~B. D'Andrea, L.~N. da~Costa, S.~Desai, J.~P.
  Dietrich, T.~F. Eifler, B.~Flaugher, P.~Fosalba, J.~Garc{\'{\i}}a-Bellido,
  E.~Gaztanaga, M.~S. Gill, D.~A. Goldstein, D.~Gruen, R.~A. Gruendl,
  J.~Gschwend, G.~Gutierrez, K.~Honscheid, D.~J. James, K.~Kuehn,
  N.~Kuropatkin, T.~S. Li, M.~Lima, M.~A.~G. Maia, M.~March, J.~L. Marshall,
  P.~Melchior, F.~Menanteau, R.~Miquel, R.~L.~C. Ogando, A.~A. Plazas, A.~K.
  Romer, E.~Sanchez, R.~Schindler, M.~Schubnell, I.~Sevilla-Noarbe, M.~Smith,
  R.~C. Smith, F.~Sobreira, E.~Suchyta, M.~E.~C. Swanson, G.~Tarle, D.~Thomas,
  and A.~R. Walker.
\newblock Models of the strongly lensed quasar {DES} j0408-5354.
\newblock \emph{\mnras}, 472\penalty0 (4):\penalty0 4038--4050, sep 2017.
\newblock \doi{10.1093/mnras/stx2242}.
\newblock URL \url{https://doi.org/10.1093%2Fmnras%2Fstx2242}.

\bibitem[{Alsing} et~al.(2019){Alsing}, {Charnock}, {Feeney}, and
  {Wandelt}]{AC2019}
Justin {Alsing}, Tom {Charnock}, Stephen {Feeney}, and Benjamin {Wandelt}.
\newblock {Fast likelihood-free cosmology with neural density estimators and
  active learning}.
\newblock \emph{\mnras}, 488\penalty0 (3):\penalty0 4440--4458, September 2019.
\newblock \doi{10.1093/mnras/stz1960}.

\bibitem[{Anau Montel} et~al.(2023){Anau Montel}, {Coogan}, {Correa},
  {Karchev}, and {Weniger}]{Anau2023}
Noemi {Anau Montel}, Adam {Coogan}, Camila {Correa}, Konstantin {Karchev}, and
  Christoph {Weniger}.
\newblock {Estimating the warm dark matter mass from strong lensing images with
  truncated marginal neural ratio estimation}.
\newblock \emph{\mnras}, 518\penalty0 (2):\penalty0 2746--2760, January 2023.
\newblock \doi{10.1093/mnras/stac3215}.

\bibitem[Angelopoulos and Bates(2022)]{angelopoulos2022gentle}
Anastasios~N. Angelopoulos and Stephen Bates.
\newblock A gentle introduction to conformal prediction and distribution-free
  uncertainty quantification, 2022.

\bibitem[Auger et~al.(2010)Auger, Treu, Bolton, Gavazzi, Koopmans, Marshall,
  Moustakas, and Burles]{Auger2010a}
M.~W. Auger, T.~Treu, A.~S. Bolton, R.~Gavazzi, L.~V.~E. Koopmans, P.~J.
  Marshall, L.~A. Moustakas, and S.~Burles.
\newblock {The Sloan Lens ACS Survey. X. Stellar, Dynamical, and Total Mass
  Correlations of Massive Early-type Galaxies}.
\newblock \emph{\apj, Volume 724, Issue 1, pp. 511-525 (2010).}, 724:\penalty0
  511--525, jul 2010.
\newblock ISSN 0004-637X.
\newblock \doi{10.1088/0004-637X/724/1/511}.

\bibitem[Barkana(1998)]{Barkana1998}
Rennan Barkana.
\newblock {Fast calculation of a family of elliptical mass gravitational lens
  models}.
\newblock \emph{\apj, Volume 502, Issue 2, pp. 531-537.}, 502:\penalty0
  531--537, jan 1998.
\newblock ISSN 0004-637X.
\newblock \doi{10.1086/305950}.

\bibitem[Barnabe et~al.(2011)Barnabe, Czoske, Koopmans, Treu, and
  Bolton]{Barnabe2011}
Matteo Barnabe, Oliver Czoske, Leon V.~E. Koopmans, Tommaso Treu, and Adam~S.
  Bolton.
\newblock {Two-dimensional kinematics of SLACS lenses: III. Mass structure and
  dynamics of early-type lens galaxies beyond z {\~{}} 0.1}.
\newblock \emph{\mnras, Volume 415, Issue 3, pp. 2215-2232.}, 415:\penalty0
  2215--2232, feb 2011.
\newblock ISSN 0035-8711.
\newblock \doi{10.1111/j.1365-2966.2011.18842.x}.

\bibitem[Bayliss et~al.(2014)Bayliss, Rigby, Sharon, Wuyts, Florian, Gladders,
  Johnson, and Oguri]{Bayliss2014}
Matthew~B. Bayliss, Jane~R. Rigby, Keren Sharon, Eva Wuyts, Michael Florian,
  Michael~D. Gladders, Traci Johnson, and Masamune Oguri.
\newblock {The Physical Conditions, Metallicity and Metal Abundance Ratios in a
  Highly Magnified Galaxy at z = 3.6252}.
\newblock \emph{\apj, Volume 790, Issue 2, article id. 144, 20 pp. (2014).},
  790, 2014.
\newblock ISSN 0004-637X.
\newblock \doi{10.1088/0004-637X/790/2/144}.

\bibitem[Beaumont et~al.(2002)Beaumont, Zhang, and Balding]{Beaumont2002}
Mark~A Beaumont, Wenyang Zhang, and David~J Balding.
\newblock {Approximate Bayesian Computation in Population Genetics}.
\newblock \emph{Genetics}, 162\penalty0 (4):\penalty0 2025--2035, 12 2002.
\newblock ISSN 1943-2631.
\newblock \doi{10.1093/genetics/162.4.2025}.
\newblock URL \url{https://doi.org/10.1093/genetics/162.4.2025}.

\bibitem[Bengio et~al.(2014)Bengio, Courville, and
  Vincent]{bengio2014representation}
Yoshua Bengio, Aaron Courville, and Pascal Vincent.
\newblock Representation learning: A review and new perspectives, 2014.
\newblock URL \url{https://arxiv.org/abs/1206.5538}.

\bibitem[{Birrer} and {Amara}(2018)]{BA2018}
Simon {Birrer} and Adam {Amara}.
\newblock {lenstronomy: Multi-purpose gravitational lens modelling software
  package}.
\newblock \emph{Physics of the Dark Universe}, 22:\penalty0 189--201, December
  2018.
\newblock \doi{10.1016/j.dark.2018.11.002}.

\bibitem[{Birrer} et~al.(2017){Birrer}, {Welschen}, {Amara}, and
  {Refregier}]{birrer2017}
Simon {Birrer}, Cyril {Welschen}, Adam {Amara}, and Alexandre {Refregier}.
\newblock {Line-of-sight effects in strong lensing: putting theory into
  practice}.
\newblock \emph{\jcap}, 2017\penalty0 (4):\penalty0 049, April 2017.
\newblock \doi{10.1088/1475-7516/2017/04/049}.

\bibitem[{Birrer} et~al.(2021){Birrer}, {Shajib}, {Gilman}, {Galan}, {Aalbers},
  {Millon}, {Morgan}, {Pagano}, {Park}, {Teodori}, {Tessore}, {Ueland}, {Van de
  Vyvere}, {Wagner-Carena}, {Wempe}, {Yang}, {Ding}, {Schmidt}, {Sluse},
  {Zhang}, and {Amara}]{BS2021}
Simon {Birrer}, Anowar {Shajib}, Daniel {Gilman}, Aymeric {Galan}, Jelle
  {Aalbers}, Martin {Millon}, Robert {Morgan}, Giulia {Pagano}, Ji~{Park}, Luca
  {Teodori}, Nicolas {Tessore}, Madison {Ueland}, Lyne {Van de Vyvere},
  Sebastian {Wagner-Carena}, Ewoud {Wempe}, Lilan {Yang}, Xuheng {Ding}, Thomas
  {Schmidt}, Dominique {Sluse}, Ming {Zhang}, and Adam {Amara}.
\newblock {lenstronomy II: A gravitational lensing software ecosystem}.
\newblock \emph{JOSS}, 6\penalty0 (62):\penalty0 3283, June 2021.
\newblock \doi{10.21105/joss.03283}.

\bibitem[Blum et~al.(2013)Blum, Nunes, Prangle, and Sisson]{Blum_2013}
M.~G.~B. Blum, M.~A. Nunes, D.~Prangle, and S.~A. Sisson.
\newblock A comparative review of dimension reduction methods in approximate
  bayesian computation.
\newblock \emph{Stat. Sci.}, 28\penalty0 (2), May 2013.
\newblock ISSN 0883-4237.
\newblock \doi{10.1214/12-sts406}.
\newblock URL \url{http://dx.doi.org/10.1214/12-STS406}.

\bibitem[Bocquet and Carter(2016)]{Bocquet2016}
Sebastian Bocquet and Faustin~W. Carter.
\newblock pygtc: beautiful parameter covariance plots (aka. giant triangle
  confusograms).
\newblock \emph{JOSS}, 1\penalty0 (6), oct 2016.
\newblock \doi{10.21105/joss.00046}.
\newblock URL \url{http://dx.doi.org/10.21105/joss.00046}.

\bibitem[{Bolton} et~al.(2008){Bolton}, {Burles}, {Koopmans}, {Treu},
  {Gavazzi}, {Moustakas}, {Wayth}, and {Schlegel}]{SLACS4}
Adam~S. {Bolton}, Scott {Burles}, L{\'e}on V.~E. {Koopmans}, Tommaso {Treu},
  Rapha{\"e}l {Gavazzi}, Leonidas~A. {Moustakas}, Randall {Wayth}, and David~J.
  {Schlegel}.
\newblock {The Sloan Lens ACS Survey. V. The Full ACS Strong-Lens Sample}.
\newblock \emph{\apj}, 682\penalty0 (2):\penalty0 964--984, August 2008.
\newblock \doi{10.1086/589327}.

\bibitem[Bom et~al.(2019)Bom, Poh, Nord, Blanco-Valentin, and
  Dias]{bom2019deep}
Clecio Bom, Jason Poh, Brian Nord, Manuel Blanco-Valentin, and Luciana Dias.
\newblock Deep learning in wide-field surveys: Fast analysis of strong lenses
  in ground-based cosmic experiments, 2019.

\bibitem[Brehmer et~al.(2018{\natexlab{a}})Brehmer, Cranmer, Louppe, and
  Pavez]{Brehmer_2018}
Johann Brehmer, Kyle Cranmer, Gilles Louppe, and Juan Pavez.
\newblock Constraining effective field theories with machine learning.
\newblock \emph{\prl}, 121\penalty0 (11), sep 2018{\natexlab{a}}.
\newblock \doi{10.1103/physrevlett.121.111801}.
\newblock URL \url{https://doi.org/10.1103%2Fphysrevlett.121.111801}.

\bibitem[Brehmer et~al.(2018{\natexlab{b}})Brehmer, Cranmer, Louppe, and
  Pavez]{brehmer2018guide}
Johann Brehmer, Kyle Cranmer, Gilles Louppe, and Juan Pavez.
\newblock A guide to constraining effective field theories with machine
  learning.
\newblock \emph{\prd}, 98\penalty0 (5), September 2018{\natexlab{b}}.
\newblock ISSN 2470-0029.
\newblock \doi{10.1103/physrevd.98.052004}.
\newblock URL \url{http://dx.doi.org/10.1103/PhysRevD.98.052004}.

\bibitem[Brehmer et~al.(2019)Brehmer, Mishra-Sharma, Hermans, Louppe, and
  Cranmer]{Brehmer_2019}
Johann Brehmer, Siddharth Mishra-Sharma, Joeri Hermans, Gilles Louppe, and Kyle
  Cranmer.
\newblock Mining for dark matter substructure: Inferring subhalo population
  properties from strong lenses with machine learning.
\newblock \emph{\apj}, 886\penalty0 (1):\penalty0 49, nov 2019.
\newblock \doi{10.3847/1538-4357/ab4c41}.
\newblock URL \url{https://doi.org/10.3847%2F1538-4357%2Fab4c41}.

\bibitem[{Butter} et~al.(2022){Butter}, {Finke}, {Keil}, {Kr{\"a}mer}, and
  {Manconi}]{BF2022}
Anja {Butter}, Thorben {Finke}, Felicitas {Keil}, Michael {Kr{\"a}mer}, and
  Silvia {Manconi}.
\newblock {Classification of Fermi-LAT blazars with Bayesian neural networks}.
\newblock \emph{\jcap}, 2022\penalty0 (4):\penalty0 023, April 2022.
\newblock \doi{10.1088/1475-7516/2022/04/023}.

\bibitem[{Cameron} and {Pettitt}(2012)]{cameron2012}
E.~{Cameron} and A.~N. {Pettitt}.
\newblock {Approximate Bayesian Computation for astronomical model analysis: a
  case study in galaxy demographics and morphological transformation at high
  redshift}.
\newblock \emph{\mnras}, 425\penalty0 (1):\penalty0 44--65, September 2012.
\newblock \doi{10.1111/j.1365-2966.2012.21371.x}.

\bibitem[Charnock et~al.(2018)Charnock, Lavaux, and Wandelt]{Charnock_2018}
Tom Charnock, Guilhem Lavaux, and Benjamin~D. Wandelt.
\newblock Automatic physical inference with information maximizing neural
  networks.
\newblock \emph{\prd}, 97\penalty0 (8), April 2018.
\newblock ISSN 2470-0029.
\newblock \doi{10.1103/physrevd.97.083004}.
\newblock URL \url{http://dx.doi.org/10.1103/PhysRevD.97.083004}.

\bibitem[Chen et~al.(2019)Chen, Li, Shu, and Cao]{Chen_2019}
Yun Chen, Ran Li, Yiping Shu, and Xiaoyue Cao.
\newblock Assessing the effect of lens mass model in cosmological application
  with updated galaxy-scale strong gravitational lensing sample.
\newblock \emph{\mnras}, 488\penalty0 (3):\penalty0 3745–3758, July 2019.
\newblock ISSN 1365-2966.
\newblock \doi{10.1093/mnras/stz1902}.
\newblock URL \url{http://dx.doi.org/10.1093/mnras/stz1902}.

\bibitem[Christensen et~al.(2012)Christensen, Laursen, Richard, Hjorth,
  Milvang-Jensen, Dessauges-Zavadsky, Limousin, Grillo, and
  Ebeling]{Christensen2012}
Lise Christensen, Peter Laursen, Johan Richard, Jens Hjorth, Bo~Milvang-Jensen,
  Miroslava Dessauges-Zavadsky, Marceau Limousin, Claudio Grillo, and Harald
  Ebeling.
\newblock {Gravitationally Lensed Galaxies at 2}.
\newblock \emph{\mnras, Volume 427, Issue 3, pp. 1973-1982.}, 427:\penalty0
  1973--1982, sep 2012.
\newblock ISSN 0035-8711.
\newblock \doi{10.1111/j.1365-2966.2012.22007.x}.

\bibitem[{Ciotti} and {Bertin}(1999)]{Ciotti1999}
L.~{Ciotti} and G.~{Bertin}.
\newblock {Analytical properties of the R$^{1/m}$ law}.
\newblock \emph{\aap}, 352:\penalty0 447--451, December 1999.

\bibitem[Collett et~al.(2012)Collett, Auger, Belokurov, Marshall, and
  Hall]{Collett2012}
T.~E. Collett, M.~W. Auger, V.~Belokurov, P.~J. Marshall, and A.~C. Hall.
\newblock {Constraining the dark energy equation of state with double-source
  plane strong lenses}.
\newblock \emph{\mnras, Volume 424, Issue 4, pp. 2864-2875.}, 424:\penalty0
  2864--2875, 2012.
\newblock ISSN 0035-8711.
\newblock \doi{10.1111/J.1365-2966.2012.21424.X}.

\bibitem[{Collett}(2015)]{Collett2015}
Thomas~E. {Collett}.
\newblock {The Population of Galaxy-Galaxy Strong Lenses in Forthcoming Optical
  Imaging Surveys}.
\newblock \emph{\apj}, 811\penalty0 (1):\penalty0 20, September 2015.
\newblock \doi{10.1088/0004-637X/811/1/20}.

\bibitem[Collett and Auger(2014)]{Collett2014}
Thomas~E. Collett and Matthew~W. Auger.
\newblock {Cosmological constraints from the double source plane lens
  SDSSJ0946+1006}.
\newblock \emph{\mnras, Volume 443, Issue 2, p.969-976}, 443:\penalty0
  969--976, 2014.
\newblock ISSN 0035-8711.
\newblock \doi{10.1093/MNRAS/STU1190}.

\bibitem[{Collett} et~al.(2017){Collett}, {Buckley-Geer}, {Lin}, {Bacon},
  {Nichol}, {Nord}, {Morice-Atkinson}, {Amara}, {Birrer}, {Kuropatkin}, {More},
  {Papovich}, {Romer}, {Tessore}, {Abbott}, {Allam}, {Annis},
  {Benoit-L{\'e}vy}, {Brooks}, {Burke}, {Carrasco Kind}, {Castander},
  {D'Andrea}, {da Costa}, {Desai}, {Diehl}, {Doel}, {Eifler}, {Flaugher},
  {Frieman}, {Gerdes}, {Goldstein}, {Gruen}, {Gschwend}, {Gutierrez}, {James},
  {Kuehn}, {Kuhlmann}, {Lahav}, {Li}, {Lima}, {Maia}, {March}, {Marshall},
  {Martini}, {Melchior}, {Miquel}, {Plazas}, {Rykoff}, {Sanchez}, {Scarpine},
  {Schindler}, {Schubnell}, {Sevilla-Noarbe}, {Smith}, {Sobreira}, {Suchyta},
  {Swanson}, {Tarle}, {Tucker}, and {Walker}]{Collett_2017}
Thomas~E. {Collett}, Elizabeth {Buckley-Geer}, Huan {Lin}, David {Bacon},
  Robert~C. {Nichol}, Brian {Nord}, Xan {Morice-Atkinson}, Adam {Amara}, Simon
  {Birrer}, Nikolay {Kuropatkin}, Anupreeta {More}, Casey {Papovich}, Kathy~K.
  {Romer}, Nicolas {Tessore}, Tim M.~C. {Abbott}, Sahar {Allam}, James {Annis},
  Aurlien {Benoit-L{\'e}vy}, David {Brooks}, David~L. {Burke}, Matias {Carrasco
  Kind}, Francisco Javier~J. {Castander}, Chris~B. {D'Andrea}, Luiz~N. {da
  Costa}, Shantanu {Desai}, H.~Thomas {Diehl}, Peter {Doel}, Tim~F. {Eifler},
  Brenna {Flaugher}, Josh {Frieman}, David~W. {Gerdes}, Daniel~A. {Goldstein},
  Daniel {Gruen}, Julia {Gschwend}, Gaston {Gutierrez}, David~J. {James}, Kyler
  {Kuehn}, Steve {Kuhlmann}, Ofer {Lahav}, Ting~S. {Li}, Marcos {Lima}, Marcio
  A.~G. {Maia}, Marisa {March}, Jennifer~L. {Marshall}, Paul {Martini}, Peter
  {Melchior}, Ramon {Miquel}, Andrs~A. {Plazas}, Eli~S. {Rykoff}, Eusebio
  {Sanchez}, Vic {Scarpine}, Rafe {Schindler}, Michael {Schubnell}, Ignacio
  {Sevilla-Noarbe}, Mathew {Smith}, Flavia {Sobreira}, Eric {Suchyta}, Molly
  E.~C. {Swanson}, Gregory {Tarle}, Douglas~L. {Tucker}, and Alistair~R.
  {Walker}.
\newblock {Core or Cusps: The Central Dark Matter Profile of a Strong Lensing
  Cluster with a Bright Central Image at Redshift 1}.
\newblock \emph{\apj}, 843\penalty0 (2):\penalty0 148, July 2017.
\newblock \doi{10.3847/1538-4357/aa76e6}.

\bibitem[{Coogan} et~al.(2022){Coogan}, {Anau Montel}, {Karchev}, {Grootes},
  {Nattino}, and {Weniger}]{Coogan2022}
Adam {Coogan}, Noemi {Anau Montel}, Konstantin {Karchev}, Meiert~W. {Grootes},
  Francesco {Nattino}, and Christoph {Weniger}.
\newblock {One never walks alone: the effect of the perturber population on
  subhalo measurements in strong gravitational lenses}.
\newblock \emph{arXiv e-prints}, art. arXiv:2209.09918, September 2022.

\bibitem[Cook et~al.(2006)Cook, Gelman, and Rubin]{Cook2006}
Samantha Cook, Andrew Gelman, and Donald Rubin.
\newblock Validation of software for bayesian models using posterior quantiles.
\newblock \emph{J. Comput. Graph. Stat.}, 15, 09 2006.
\newblock \doi{10.1198/106186006X136976}.

\bibitem[Cranmer et~al.(2020)Cranmer, Brehmer, and Louppe]{CB2019}
Kyle Cranmer, Johann Brehmer, and Gilles Louppe.
\newblock The frontier of simulation-based inference.
\newblock \emph{Proceedings of the National Academy of Sciences}, 117\penalty0
  (48):\penalty0 30055–30062, May 2020.
\newblock ISSN 1091-6490.
\newblock \doi{10.1073/pnas.1912789117}.
\newblock URL \url{http://dx.doi.org/10.1073/pnas.1912789117}.

\bibitem[{Dark Energy Survey Collaboration} et~al.(2016){Dark Energy Survey
  Collaboration}, {Abbott}, {Abdalla}, {Aleksi{\'c}}, {Allam}, {Amara},
  {Bacon}, {Balbinot}, {Banerji}, {Bechtol}, {Benoit-L{\'e}vy}, and
  et~al.]{DES2016}
{Dark Energy Survey Collaboration}, T.~{Abbott}, F.~B. {Abdalla},
  J.~{Aleksi{\'c}}, S.~{Allam}, A.~{Amara}, D.~{Bacon}, E.~{Balbinot},
  M.~{Banerji}, K.~{Bechtol}, A.~{Benoit-L{\'e}vy}, and et~al.
\newblock {The Dark Energy Survey: more than dark energy - an overview}.
\newblock \emph{\mnras}, 460\penalty0 (2):\penalty0 1270--1299, August 2016.
\newblock \doi{10.1093/mnras/stw641}.

\bibitem[Deistler et~al.(2022)Deistler, Goncalves, and
  Macke]{deistler2022truncated}
Michael Deistler, Pedro~J Goncalves, and Jakob~H Macke.
\newblock Truncated proposals for scalable and hassle-free simulation-based
  inference, 2022.

\bibitem[Dessauges-Zavadsky et~al.(2009)Dessauges-Zavadsky, D'Odorico,
  Schaerer, Modigliani, Tapken, and Vernet]{Dessauges-Zavadsky2009}
M.~Dessauges-Zavadsky, S.~D'Odorico, D.~Schaerer, A.~Modigliani, C.~Tapken, and
  J.~Vernet.
\newblock {Rest-frame ultraviolet spectrum of the gravitationally lensed galaxy
  `the 8 o'clock arc': stellar and interstellar medium properties}.
\newblock \emph{\aap, Volume 510, id.A26}, 510, dec 2009.
\newblock ISSN 0004-6361.
\newblock \doi{10.1051/0004-6361/200913337}.

\bibitem[Durkan et~al.(2020)Durkan, Murray, and
  Papamakarios]{durkan2020contrastive}
Conor Durkan, Iain Murray, and George Papamakarios.
\newblock On contrastive learning for likelihood-free inference, 2020.

\bibitem[{Escamilla-Rivera} et~al.(2022){Escamilla-Rivera}, {Carvajal},
  {Zamora}, and {Hendry}]{ER2022}
Celia {Escamilla-Rivera}, Maryi {Carvajal}, Cristian {Zamora}, and Martin
  {Hendry}.
\newblock {Neural networks and standard cosmography with newly calibrated high
  redshift GRB observations}.
\newblock \emph{\jcap}, 2022\penalty0 (4):\penalty0 016, April 2022.
\newblock \doi{10.1088/1475-7516/2022/04/016}.

\bibitem[Finkelstein et~al.(2009)Finkelstein, Papovich, Rudnick, Egami, Floc'h,
  Rieke, Rigby, and Willmer]{Finkelstein2009}
Steven~L. Finkelstein, Casey Papovich, Gregory Rudnick, Eiichi Egami, Emeric~Le
  Floc'h, Marcia~J. Rieke, Jane Rigby, and Christopher N.~A. Willmer.
\newblock {Turning Back the Clock: Inferring the History of the Eight O'clock
  Arc}.
\newblock \emph{\apj, Volume 700, Issue 1, pp. 376-386 (2009).}, 700:\penalty0
  376--386, may 2009.
\newblock ISSN 0004-637X.
\newblock \doi{10.1088/0004-637X/700/1/376}.

\bibitem[Foreman-Mackey(2016)]{corner}
Daniel Foreman-Mackey.
\newblock corner.py: Scatterplot matrices in python.
\newblock \emph{JOSS}, 1\penalty0 (2):\penalty0 24, jun 2016.
\newblock \doi{10.21105/joss.00024}.
\newblock URL \url{https://doi.org/10.21105/joss.00024}.

\bibitem[Gabry et~al.(2019)Gabry, Simpson, Vehtari, Betancourt, and
  Gelman]{PPC}
Jonah Gabry, Daniel Simpson, Aki Vehtari, Michael Betancourt, and Andrew
  Gelman.
\newblock {Visualization in Bayesian Workflow}.
\newblock \emph{Journal of the Royal Statistical Society Series A: Statistics
  in Society}, 182\penalty0 (2):\penalty0 389--402, 01 2019.
\newblock ISSN 0964-1998.
\newblock \doi{10.1111/rssa.12378}.
\newblock URL \url{https://doi.org/10.1111/rssa.12378}.

\bibitem[Gavazzi et~al.(2008)Gavazzi, Treu, Koopmans, Bolton, Moustakas,
  Burles, and Marshall]{Gavazzi2008}
Raphael Gavazzi, Tommaso Treu, Leon V.~E. Koopmans, Adam~S. Bolton, Leonidas~A.
  Moustakas, Scott Burles, and Philip~J. Marshall.
\newblock {The Sloan Lens ACS Survey. VI: Discovery and analysis of a double
  Einstein ring}.
\newblock \emph{\apj, Volume 677, Issue 2, article id. 1046-1059, pp. (2008).},
  677, jan 2008.
\newblock ISSN 0004-637X.
\newblock \doi{10.1086/529541}.

\bibitem[Gentile et~al.(2023)Gentile, Tortora, Covone, Koopmans, Li, Leuzzi,
  and Napolitano]{Gentile_2023}
Fabrizio Gentile, Crescenzo Tortora, Giovanni Covone, Léon V~E Koopmans, Rui
  Li, Laura Leuzzi, and Nicola~R Napolitano.
\newblock <scp>lemon</scp>: Lens modelling with neural networks – i.
  automated modelling of strong gravitational lenses with bayesian neural
  networks.
\newblock \emph{\mnras}, 522\penalty0 (4):\penalty0 5442–5455, May 2023.
\newblock ISSN 1365-2966.
\newblock \doi{10.1093/mnras/stad1325}.
\newblock URL \url{http://dx.doi.org/10.1093/mnras/stad1325}.

\bibitem[{Gerardi} et~al.(2021){Gerardi}, {Feeney}, and {Alsing}]{GF2021}
Francesca {Gerardi}, Stephen~M. {Feeney}, and Justin {Alsing}.
\newblock {Unbiased likelihood-free inference of the Hubble constant from light
  standard sirens}.
\newblock \emph{\prd}, 104\penalty0 (8):\penalty0 083531, October 2021.
\newblock \doi{10.1103/PhysRevD.104.083531}.

\bibitem[Germain et~al.(2015)Germain, Gregor, Murray, and Larochelle]{GG2015}
Mathieu Germain, Karol Gregor, Iain Murray, and Hugo Larochelle.
\newblock Made: Masked autoencoder for distribution estimation.
\newblock In Francis Bach and David Blei, editors, \emph{Proceedings of the
  32nd International Conference on Machine Learning}, volume~37 of
  \emph{Proceedings of Machine Learning Research}, pages 881--889, Lille,
  France, 07--09 Jul 2015. PMLR.
\newblock URL \url{https://proceedings.mlr.press/v37/germain15.html}.

\bibitem[Goncalves et~al.(2020)Goncalves, Lueckmann, Deistler, Nonnenmacher,
  Öcal, Bassetto, Chintaluri, Podlaski, Haddad, Vogels, Greenberg, and
  Macke]{goncalves2020}
Pedro Goncalves, Jan-Matthis Lueckmann, Michael Deistler, Marcel Nonnenmacher,
  Kaan Öcal, Giacomo Bassetto, Chaitanya Chintaluri, William Podlaski, Sara
  Haddad, Tim Vogels, David Greenberg, and Jakob Macke.
\newblock Training deep neural density estimators to identify mechanistic
  models of neural dynamics.
\newblock \emph{eLife}, 9, 09 2020.
\newblock \doi{10.7554/eLife.56261}.

\bibitem[{Graves}(2011)]{GR2011}
Alex {Graves}.
\newblock Practical variational inference for neural networks.
\newblock In J.~Shawe-Taylor, R.~Zemel, P.~Bartlett, F.~Pereira, and K.Q.
  Weinberger, editors, \emph{Advances in Neural Information Processing
  Systems}, volume~24. Curran Associates, Inc., 2011.
\newblock URL
  \url{https://proceedings.neurips.cc/paper/2011/file/7eb3c8be3d411e8ebfab08eba5f49632-Paper.pdf}.

\bibitem[Greenberg et~al.(2019)Greenberg, Nonnenmacher, and
  Macke]{greenberg2019automatic}
David Greenberg, Marcel Nonnenmacher, and Jakob Macke.
\newblock Automatic posterior transformation for likelihood-free inference.
\newblock In \emph{International Conference on Machine Learning}, pages
  2404--2414. PMLR, 2019.

\bibitem[Guo et~al.(2017)Guo, Pleiss, Sun, and Weinberger]{guo2017calibration}
Chuan Guo, Geoff Pleiss, Yu~Sun, and Kilian~Q. Weinberger.
\newblock On calibration of modern neural networks, 2017.

\bibitem[Hainline et~al.(2009)Hainline, Shapley, Kornei, Pettini, Buckley-Geer,
  Allam, and Tucker]{Hainline2009}
Kevin~N. Hainline, Alice~E. Shapley, Katherine~A. Kornei, Max Pettini,
  Elizabeth Buckley-Geer, Sahar~S. Allam, and Douglas~L. Tucker.
\newblock {Rest-Frame Optical Spectra of Three Strongly Lensed Galaxies at
  z{\~{}}2}.
\newblock \emph{arxiv}, jun 2009.
\newblock \doi{10.1088/0004-637X/701/1/52}.

\bibitem[Heinrich et~al.(2024)Heinrich, Mishra-Sharma, Pollard, and
  Windischhofer]{heinrich2024hierarchicalneuralsimulationbasedinference}
Lukas Heinrich, Siddharth Mishra-Sharma, Chris Pollard, and Philipp
  Windischhofer.
\newblock Hierarchical neural simulation-based inference over event ensembles,
  2024.
\newblock URL \url{https://arxiv.org/abs/2306.12584}.

\bibitem[Hermans et~al.(2022)Hermans, Delaunoy, Rozet, Wehenkel, Begy, and
  Louppe]{hermans2022trust}
Joeri Hermans, Arnaud Delaunoy, François Rozet, Antoine Wehenkel, Volodimir
  Begy, and Gilles Louppe.
\newblock A trust crisis in simulation-based inference? your posterior
  approximations can be unfaithful, 2022.

\bibitem[Hezaveh et~al.(2017)Hezaveh, Levasseur, and Marshall]{Hezaveh2017}
Yashar~D. Hezaveh, Laurence~Perreault Levasseur, and Philip~J. Marshall.
\newblock Fast automated analysis of strong gravitational lenses with
  convolutional neural networks.
\newblock \emph{Nature}, 548\penalty0 (7669):\penalty0 555--557, aug 2017.
\newblock \doi{10.1038/nature23463}.
\newblock URL \url{https://doi.org/10.1038%2Fnature23463}.

\bibitem[Hort{\'{u}}a et~al.(2020)Hort{\'{u}}a, Volpi, Marinelli, and
  Malag{\`{o}}]{Hort_a_2020}
H{\'{e} }ctor~J. Hort{\'{u}}a, Riccardo Volpi, Dimitri Marinelli, and Luigi
  Malag{\`{o}}.
\newblock Parameter estimation for the cosmic microwave background with
  bayesian neural networks.
\newblock \emph{\prd}, 102\penalty0 (10), nov 2020.
\newblock \doi{10.1103/physrevd.102.103509}.
\newblock URL \url{https://doi.org/10.1103%2Fphysrevd.102.103509}.

\bibitem[Huang et~al.(2022)Huang, Chen, Chang, Lin, Hsu, Thengane, and
  Lin]{huang2022strong}
Kuan-Wei Huang, Geoff Chih-Fan Chen, Po-Wen Chang, Sheng-Chieh Lin, Chia-Jung
  Hsu, Vishal Thengane, and Joshua Yao-Yu Lin.
\newblock Strong gravitational lensing parameter estimation with vision
  transformer.
\newblock \emph{arXiv preprint arXiv:2210.04143}, 2022.

\bibitem[{Huppenkothen} and {Bachetti}(2021)]{HB2021}
D.~{Huppenkothen} and M.~{Bachetti}.
\newblock {Accurate X-ray Timing in the Presence of Systematic Biases With
  Simulation-Based Inference}.
\newblock \emph{arXiv e-prints}, art. arXiv:2104.03278, April 2021.

\bibitem[{Ivezi{\'c}} et~al.(2019){Ivezi{\'c}}, {Kahn}, {Tyson}, {Abel},
  {Acosta}, {Allsman}, {Alonso}, {AlSayyad}, and et~al.]{IK2019}
{\v{Z}}eljko {Ivezi{\'c}}, Steven~M. {Kahn}, J.~Anthony {Tyson}, Bob {Abel},
  Emily {Acosta}, Robyn {Allsman}, David {Alonso}, Yusra {AlSayyad}, and et~al.
\newblock {LSST}: From science drivers to reference design and anticipated data
  products.
\newblock \emph{ApJ}, 873\penalty0 (2):\penalty0 111, March 2019.
\newblock \doi{10.3847/1538-4357/ab042c}.

\bibitem[{Iwata} et~al.(2009){Iwata}, {Inoue}, {Matsuda}, {Furusawa},
  {Hayashino}, {Kousai}, {Akiyama}, {Yamada}, {Burgarella}, and
  {Deharveng}]{Iwata2008}
I.~{Iwata}, A.~K. {Inoue}, Y.~{Matsuda}, H.~{Furusawa}, T.~{Hayashino},
  K.~{Kousai}, M.~{Akiyama}, T.~{Yamada}, D.~{Burgarella}, and J.~M.
  {Deharveng}.
\newblock {Detections of Lyman Continuum from Star-Forming Galaxies at z
  \raisebox{-0.5ex}\textasciitilde 3 through Subaru/Suprime-Cam Narrow-Band
  Imaging}.
\newblock \emph{\apj}, 692\penalty0 (2):\penalty0 1287--1293, February 2009.
\newblock \doi{10.1088/0004-637X/692/2/1287}.

\bibitem[Jacobs et~al.(2017)Jacobs, Glazebrook, Collett, More, and
  McCarthy]{Jacobs_2017}
C.~Jacobs, K.~Glazebrook, T.~Collett, A.~More, and C.~McCarthy.
\newblock Finding strong lenses in cfhtls using convolutional neural networks.
\newblock \emph{\mnras}, 471\penalty0 (1):\penalty0 167–181, June 2017.
\newblock ISSN 1365-2966.
\newblock \doi{10.1093/mnras/stx1492}.
\newblock URL \url{http://dx.doi.org/10.1093/mnras/stx1492}.

\bibitem[Jacobs et~al.(2019)Jacobs, Collett, Glazebrook, Buckley-Geer, Diehl,
  Lin, McCarthy, Qin, Odden, Escudero, Dial, Yung, Gaitsch, Pellico, Lindgren,
  Abbott, Annis, Avila, Brooks, Burke, Rosell, Kind, Carretero, Costa, Vicente,
  Fosalba, Frieman, García-Bellido, Gaztanaga, Goldstein, Gruen, Gruendl,
  Gschwend, Hollowood, Honscheid, Hoyle, James, Krause, Kuropatkin, Lahav,
  Lima, Maia, Marshall, Miquel, Plazas, Roodman, Sanchez, Scarpine, Serrano,
  Sevilla-Noarbe, Smith, Sobreira, Suchyta, Swanson, Tarle, Vikram, Walker, and
  Zhang]{Jacobs_2019}
C.~Jacobs, T.~Collett, K.~Glazebrook, E.~Buckley-Geer, H.~T. Diehl, H.~Lin,
  C.~McCarthy, A.~K. Qin, C.~Odden, M.~Caso Escudero, P.~Dial, V.~J. Yung,
  S.~Gaitsch, A.~Pellico, K.~A. Lindgren, T.~M.~C. Abbott, J.~Annis, S.~Avila,
  D.~Brooks, D.~L. Burke, A.~Carnero Rosell, M.~Carrasco Kind, J.~Carretero,
  L.~N.~da Costa, J.~De Vicente, P.~Fosalba, J.~Frieman, J.~García-Bellido,
  E.~Gaztanaga, D.~A. Goldstein, D.~Gruen, R.~A. Gruendl, J.~Gschwend, D.~L.
  Hollowood, K.~Honscheid, B.~Hoyle, D.~J. James, E.~Krause, N.~Kuropatkin,
  O.~Lahav, M.~Lima, M.~A.~G. Maia, J.~L. Marshall, R.~Miquel, A.~A. Plazas,
  A.~Roodman, E.~Sanchez, V.~Scarpine, S.~Serrano, I.~Sevilla-Noarbe, M.~Smith,
  F.~Sobreira, E.~Suchyta, M.~E.~C. Swanson, G.~Tarle, V.~Vikram, A.~R. Walker,
  and Y.~Zhang.
\newblock An extended catalog of galaxy–galaxy strong gravitational lenses
  discovered in des using convolutional neural networks.
\newblock \emph{\apjs}, 243\penalty0 (1):\penalty0 17, July 2019.
\newblock ISSN 1538-4365.
\newblock \doi{10.3847/1538-4365/ab26b6}.
\newblock URL \url{http://dx.doi.org/10.3847/1538-4365/ab26b6}.

\bibitem[James et~al.(2013)James, Pettini, Christensen, Auger, Becker, King,
  Quider, Shapley, and Steidel]{James2013}
Bethan~L. James, Max Pettini, Lise Christensen, Matthew~W. Auger, George~D.
  Becker, Lindsay~J. King, Anna~M. Quider, Alice~E. Shapley, and Charles~C.
  Steidel.
\newblock {Testing metallicity indicators at z{\~{}}1.4 with the
  gravitationally lensed galaxy CASSOWARY 20}.
\newblock \emph{\mnras, Volume 440, Issue 2, p.1794-1809}, 440:\penalty0
  1794--1809, nov 2013.
\newblock ISSN 0035-8711.
\newblock \doi{10.1093/mnras/stu287}.

\bibitem[Jarugula et~al.(2024)Jarugula, Nord, Gandrakota, and
  Ciprijanovic]{jarugula2024}
Sreevani Jarugula, Brian~D. Nord, Abhijith Gandrakota, and Alexandra
  Ciprijanovic.
\newblock {Population-level Dark Energy Constraints from Strong Lensing using
  Neural Ratio Estimation}.
\newblock \emph{Accepted to ICML AI for Science Workshop}, 2024.

\bibitem[Jee et~al.(2016)Jee, Komatsu, Suyu, and Huterer]{Jee_2016}
I.~Jee, E.~Komatsu, S.H. Suyu, and D.~Huterer.
\newblock Time-delay cosmography: increased leverage with angular diameter
  distances.
\newblock \emph{\jcap}, 2016\penalty0 (04):\penalty0 031–031, April 2016.
\newblock ISSN 1475-7516.
\newblock \doi{10.1088/1475-7516/2016/04/031}.
\newblock URL \url{http://dx.doi.org/10.1088/1475-7516/2016/04/031}.

\bibitem[Jee et~al.(2019)Jee, Suyu, Komatsu, Fassnacht, Hilbert, and
  Koopmans]{Jee_2019}
Inh Jee, Sherry~H. Suyu, Eiichiro Komatsu, Christopher~D. Fassnacht, Stefan
  Hilbert, and Léon V.~E. Koopmans.
\newblock A measurement of the hubble constant from angular diameter distances
  to two gravitational lenses.
\newblock \emph{Science}, 365\penalty0 (6458):\penalty0 1134–1138, September
  2019.
\newblock ISSN 1095-9203.
\newblock \doi{10.1126/science.aat7371}.
\newblock URL \url{http://dx.doi.org/10.1126/science.aat7371}.

\bibitem[Jennings and Madigan(2017)]{Jennings_2017}
E.~Jennings and M.~Madigan.
\newblock astroabc: An approximate bayesian computation sequential monte carlo
  sampler for cosmological parameter estimation.
\newblock \emph{Astronomy and Computing}, 19:\penalty0 16–22, April 2017.
\newblock ISSN 2213-1337.
\newblock \doi{10.1016/j.ascom.2017.01.001}.
\newblock URL \url{http://dx.doi.org/10.1016/j.ascom.2017.01.001}.

\bibitem[{Jones} et~al.(2013){Jones}, {Ellis}, {Schenker}, and
  {Stark}]{Jones2013a}
Tucker~A. {Jones}, Richard~S. {Ellis}, Matthew~A. {Schenker}, and Daniel~P.
  {Stark}.
\newblock {Keck Spectroscopy of Gravitationally Lensed z
  \raisebox{-0.5ex}\textasciitilde= 4 Galaxies: Improved Constraints on the
  Escape Fraction of Ionizing Photons}.
\newblock \emph{\apj}, 779\penalty0 (1):\penalty0 52, December 2013.
\newblock \doi{10.1088/0004-637X/779/1/52}.

\bibitem[Jospin et~al.(2022)Jospin, Laga, Boussaid, Buntine, and
  Bennamoun]{BNN2022}
Laurent~Valentin Jospin, Hamid Laga, Farid Boussaid, Wray Buntine, and Mohammed
  Bennamoun.
\newblock Hands-on bayesian neural networks—a tutorial for deep learning
  users.
\newblock \emph{IEEE Computational Intelligence Magazine}, 17\penalty0
  (2):\penalty0 29--48, 2022.
\newblock \doi{10.1109/MCI.2022.3155327}.

\bibitem[Kassiola and Kovner(1993)]{Kassiola1993}
Aggeliki Kassiola and Israel Kovner.
\newblock {Elliptic Mass Distributions versus Elliptic Potentials in
  Gravitational Lenses}.
\newblock \emph{\apj}, 417:\penalty0 450, nov 1993.
\newblock ISSN 0004-637X.
\newblock \doi{10.1086/173325}.

\bibitem[Keeton et~al.(1997)Keeton, Kochanek, and Seljak]{Keeton1997}
C.~R. Keeton, C.~S. Kochanek, and U.~Seljak.
\newblock Shear and ellipticity in gravitational lenses.
\newblock \emph{\apj}, 482\penalty0 (2):\penalty0 604--620, jun 1997.
\newblock \doi{10.1086/304172}.
\newblock URL \url{https://doi.org/10.1086%2F304172}.

\bibitem[Kingma and Ba(2017)]{kingma2017adam}
Diederik~P. Kingma and Jimmy Ba.
\newblock Adam: A method for stochastic optimization, 2017.

\bibitem[Knabel et~al.(2023)Knabel, Holwerda, Nightingale, Treu, Bilicki,
  Brough, Driver, Finnerty, Haberzettl, Hegde, Hopkins, Kuijken, Liske,
  Pimblett, Steele, and Wright]{Knabel_2023}
Shawn Knabel, B~W Holwerda, J~Nightingale, T~Treu, M~Bilicki, S~Brough,
  S~Driver, L~Finnerty, L~Haberzettl, S~Hegde, A~M Hopkins, K~Kuijken, J~Liske,
  A~K Pimblett, R~C Steele, and A~H Wright.
\newblock Modelling strong lenses from wide-field ground-based observations in
  {KiDS} and {GAMA}.
\newblock \emph{\mnras}, 520\penalty0 (1):\penalty0 804--827, jan 2023.
\newblock \doi{10.1093/mnras/stad133}.
\newblock URL \url{https://doi.org/10.1093%2Fmnras%2Fstad133}.

\bibitem[{Kneib} et~al.(2011){Kneib}, {Bonnet}, {Golse}, {Sand}, {Jullo}, and
  {Marshall}]{Kneib2011}
Jean-Paul {Kneib}, Henri {Bonnet}, Ghyslain {Golse}, David {Sand}, Eric
  {Jullo}, and Phil {Marshall}.
\newblock {LENSTOOL: A Gravitational Lensing Software for Modeling Mass
  Distribution of Galaxies and Clusters (strong and weak regime)}.
\newblock Astrophysics Source Code Library, record ascl:1102.004, February
  2011.

\bibitem[Kormann et~al.(1994)Kormann, Schneider, and Bartelmann]{Kormann1994}
R.~Kormann, P.~Schneider, and M.~Bartelmann.
\newblock {\aap.}
\newblock \emph{\aap (ISSN 0004-6361), vol. 284, no. 1, p. 285-299},
  284\penalty0 (1):\penalty0 285--299, 1994.

\bibitem[Kull et~al.(2017)Kull, Filho, and Flach]{pmlr-v54-kull17a}
Meelis Kull, Telmo~Silva Filho, and Peter Flach.
\newblock {Beta calibration: a well-founded and easily implemented improvement
  on logistic calibration for binary classifiers}.
\newblock In Aarti Singh and Jerry Zhu, editors, \emph{Proceedings of the 20th
  International Conference on Artificial Intelligence and Statistics},
  volume~54 of \emph{Proceedings of Machine Learning Research}, pages 623--631.
  PMLR, 20--22 Apr 2017.
\newblock URL \url{https://proceedings.mlr.press/v54/kull17a.html}.

\bibitem[Kullback and Leibler(1951)]{KL1951}
Solomon Kullback and Richard~A Leibler.
\newblock On information and sufficiency.
\newblock \emph{The annals of mathematical statistics}, 22\penalty0
  (1):\penalty0 79--86, 1951.

\bibitem[Lefor et~al.(2013)Lefor, Futamase, and Akhlaghi]{Lefor_2013}
Alan~T. Lefor, Toshifumi Futamase, and Mohammad Akhlaghi.
\newblock A systematic review of strong gravitational lens modeling software.
\newblock \emph{New Astron. Rev.}, 57\penalty0 (1–2):\penalty0 1–13, July
  2013.
\newblock ISSN 1387-6473.
\newblock \doi{10.1016/j.newar.2013.05.001}.
\newblock URL \url{http://dx.doi.org/10.1016/j.newar.2013.05.001}.

\bibitem[{Legin} et~al.(2021){Legin}, {Hezaveh}, {Perreault Levasseur}, and
  {Wandelt}]{Legin2021}
Ronan {Legin}, Yashar {Hezaveh}, Laurence {Perreault Levasseur}, and Benjamin
  {Wandelt}.
\newblock {Simulation-Based Inference of Strong Gravitational Lensing
  Parameters}.
\newblock \emph{arXiv e-prints}, art. arXiv:2112.05278, December 2021.

\bibitem[{Legin} et~al.(2023){Legin}, {Hezaveh}, {Perreault-Levasseur}, and
  {Wandelt}]{Legin2023}
Ronan {Legin}, Yashar {Hezaveh}, Laurence {Perreault-Levasseur}, and Benjamin
  {Wandelt}.
\newblock {A Framework for Obtaining Accurate Posteriors of Strong
  Gravitational Lensing Parameters with Flexible Priors and Implicit
  Likelihoods Using Density Estimation}.
\newblock \emph{\apj}, 943\penalty0 (1):\penalty0 4, January 2023.
\newblock \doi{10.3847/1538-4357/aca7c2}.

\bibitem[Lemon(2017)]{Lemon}
Cameron Lemon.
\newblock Gravitationally lensed quasar database, 2017.
\newblock URL \url{https://research.ast.cam.ac.uk/lensedquasars/}.

\bibitem[Levasseur et~al.(2017)Levasseur, Hezaveh, and
  Wechsler]{levasseur2017uncertainties}
Laurence~Perreault Levasseur, Yashar~D Hezaveh, and Risa~H Wechsler.
\newblock Uncertainties in parameters estimated with neural networks:
  Application to strong gravitational lensing.
\newblock \emph{\apjl}, 850\penalty0 (1):\penalty0 L7, 2017.

\bibitem[Lewis(2019)]{lewis2019getdist}
Antony Lewis.
\newblock Getdist: a python package for analysing monte carlo samples, 2019.

\bibitem[Li et~al.(2022)Li, Liu, Yang, Peng, and Zhou]{li2022cnn}
Zewen Li, Fan Liu, Wenjie Yang, Shouheng Peng, and Jun Zhou.
\newblock A survey of convolutional neural networks: Analysis, applications,
  and prospects.
\newblock \emph{IEEE Transactions on Neural Networks and Learning Systems},
  33\penalty0 (12):\penalty0 6999--7019, 2022.
\newblock \doi{10.1109/TNNLS.2021.3084827}.

\bibitem[Lin et~al.(2021)Lin, Trivedi, and Sun]{lin2021locally}
Zhen Lin, Shubhendu Trivedi, and Jimeng Sun.
\newblock Locally valid and discriminative prediction intervals for deep
  learning models, 2021.

\bibitem[{Lueckmann} et~al.(2018){Lueckmann}, {Bassetto}, {Karaletsos}, and
  {Macke}]{NPE-B}
Jan-Matthis {Lueckmann}, Giacomo {Bassetto}, Theofanis {Karaletsos}, and
  Jakob~H. {Macke}.
\newblock {Likelihood-free inference with emulator networks}.
\newblock \emph{arXiv e-prints}, art. arXiv:1805.09294, May 2018.
\newblock \doi{10.48550/arXiv.1805.09294}.

\bibitem[Lueckmann et~al.(2021)Lueckmann, Boelts, Greenberg, Gonçalves, and
  Macke]{lueckmann2021benchmarking}
Jan-Matthis Lueckmann, Jan Boelts, David~S. Greenberg, Pedro~J. Gonçalves, and
  Jakob~H. Macke.
\newblock Benchmarking simulation-based inference, 2021.

\bibitem[Mahler et~al.(2023)Mahler, Jauzac, Richard, Beauchesne, Ebeling,
  Lagattuta, Natarajan, Sharon, Atek, Claeyssens, Clément, Eckert, Edge,
  Kneib, and Niemiec]{mahler2023precision}
Guillaume Mahler, Mathilde Jauzac, Johan Richard, Benjamin Beauchesne, Harald
  Ebeling, David Lagattuta, Priyamvada Natarajan, Keren Sharon, Hakim Atek,
  Adélaïde Claeyssens, Benjamin Clément, Dominique Eckert, Alastair Edge,
  Jean-Paul Kneib, and Anna Niemiec.
\newblock Precision modeling of {JWST}'s first cluster lens
  {SMACSJ}0723.3-7327, 2023.

\bibitem[Mancini et~al.(2022)Mancini, Docherty, Price, and
  McEwen]{mancini2022bayesian}
A.~Spurio Mancini, M.~M. Docherty, M.~A. Price, and J.~D. McEwen.
\newblock Bayesian model comparison for simulation-based inference, 2022.

\bibitem[Modrák et~al.(2023)Modrák, Moon, Kim, Bürkner, Huurre,
  Faltejsková, Gelman, and Vehtari]{stanhyunjimoon}
Martin Modrák, Angie~H. Moon, Shinyoung Kim, Paul Bürkner, Niko Huurre,
  Kateřina Faltejsková, Andrew Gelman, and Aki Vehtari.
\newblock Simulation-based calibration checking for bayesian computation: The
  choice of test quantities shapes sensitivity.
\newblock \emph{Bayesian Analysis}, Advance publication, 2023.
\newblock \doi{10.1214/23-BA1404}.

\bibitem[{Morgan} et~al.(2021){Morgan}, {Nord}, {Birrer}, {Lin}, and
  {Poh}]{MN2021}
Robert {Morgan}, Brian {Nord}, Simon {Birrer}, Joshua {Lin}, and Jason {Poh}.
\newblock {deeplenstronomy: A dataset simulation package for strong
  gravitational lensing}.
\newblock \emph{JOSS}, 6\penalty0 (58):\penalty0 2854, February 2021.
\newblock \doi{10.21105/joss.02854}.

\bibitem[{Moustakas}(2012)]{Moustakas2012}
Leonidas {Moustakas}.
\newblock {The Master Lens Database and The Orphan Lenses Project}.
\newblock HST Proposal ID 12833. Cycle 20, October 2012.

\bibitem[Narayan and Bartelmann(1996)]{Narayan1996}
Ramesh Narayan and Matthias Bartelmann.
\newblock {Lectures on Gravitational Lensing}.
\newblock \emph{eprint arXiv:astro-ph/9606001}, jun 1996.

\bibitem[{Neal}(1996)]{NR1996}
R.~M {Neal}.
\newblock \emph{Bayesian Learning for Neural Networks}, volume 118.
\newblock Springer Science \& Business Media, 1996.
\newblock ISBN 978-0-387-94724-2.

\bibitem[Newman et~al.(2012{\natexlab{a}})Newman, Treu, Ellis, and
  Sand]{Newman2012}
Andrew~B. Newman, Tommaso Treu, Richard~S. Ellis, and David~J. Sand.
\newblock {The Density Profiles of Massive, Relaxed Galaxy Clusters. II.
  Separating Luminous and Dark Matter in Cluster Cores}.
\newblock \emph{\apj}, 765, sep 2012{\natexlab{a}}.
\newblock ISSN 0004-637X.
\newblock \doi{10.1088/0004-637X/765/1/25}.

\bibitem[Newman et~al.(2012{\natexlab{b}})Newman, Treu, Ellis, Sand, Nipoti,
  Richard, and Jullo]{Newman2012a}
Andrew~B. Newman, Tommaso Treu, Richard~S. Ellis, David~J. Sand, Carlo Nipoti,
  Johan Richard, and Eric Jullo.
\newblock {The Density Profiles of Massive, Relaxed Galaxy Clusters. I. The
  Total Density Over Three Decades in Radius}.
\newblock \emph{\apj, Volume 765, Issue 1, article id. 24, 35 pp. (2013).},
  765, sep 2012{\natexlab{b}}.
\newblock ISSN 0004-637X.
\newblock \doi{10.1088/0004-637X/765/1/24}.

\bibitem[{Newman} et~al.(2015){Newman}, {Ellis}, and {Treu}]{Newman2015}
Andrew~B. {Newman}, Richard~S. {Ellis}, and Tommaso {Treu}.
\newblock {Luminous and Dark Matter Profiles from Galaxies to Clusters:
  Bridging the Gap with Group-scale Lenses}.
\newblock \emph{\apj}, 814\penalty0 (1):\penalty0 26, November 2015.
\newblock \doi{10.1088/0004-637X/814/1/26}.

\bibitem[Nightingale(2016)]{Nightingale2016}
James~J.N. Nightingale.
\newblock \emph{{AutoLens: automated modeling of a strong lens's light, mass
  and source}}.
\newblock PhD thesis, School of Physics and Astronomy, Nottingham University,
  University Park, Nottingham, NG7 2RD, UK, dec 2016.
\newblock URL \url{http://eprints.nottingham.ac.uk/38507/}.

\bibitem[Nord et~al.(2016)Nord, Buckley-Geer, Lin, Diehl, Helsby, Kuropatkin,
  Amara, Collett, Allam, Caminha, {De Bom}, Desai, D{\'{u}}met-Montoya,
  Pereira, Finley, Flaugher, Furlanetto, Gaitsch, Gill, Merritt, More, Tucker,
  Saro, Rykoff, Rozo, Birrer, Abdalla, Agnello, Auger, Brunner, {Carrasco
  Kind}, Castander, Cunha, da~Costa, Foley, Gerdes, Glazebrook, Gschwend,
  Hartley, Kessler, Lagattuta, Lewis, Maia, Makler, Menanteau, Niernberg,
  Scolnic, Vieira, Gramillano, Abbott, Banerji, Benoit-L{\'{e}}vy, Brooks,
  Burke, Capozzi, {Carnero Rosell}, Carretero, D'Andrea, Dietrich, Doel,
  Evrard, Frieman, Gaztanaga, Gruen, Honscheid, James, Kuehn, Li, Lima,
  Marshall, Martini, Melchior, Miquel, Neilsen, Nichol, Ogando, Plazas, Romer,
  Sako, Sanchez, Scarpine, Schubnell, Sevilla-Noarbe, Smith, Soares-Santos,
  Sobreira, Suchyta, Swanson, Tarle, Thaler, Walker, Wester, Zhang, and
  Collaboration]{Nord2016}
B.~Nord, E.~Buckley-Geer, H.~Lin, H.~T. Diehl, J.~Helsby, N.~Kuropatkin,
  A.~Amara, T.~Collett, S.~Allam, G.~B. Caminha, C.~{De Bom}, S.~Desai,
  H.~D{\'{u}}met-Montoya, M.~Elidaiana da~S. Pereira, D.~A. Finley,
  B.~Flaugher, C.~Furlanetto, H.~Gaitsch, M.~Gill, K.~W. Merritt, A.~More,
  D.~Tucker, A.~Saro, E.~S. Rykoff, E.~Rozo, S.~Birrer, F.~B. Abdalla,
  A.~Agnello, M.~Auger, R.~J. Brunner, M.~{Carrasco Kind}, F.~J. Castander,
  C.~E. Cunha, L.~N. da~Costa, R.~J. Foley, D.~W. Gerdes, K.~Glazebrook,
  J.~Gschwend, W.~Hartley, R.~Kessler, D.~Lagattuta, G.~Lewis, M.~A.~G. Maia,
  M.~Makler, F.~Menanteau, A.~Niernberg, D.~Scolnic, J.~D. Vieira,
  R.~Gramillano, T.~M.~C. Abbott, M.~Banerji, A.~Benoit-L{\'{e}}vy, D.~Brooks,
  D.~L. Burke, D.~Capozzi, A.~{Carnero Rosell}, J.~Carretero, C.~B. D'Andrea,
  J.~P. Dietrich, P.~Doel, A.~E. Evrard, J.~Frieman, E.~Gaztanaga, D.~Gruen,
  K.~Honscheid, D.~J. James, K.~Kuehn, T.~S. Li, M.~Lima, J.~L. Marshall,
  P.~Martini, P.~Melchior, R.~Miquel, E.~Neilsen, R.~C. Nichol, R.~Ogando,
  A.~A. Plazas, A.~K. Romer, M.~Sako, E.~Sanchez, V.~Scarpine, M.~Schubnell,
  I.~Sevilla-Noarbe, R.~C. Smith, M.~Soares-Santos, F.~Sobreira, E.~Suchyta,
  M.~E.~C. Swanson, G.~Tarle, J.~Thaler, A.~R. Walker, W.~Wester, Y.~Zhang, and
  DES Collaboration.
\newblock {Observation and Confirmation of Six Strong-lensing Systems in the
  Dark Energy Survey Science Verification Data}.
\newblock \emph{\apj}, 827, 2016.
\newblock ISSN 0004-637X.
\newblock \doi{10.3847/0004-637X/827/1/51}.

\bibitem[Oguri and Marshall(2010)]{Oguri2010}
Masamune Oguri and Philip~J. Marshall.
\newblock {Gravitationally lensed quasars and supernovae in future wide-field
  optical imaging surveys}.
\newblock \emph{\mnras, Volume 405, Issue 4, pp. 2579-2593.}, 405:\penalty0
  2579--2593, 2010.
\newblock ISSN 0035-8711.
\newblock \doi{10.1111/J.1365-2966.2010.16639.X}.

\bibitem[Papamakarios and Murray(2018)]{NPE-A}
George Papamakarios and Iain Murray.
\newblock Fast $\epsilon$-free inference of simulation models with bayesian
  conditional density estimation, 2018.

\bibitem[Papamakarios et~al.(2017)Papamakarios, Pavlakou, and
  Murray]{papamakarios2017masked}
George Papamakarios, Theo Pavlakou, and Iain Murray.
\newblock Masked autoregressive flow for density estimation.
\newblock \emph{arXiv preprint arXiv:1705.07057}, 2017.

\bibitem[Papamakarios et~al.(2021)Papamakarios, Nalisnick, Rezende, Mohamed,
  and Lakshminarayanan]{papamakarios2021}
George Papamakarios, Eric Nalisnick, Danilo~Jimenez Rezende, Shakir Mohamed,
  and Balaji Lakshminarayanan.
\newblock Normalizing flows for probabilistic modeling and inference.
\newblock \emph{JMLR}, 22\penalty0 (57):\penalty0 1--64, 2021.
\newblock URL \url{http://jmlr.org/papers/v22/19-1028.html}.

\bibitem[Park et~al.(2021)Park, Wagner-Carena, Birrer, Marshall, Lin, Roodman,
  Collaboration, et~al.]{park2021large}
Ji~Won Park, Sebastian Wagner-Carena, Simon Birrer, Philip~J Marshall, Joshua
  Yao-Yu Lin, Aaron Roodman, LSST Dark Energy~Science Collaboration, et~al.
\newblock Large-scale gravitational lens modeling with bayesian neural networks
  for accurate and precise inference of the hubble constant.
\newblock \emph{\apj}, 910\penalty0 (1):\penalty0 39, 2021.

\bibitem[Pearson et~al.(2021)Pearson, Maresca, Li, and Dye]{pearson2021strong}
James Pearson, Jacob Maresca, Nan Li, and Simon Dye.
\newblock Strong lens modelling: comparing and combining bayesian neural
  networks and parametric profile fitting.
\newblock \emph{\mnras}, 505\penalty0 (3):\penalty0 4362--4382, 2021.

\bibitem[{Picard}(2021)]{Picard2021}
David {Picard}.
\newblock {Torch.manual\_seed(3407) is all you need: On the influence of random
  seeds in deep learning architectures for computer vision}.
\newblock \emph{arXiv e-prints}, art. arXiv:2109.08203, September 2021.
\newblock \doi{10.48550/arXiv.2109.08203}.

\bibitem[Poh et~al.(2022)Poh, Samudre, Ćiprijanović, Nord, Khullar,
  Tanoglidis, and Frieman]{poh2022stronglensingparameterestimation}
Jason Poh, Ashwin Samudre, Aleksandra Ćiprijanović, Brian Nord, Gourav
  Khullar, Dimitrios Tanoglidis, and Joshua~A. Frieman.
\newblock Strong lensing parameter estimation on ground-based imaging data
  using simulation-based inference, 2022.
\newblock URL \url{https://arxiv.org/abs/2211.05836}.

\bibitem[Prangle(2015)]{prangle2015summary}
Dennis Prangle.
\newblock Summary statistics in approximate bayesian computation, 2015.

\bibitem[Quider et~al.(2009)Quider, Pettini, Shapley, and Steidel]{Quider2009}
Anna~M. Quider, Max Pettini, Alice~E. Shapley, and Charles~C. Steidel.
\newblock {The Ultraviolet Spectrum of the Gravitationally Lensed Galaxy `The
  Cosmic Horseshoe': A Close-up of a Star-forming Galaxy at z = 2}.
\newblock \emph{\mnras, Volume 398, Issue 3, pp. 1263-1278.}, 398:\penalty0
  1263--1278, jun 2009.
\newblock ISSN 0035-8711.
\newblock \doi{10.1111/j.1365-2966.2009.15234.x}.

\bibitem[Racca et~al.(2016)Racca, Laureijs, Stagnaro, Salvignol,
  Lorenzo~Alvarez, Saavedra~Criado, Gaspar~Venancio, Short, Strada, Bönke,
  Colombo, Calvi, Maiorano, Piersanti, Prezelus, Rosato, Pinel, Rozemeijer,
  Lesna, Musi, Sias, Anselmi, Cazaubiel, Vaillon, Mellier, Amiaux, Berthé,
  Sauvage, Azzollini, Cropper, Pottinger, Jahnke, Ealet, Maciaszek, Pasian,
  Zacchei, Scaramella, Hoar, Kohley, Vavrek, Rudolph, and Schmidt]{Racca_2016}
Giuseppe~D. Racca, René Laureijs, Luca Stagnaro, Jean-Christophe Salvignol,
  José Lorenzo~Alvarez, Gonzalo Saavedra~Criado, Luis Gaspar~Venancio, Alex
  Short, Paolo Strada, Tobias Bönke, Cyril Colombo, Adriano Calvi, Elena
  Maiorano, Osvaldo Piersanti, Sylvain Prezelus, Pierluigi Rosato, Jacques
  Pinel, Hans Rozemeijer, Valentina Lesna, Paolo Musi, Marco Sias, Alberto
  Anselmi, Vincent Cazaubiel, Ludovic Vaillon, Yannick Mellier, Jérôme
  Amiaux, Michel Berthé, Marc Sauvage, Ruyman Azzollini, Mark Cropper, Sabrina
  Pottinger, Knud Jahnke, Anne Ealet, Thierry Maciaszek, Fabio Pasian, Andrea
  Zacchei, Roberto Scaramella, John Hoar, Ralf Kohley, Roland Vavrek, Andreas
  Rudolph, and Micha Schmidt.
\newblock The euclid mission design.
\newblock In Howard~A. MacEwen, Giovanni~G. Fazio, Makenzie Lystrup, Natalie
  Batalha, Nicholas Siegler, and Edward~C. Tong, editors, \emph{Space
  Telescopes and Instrumentation 2016: Optical, Infrared, and Millimeter Wave}.
  SPIE, July 2016.
\newblock \doi{10.1117/12.2230762}.
\newblock URL \url{http://dx.doi.org/10.1117/12.2230762}.

\bibitem[Rezende and Mohamed(2015)]{rezende2015variational}
Danilo Rezende and Shakir Mohamed.
\newblock Variational inference with normalizing flows.
\newblock In \emph{International conference on machine learning}, pages
  1530--1538. PMLR, 2015.

\bibitem[Rodrigues et~al.(2021)Rodrigues, Moreau, Louppe, and
  Gramfort]{rodrigues2021hnpeleveragingglobalparameters}
Pedro L.~C. Rodrigues, Thomas Moreau, Gilles Louppe, and Alexandre Gramfort.
\newblock Hnpe: Leveraging global parameters for neural posterior estimation,
  2021.
\newblock URL \url{https://arxiv.org/abs/2102.06477}.

\bibitem[Rubin(1984)]{Rubin1984}
Donald~B. Rubin.
\newblock Bayesianly justifiable and relevant frequency calculations for the
  applied statistician.
\newblock \emph{The Annals of Statistics}, 12\penalty0 (4):\penalty0
  1151--1172, 1984.
\newblock ISSN 00905364, 21688966.
\newblock URL \url{http://www.jstor.org/stable/2240995}.

\bibitem[Ruff et~al.(2011)Ruff, Gavazzi, Marshall, Treu, Auger, and
  Brault]{Ruff_2011}
Andrea~J. Ruff, Raphaël Gavazzi, Philip~J. Marshall, Tommaso Treu, Matthew~W.
  Auger, and Florence Brault.
\newblock {THE} {SL}2s {GALAXY}-{SCALE} {LENS} {SAMPLE}. {II}. {COSMIC}
  {EVOLUTION} {OF} {DARK} {AND} {LUMINOUS} {MASS} {IN} {EARLY}-{TYPE}
  {GALAXIES}.
\newblock \emph{\apj}, 727\penalty0 (2):\penalty0 96, jan 2011.
\newblock \doi{10.1088/0004-637x/727/2/96}.
\newblock URL \url{https://doi.org/10.1088%2F0004-637x%2F727%2F2%2F96}.

\bibitem[Sanderson et~al.(2024)Sanderson, Hickox, Hirata, Holman, Lu, and
  Villar]{sanderson2024recommendationsearlydefinitionscience}
Robyn~E. Sanderson, Ryan Hickox, Christopher~M. Hirata, Matthew~J. Holman,
  Jessica~R. Lu, and Ashley Villar.
\newblock Recommendations for early definition science with the nancy grace
  roman space telescope, 2024.
\newblock URL \url{https://arxiv.org/abs/2404.14342}.

\bibitem[{Schneider} et~al.(2006){Schneider}, {Kochanek}, and
  {Wambsganss}]{Kochanek2004}
Peter {Schneider}, Christopher~S. {Kochanek}, and Joachim {Wambsganss}.
\newblock \emph{{Gravitational Lensing: Strong, Weak and Micro}}.
\newblock Springer Berlin, Heidelberg, 2006.
\newblock \doi{https://doi.org/10.1007/978-3-540-30310-7}.

\bibitem[Scot(1992)]{scot1992multivariate}
David~W Scot.
\newblock Multivariate density estimation, 1992.

\bibitem[{Sersic}(1968)]{Sersic1968}
Jose~Luis {Sersic}.
\newblock \emph{{Atlas de Galaxias Australes}}.
\newblock Cordoba, Argentina: Observatorio Astronomico, 1968.

\bibitem[{Shajib} et~al.(2020){Shajib}, {Birrer}, {Treu}, {Agnello},
  {Buckley-Geer}, {Chan}, {Christensen}, {Lemon}, and et~al.]{Shajib2020}
A.~J. {Shajib}, S.~{Birrer}, T.~{Treu}, A.~{Agnello}, E.~J. {Buckley-Geer},
  J.~H.~H. {Chan}, L.~{Christensen}, C.~{Lemon}, and et~al.
\newblock {STRIDES: a 3.9 per cent measurement of the Hubble constant from the
  strong lens system DES J0408-5354}.
\newblock \emph{\mnras}, 494\penalty0 (4):\penalty0 6072--6102, June 2020.
\newblock \doi{10.1093/mnras/staa828}.

\bibitem[Shridhar et~al.(2019)Shridhar, Laumann, and Liwicki]{SL2019}
Kumar Shridhar, Felix Laumann, and Marcus Liwicki.
\newblock A comprehensive guide to bayesian convolutional neural network with
  variational inference.
\newblock \emph{arXiv preprint arXiv:1901.02731}, 2019.

\bibitem[Smit et~al.(2017)Smit, Swinbank, Massey, Richard, Smail, and
  Kneib]{Smit2017}
Renske Smit, A.~M. Swinbank, Richard Massey, Johan Richard, Ian Smail, and
  J.~P. Kneib.
\newblock {A gravitationally-boosted MUSE survey for emission-line galaxies at
  z{\textgreater}{\~{}}5 behind the massive cluster RCS 0224}.
\newblock \emph{\mnras, Volume 467, Issue 3, p.3306-3323}, 467:\penalty0
  3306--3323, jan 2017.
\newblock ISSN 0035-8711.
\newblock \doi{10.1093/mnras/stx245}.

\bibitem[Suyu et~al.(2010)Suyu, Marshall, Auger, Hilbert, Blandford, Koopmans,
  Fassnacht, and Treu]{Suyu2010}
S.~H. Suyu, P.~J. Marshall, M.~W. Auger, S.~Hilbert, R.~D. Blandford, L.~V.~E.
  Koopmans, C.~D. Fassnacht, and T.~Treu.
\newblock {Dissecting the Gravitational lens B1608+656. II. Precision
  Measurements of the Hubble Constant, Spatial Curvature, and the Dark Energy
  Equation of State}.
\newblock \emph{\apj}, 711:\penalty0 201--221, 2010.
\newblock ISSN 0004-637X.
\newblock \doi{10.1088/0004-637X/711/1/201}.

\bibitem[Suyu et~al.(2013)Suyu, Auger, Hilbert, Marshall, Tewes, Treu,
  Fassnacht, Koopmans, Sluse, Blandford, Courbin, and Meylan]{Suyu2013}
S.~H. Suyu, M.~W. Auger, S.~Hilbert, P.~J. Marshall, M.~Tewes, T.~Treu, C.~D.
  Fassnacht, L.~V.~E. Koopmans, D.~Sluse, R.~D. Blandford, F.~Courbin, and
  G.~Meylan.
\newblock {Two Accurate Time-delay Distances from Strong Lensing: Implications
  for Cosmology}.
\newblock \emph{\apj, Volume 766, Issue 2, article id. 70, 19 pp. (2013).},
  766, 2013.
\newblock ISSN 0004-637X.
\newblock \doi{10.1088/0004-637X/766/2/70}.

\bibitem[Swierc et~al.(2024)Swierc, Tamargo-Arizmendi, Ćiprijanović, and
  Nord]{swierc2024domainadaptiveneuralposteriorestimation}
Paxson Swierc, Marcos Tamargo-Arizmendi, Aleksandra Ćiprijanović, and
  Brian~D. Nord.
\newblock Domain-adaptive neural posterior estimation for strong gravitational
  lens analysis, 2024.
\newblock URL \url{https://arxiv.org/abs/2410.16347}.

\bibitem[Talts et~al.(2018)Talts, Betancourt, Simpson, Vehtari, and
  Gelman]{Talts2018}
Sean Talts, Michael Betancourt, Daniel Simpson, Aki Vehtari, and Andrew Gelman.
\newblock Validating bayesian inference algorithms with simulation-based
  calibration, 2018.
\newblock URL \url{https://arxiv.org/abs/1804.06788}.

\bibitem[Tan et~al.(2023)Tan, Shajib, Birrer, Sonnenfeld, Treu, Wells,
  Williams, Buckley-Geer, Drlica-Wagner, and Frieman]{tan2023project}
Chin~Yi Tan, Anowar~J. Shajib, Simon Birrer, Alessandro Sonnenfeld, Tommaso
  Treu, Patrick Wells, Devon Williams, Elizabeth~J. Buckley-Geer, Alex
  Drlica-Wagner, and Joshua Frieman.
\newblock Project dinos i: A joint lensing-dynamics constraint on the deviation
  from the power law in the mass profile of massive ellipticals, 2023.

\bibitem[{Tanoglidis} et~al.(2022){Tanoglidis}, {{\'C}iprijanovi{\'c}}, and
  {Drlica-Wagner}]{TC2022}
Dimitrios {Tanoglidis}, Aleksandra {{\'C}iprijanovi{\'c}}, and Alex
  {Drlica-Wagner}.
\newblock {Inferring Structural Parameters of Low-Surface-Brightness-Galaxies
  with Uncertainty Quantification using Bayesian Neural Networks}.
\newblock \emph{arXiv e-prints}, art. arXiv:2207.03471, July 2022.

\bibitem[Tavar{\'e} et~al.(1997)Tavar{\'e}, Balding, Griffiths, and
  Donnelly]{tavare1997inferring}
Simon Tavar{\'e}, David~J Balding, Robert~C Griffiths, and Peter Donnelly.
\newblock Inferring coalescence times from dna sequence data.
\newblock \emph{Genetics}, 145\penalty0 (2):\penalty0 505--518, 1997.

\bibitem[Tejero-Cantero et~al.(2020)Tejero-Cantero, Boelts, Deistler,
  Lueckmann, Durkan, Gonçalves, Greenberg, and Macke]{tejero-cantero2020sbi}
Alvaro Tejero-Cantero, Jan Boelts, Michael Deistler, Jan-Matthis Lueckmann,
  Conor Durkan, Pedro~J. Gonçalves, David~S. Greenberg, and Jakob~H. Macke.
\newblock sbi: A toolkit for simulation-based inference.
\newblock \emph{JOSS}, 5\penalty0 (52):\penalty0 2505, 2020.
\newblock \doi{10.21105/joss.02505}.
\newblock URL \url{https://doi.org/10.21105/joss.02505}.

\bibitem[{Treu}(2010)]{2010ARA&A..48...87T}
Tommaso {Treu}.
\newblock {Strong Lensing by Galaxies}.
\newblock \emph{\araa}, 48:\penalty0 87--125, September 2010.
\newblock \doi{10.1146/annurev-astro-081309-130924}.

\bibitem[Troja et~al.(2022)Troja, Tutusaus, and Sorce]{troja2022euclidnutshell}
Antonino Troja, Isaac Tutusaus, and Jenny~G. Sorce.
\newblock Euclid in a nutshell, 2022.
\newblock URL \url{https://arxiv.org/abs/2211.09668}.

\bibitem[{Unruh} et~al.(2017){Unruh}, {Schneider}, and {Sluse}]{Unruh2017}
Sandra {Unruh}, Peter {Schneider}, and Dominique {Sluse}.
\newblock {Ambiguities in gravitational lens models: the density field from the
  source position transformation}.
\newblock \emph{\aap}, 601:\penalty0 A77, May 2017.
\newblock \doi{10.1051/0004-6361/201629048}.

\bibitem[Uria et~al.(2016)Uria, C{\^o}t{\'e}, Gregor, Murray, and
  Larochelle]{uria2016neural}
Benigno Uria, Marc-Alexandre C{\^o}t{\'e}, Karol Gregor, Iain Murray, and Hugo
  Larochelle.
\newblock Neural autoregressive distribution estimation.
\newblock \emph{JMLR}, 17\penalty0 (1):\penalty0 7184--7220, 2016.

\bibitem[Vovk et~al.(1999)Vovk, Gammerman, and Saunders]{vovk1999}
V.~Vovk, A.~Gammerman, and C.~Saunders.
\newblock Machine-learning applications of algorithmic randomness.
\newblock In \emph{Sixteenth International Conference on Machine Learning
  (ICML-1999) (01/01/99)}, pages 444--453, 1999.
\newblock URL \url{https://eprints.soton.ac.uk/258960/}.

\bibitem[Wagner-Carena et~al.(2021)Wagner-Carena, Park, Birrer, Marshall,
  Roodman, Wechsler, Collaboration, et~al.]{wagner2021hierarchical}
Sebastian Wagner-Carena, Ji~Won Park, Simon Birrer, Philip~J Marshall, Aaron
  Roodman, Risa~H Wechsler, LSST Dark Energy~Science Collaboration, et~al.
\newblock Hierarchical inference with bayesian neural networks: An application
  to strong gravitational lensing.
\newblock \emph{\apj}, 909\penalty0 (2):\penalty0 187, 2021.

\bibitem[{Wagner-Carena} et~al.(2023){Wagner-Carena}, {Aalbers}, {Birrer},
  {Nadler}, {Darragh-Ford}, {Marshall}, and {Wechsler}]{WagnerCarena2022}
Sebastian {Wagner-Carena}, Jelle {Aalbers}, Simon {Birrer}, Ethan~O. {Nadler},
  Elise {Darragh-Ford}, Philip~J. {Marshall}, and Risa~H. {Wechsler}.
\newblock {From Images to Dark Matter: End-to-end Inference of Substructure
  from Hundreds of Strong Gravitational Lenses}.
\newblock \emph{\apj}, 942\penalty0 (2):\penalty0 75, January 2023.
\newblock \doi{10.3847/1538-4357/aca525}.

\bibitem[{Walmsley} et~al.(2022){Walmsley}, {Lintott}, {G{\'e}ron}, {Kruk},
  {Krawczyk}, {Willett}, {Bamford}, {Kelvin}, {Fortson}, {Gal}, {Keel},
  {Masters}, {Mehta}, {Simmons}, {Smethurst}, {Smith}, {Baeten}, and
  {Macmillan}]{WL2022}
Mike {Walmsley}, Chris {Lintott}, Tobias {G{\'e}ron}, Sandor {Kruk}, Coleman
  {Krawczyk}, Kyle~W. {Willett}, Steven {Bamford}, Lee~S. {Kelvin}, Lucy
  {Fortson}, Yarin {Gal}, William {Keel}, Karen~L. {Masters}, Vihang {Mehta},
  Brooke~D. {Simmons}, Rebecca {Smethurst}, Lewis {Smith}, Elisabeth~M.
  {Baeten}, and Christine {Macmillan}.
\newblock {Galaxy Zoo DECaLS: Detailed visual morphology measurements from
  volunteers and deep learning for 314 000 galaxies}.
\newblock \emph{\mnras}, 509\penalty0 (3):\penalty0 3966--3988, January 2022.
\newblock \doi{10.1093/mnras/stab2093}.

\bibitem[Weiner et~al.(2020)Weiner, Serjeant, and Sedgwick]{Weiner_2020}
Charles Weiner, Stephen Serjeant, and Chris Sedgwick.
\newblock Predictions for strong-lens detections with the nancy grace roman
  space telescope.
\newblock \emph{Research Notes of the AAS}, 4\penalty0 (10):\penalty0 190,
  October 2020.
\newblock ISSN 2515-5172.
\newblock \doi{10.3847/2515-5172/abc4ea}.
\newblock URL \url{http://dx.doi.org/10.3847/2515-5172/abc4ea}.

\bibitem[Wen et~al.(2018)Wen, Vicol, Ba, Tran, and Grosse]{wen2018flipout}
Yeming Wen, Paul Vicol, Jimmy Ba, Dustin Tran, and Roger Grosse.
\newblock Flipout: Efficient pseudo-independent weight perturbations on
  mini-batches.
\newblock \emph{arXiv preprint arXiv:1803.04386}, 2018.

\bibitem[Wenzl et~al.(2022)Wenzl, Doux, Heinrich, Bean, Jain, Doré, Eifler,
  and Fang]{Wenzl_2022}
Lukas Wenzl, Cyrille Doux, Chen Heinrich, Rachel Bean, Bhuvnesh Jain, Olivier
  Doré, Tim Eifler, and Xiao Fang.
\newblock Cosmology with the roman space telescope – synergies with cmb
  lensing.
\newblock \emph{\mnras}, 512\penalty0 (4):\penalty0 5311–5328, March 2022.
\newblock ISSN 1365-2966.
\newblock \doi{10.1093/mnras/stac790}.
\newblock URL \url{http://dx.doi.org/10.1093/mnras/stac790}.

\bibitem[Yuan and Kewley(2009)]{Yuan2009}
T.~T. Yuan and L.~J. Kewley.
\newblock {First direct metallicity measurement of a lensed star-forming galaxy
  at z=1.7}.
\newblock \emph{\apjl, Volume 699, Issue 2, pp. L161-L164 (2009).},
  699:\penalty0 L161--L164, jun 2009.
\newblock ISSN 0004-637X.
\newblock \doi{10.1088/0004-637X/699/2/L161}.

\bibitem[Zaborowski et~al.(2023)Zaborowski, Drlica-Wagner, Ashmead, Wu, Morgan,
  Bom, Shajib, Birrer, Cerny, Buckley-Geer, Mutlu-Pakdil, Ferguson, Glazebrook,
  Lozano, Gordon, Martinez, Manwadkar, O’Donnell, Poh, Riley, Sakowska,
  Santana-Silva, Santiago, Sluse, Tan, Tollerud, Verma, Carballo-Bello, Choi,
  James, Kuropatkin, Martínez-Vázquez, Nidever, Castellon, Noël, Olsen,
  Pace, Mau, Yanny, Zenteno, Abbott, Aguena, Alves, Andrade-Oliveira, Bocquet,
  Brooks, Burke, Carnero~Rosell, Carrasco~Kind, Carretero, Castander,
  Conselice, Costanzi, Pereira, De~Vicente, Desai, Dietrich, Doel, Everett,
  Ferrero, Flaugher, Friedel, Frieman, García-Bellido, Gruen, Gruendl,
  Gutierrez, Hinton, Hollowood, Honscheid, Kuehn, Lin, Marshall, Melchior,
  Mena-Fernández, Menanteau, Miquel, Palmese, Paz-Chinchón, Pieres, Malagón,
  Prat, Rodriguez-Monroy, Romer, Sanchez, Scarpine, Sevilla-Noarbe, Smith,
  Suchyta, To, and Weaverdyck]{Zaborowski_2023}
E.~A. Zaborowski, A.~Drlica-Wagner, F.~Ashmead, J.~F. Wu, R.~Morgan, C.~R. Bom,
  A.~J. Shajib, S.~Birrer, W.~Cerny, E.~J. Buckley-Geer, B.~Mutlu-Pakdil, P.~S.
  Ferguson, K.~Glazebrook, S.~J.~Gonzalez Lozano, Y.~Gordon, M.~Martinez,
  V.~Manwadkar, J.~O’Donnell, J.~Poh, A.~Riley, J.~D. Sakowska,
  L.~Santana-Silva, B.~X. Santiago, D.~Sluse, C.~Y. Tan, E.~J. Tollerud,
  A.~Verma, J.~A. Carballo-Bello, Y.~Choi, D.~J. James, N.~Kuropatkin, C.~E.
  Martínez-Vázquez, D.~L. Nidever, J.~L.~Nilo Castellon, N.~E.~D. Noël,
  K.~A.~G. Olsen, A.~B. Pace, S.~Mau, B.~Yanny, A.~Zenteno, T.~M.~C. Abbott,
  M.~Aguena, O.~Alves, F.~Andrade-Oliveira, S.~Bocquet, D.~Brooks, D.~L. Burke,
  A.~Carnero~Rosell, M.~Carrasco~Kind, J.~Carretero, F.~J. Castander, C.~J.
  Conselice, M.~Costanzi, M.~E.~S. Pereira, J.~De~Vicente, S.~Desai, J.~P.
  Dietrich, P.~Doel, S.~Everett, I.~Ferrero, B.~Flaugher, D.~Friedel,
  J.~Frieman, J.~García-Bellido, D.~Gruen, R.~A. Gruendl, G.~Gutierrez, S.~R.
  Hinton, D.~L. Hollowood, K.~Honscheid, K.~Kuehn, H.~Lin, J.~L. Marshall,
  P.~Melchior, J.~Mena-Fernández, F.~Menanteau, R.~Miquel, A.~Palmese,
  F.~Paz-Chinchón, A.~Pieres, A.~A.~Plazas Malagón, J.~Prat,
  M.~Rodriguez-Monroy, A.~K. Romer, E.~Sanchez, V.~Scarpine, I.~Sevilla-Noarbe,
  M.~Smith, E.~Suchyta, C.~To, and N.~Weaverdyck.
\newblock Identification of galaxy–galaxy strong lens candidates in the decam
  local volume exploration survey using machine learning.
\newblock \emph{\apj}, 954\penalty0 (1):\penalty0 68, August 2023.
\newblock ISSN 1538-4357.
\newblock \doi{10.3847/1538-4357/ace4ba}.
\newblock URL \url{http://dx.doi.org/10.3847/1538-4357/ace4ba}.

\bibitem[{Zeiler}(2012)]{Z2012}
Matthew~D. {Zeiler}.
\newblock {ADADELTA: An Adaptive Learning Rate Method}.
\newblock \emph{arXiv e-prints}, art. arXiv:1212.5701, December 2012.

\end{thebibliography}

\end{document}